\documentclass[pr1,onecolumn,superscriptaddress,shdowpacs,amsmath,amssymb]{revtex4-1}
\usepackage{graphicx}
\usepackage{indentfirst}
\usepackage{psfrag}
\usepackage{epsfig}
\usepackage{amsmath}
\usepackage{amssymb}
\usepackage{bm}
\usepackage{color}
\usepackage{mathrsfs}
\newcommand{\be}{\begin{eqnarray}}
\newcommand{\ee}{\end{eqnarray}}

\newcommand{\rr}{\color{red}}
\newcommand{\bb}{\color{black}}
\newcommand{\ii}{\rr{\rm i}}
\newcommand{\tao}{{\rr\tau}}

\begin{document}

\title{The complex Maxwell stress tensor theorem: The imaginary  stress tensor and the reactive strength of orbital momentum. A novel scenery underlying   optical forces}    
\author{Manuel Nieto-Vesperinas}
\email{mnieto@icmm.csic.es}
\affiliation{Instituto de Ciencia de Materiales de Madrid, Consejo Superior de
Investigaciones Cient\'{i}ficas.\\
 Campus de Cantoblanco, Madrid 28049, Spain. }
\author{Xiaohao Xu}
\email{xuxhao\_dakuren@163.com}
\affiliation{\rr State Key Laboratory of Transient Optics and Photonics, Xi’an Institute of Optics and Precision Mechanics, Chinese Academy of Sciences, Xi’an 710119, China\bb}
\affiliation{Institute of Nanophotonics, Jinan University, Guangzhou 511443, China}

\begin{abstract}
We uncover the  existence  of a universal phenomenon   concerning  the  electromagnetic optical force exerted by light or other electromagnetic waves on a distribution of charges and currents in general, and of particles in particular. This conveys the appearence of underlying reactive quantities that  hinder radiation pressure and currently observed time-averaged  forces. This   constitutes a novel paradigm of the mechanical efficiency of light on matter, and  completes the  landscape of the optical, and generally electromagnetic, force in photonics and classical electrodynamics; widening our understanding in the design of both illumination and particles in optical manipulation without the need of increasing the illuminating power, and thus lowering dissipation and heating. We show that this may be accomplished through the minimization of what we establish as the reactive strength of orbital (or canonical)  momentum, which plays against the optical force  a role analogous to that of the reactive power versus the radiation efficiency of an antenna. This long time overlooked quantity, important for current progress of optical manipulation, and  that stems from the complex Maxwell theorem of conservation of complex momentum that we put forward, as well as its alternating flow associated to the imaginary part of the complex Maxwell stress tensor, conform the imaginary Lorentz force that we introduce in this work, and that like the reactive strength of orbital momentum, is \color{red} antagonistic \color{black}  to the well-known time-averaged force; thus making this reactive Lorentz force  indirectly observable near wavelengths at which the time-averaged force is  lowered.
\end{abstract}
\maketitle
 
\section{Introduction}
The Maxwell  stress  tensor in whose terms the conservation of linear and angular momentum is expressed \cite{jackson,griffiths}, is at the root of electromagnetic forces in general and optical manipulation in particular \cite{ashkin1,ashkin2,patric2000,chaumet1,nieminen,qiu1}. When the fields are characterized by complex functions, this conservation law is obtained from the real parts which yield time-averaged, or real, Lorentz forces (RLF) and torques. This is extensively employed, in particular, for time-harmonic (i.e. monochromatic) fields \cite{qiu1,qiu2,qiu3,chaumet1}.

In this context, it is well-known that the  RLF on a volume $V_0$ of charges and currents is given by the momentum flux whose density is  the real part of the Maxwell stress tensor (RMST) across any contour $\partial V$ enclosing $V_0$. In consequence this RLF is the flow, characterized by the  RMST,  into the surface of a sphere in the far-field, i.e. in the radiation zone of $V_0$ and, as such, it may be considered a “radiation force”.

In this paper we demonstrate that this theory through the RMST describes only half the physics of the electromagnetic optical force. The other half, so far ignored and that we uncover here by establishing the complex Maxwell stress tensor theorem, is characterized by the imaginary part of the complex Maxwell stress tensor (CMST), related to the exchange of reactive  (i.e. imaginary Poynting) momentum (IPM) \cite{nietoPRR}, and acquires importance as optical manipulation of matter  progresses and expands its scope incorporating reactive concepts  \cite{nietoPRR, bliokhreac,kamenetskii}. The imaginary Maxwell stress tensor (IMST) builds-up in and around $V_0$ what we find and put forward here: the reactive strength of orbital (or canonical) momentum (ROM); so that this storage of ROM  contributes to the imaginary Lorentz force (ILF) on $V_0$ which, as we shall show, may also be envisaged as a reactive strength of Poynting momentum. This reactive force is not observable on time averaging since its net value is zero, but it exists instantaneously due to the transfer of  the reactive momentum, which alternates with time, between the wave and the body.

Hence, the ILF is a basically fundamental dynamic phenomenon, inherent to the emergence of  electromagnetic optical forces, being also associated to the appearance of reactive energy, reactive work, and reactive helicity \cite{nietoPRR, bliokhreac,kamenetskii}. The former having been for many years a well-known workhorse in the design of RF antennas \cite{collin,geyi}, and recently studied in nano and micro-antennas \cite{nietoPRR,bliokhreac,kamenetskii,ziolkowski}.
\rr Therefore, like in RF antenna design  one aims to diminish the reactive power and reactive work to increase the radiation efficiency, the theory put forward in this work constitutes a tool to act on the ROM and ILF in order to optimize a desired radiation pressure in optical manipulation.  \bb

Consequently, here we show that, as such, the ILF and ROM play an antagonic role with respect to the standard RLF, so that a strong ILF, and thus  a large ROM storage, amounts to a  loss of radiative force,  RLF, and vice-versa.  This makes the ROM and ILF  indirectly observable.

 It is somewhat striking that having existed for decades the complex Poynting theorem and its consequent reactive quantities: the IPM, reactive work and reactive energy; to our knowledge, the complex Maxwell stress tensor theorem, and the reactive entities it conveys, had not been established. This might be due to the practical difficulties  involved in optical manipulation. However the fast advances and present maturity of the optical handling of matter, now warrant their formulation.

In our view, this novel scenario completes  an interpretative panorama of forces  in the science of light and classical electrodynamics,  e.g. in the design of particles and of structured beam illumination that, as done with  their radiative power and emitted field helicity \cite{nietoPRR,nietoRHELT}, the efficiency of the time-averaged force, i.e. of the RLF acting on them, be optimized  by either enhancing  or weakening it.

\rr
The outline of this paper is as follows: 

First, we establish the complex stress tensor theorem in an embedding vacuum or air; defining the ILF, IMST and ROM for general time-dependent light fields, and discussing their respective physical meaning. Then we address these concepts for time-harmonic (or monochromatic) wavefields to which the rest of this work is devoted. 

After, we shall characterize the IMST flow in terms of the magnetic and electric spin momenta of the total (i.e. incident plus scattered) field, which we introduce from first Lagrangian principles. We then express the ILF by what we put forward as the {\it reactive strength of Poynting momentum}, obtained from the electric and magnetic spin and orbital momenta, while we show that the RLF may be written as the sum of the imaginary spin and orbital momenta. 

After demonstrating the near-field nature of the IMST, we consider the extensively studied case of a dipolar particle, deriving the alternating imaginary momentum flow IMST across the surface of a surrounding sphere in the near-field, along with the ROM and ILF; showing that, in contrast with the field (i.e. Poynting) momentum flow RMST, and the RLF, these quantities  depend on the sphere radius.  

Examples are given, comparing numerical results and theory, for three dipolar archetypical particles: a low index dielectric,  a high index magnetoelectric one,  and a plasmonic sphere. It is also shown that an  heuristic direct derivation of the ILF, analogous to that employed in \cite{patric2000} for the RLF, works well for low index dielectric particles, but not for resonant ones, which first require the use of the above mentioned IMST calculation.

Finally, a recapitulation of the CMST and the reactive force is given when one considers a homogeneous, linear, isotropic dielectric as the embedding medium; establishing the time-dependent CMST  theorem, reactive force and ROM  according to whether one chooses a complex value generalization of the Minkowski or  Abraham field momentum. Demonstrating that in the case of time-harmonic waves,  the reactive force, like the time-averaged force, is independent of the choice of a Minkowski or an Abraham complex Poynting momentum.
\bb

\section{Time-dependent fields: The complex Maxwell stress tensor theorem. The imaginary  stress tensor and   the reactive strength of orbital  momentum}
\textcolor{red}{In our study we use Gaussian units and assume a homogeneous  medium with relative permittivity and permeability: $\epsilon = \mu = 1$, (i.e. vacuum), embedding the illuminated body.} A  convenient way to frame the following theory is to start with analytic signals \cite{born,mandel} as done with the complex Poynting theorem \cite{kaiser}. These are
 $\bm{\mathcal E}({\bf r},\tao)$,
 $\bm{\mathcal H}({\bf r},\tao)$ and   $\bm{\mathcal J}({\bf r},\tao)$, associated to the real vectors $\bm{\mathfrak E}({\bf r},t)$, $\bm{\mathfrak H}({\bf r},t)$ and $\bm{\mathfrak  J}({\bf r},t)$ which are  analytically continued  into the lower half complex plane $\tao=t-\rr{\rm i}\bb s$. Generically denoting each of these analytic functions  as $\bm{\mathcal V}({\bf r},t,s)$, they are expressed by the Fourier integral \cite{born,mandel}:
\be
\bm{\mathcal V}({\bf r},t,s)\equiv\bm{\mathcal V}({\bf r},\tao)=\int_{0}^{\infty}d\omega \exp(-\rr{\rm i}\bb \omega \tao)
\bm{\mathfrak V}_{\omega}({\bf r},\omega). \,\,\,\label{tfor21a}
\ee
$\bm{\mathfrak V}_{\omega}({\bf r},\omega)$ being the $\omega$-Fourier spectrum of the real function  $\bm{\mathfrak V}({\bf r},t)$ which generically denotes either  $\bm{\mathfrak E}({\bf r},t)$,  $\bm{\mathfrak  B}({\bf r},t)$ or  $\bm{\mathfrak  J}({\bf r},t)$.  Then,  (\ref{tfor21a}) allows us to write the Hilbert transformation:
\be
\bm{\mathcal V}({\bf r},t,s)\equiv\bm{\mathcal V}({\bf r},\tao)=\frac{\ii}{2\pi}\int_{-\infty}^{\infty}dt' \,\frac{\bm{\mathfrak  V}({\bf r},t')}{t' - \tao}\nonumber \\
=\bm{\mathfrak V}({\bf r},t)*C_s(t), \,\,\,\,
 C_s(t)=\frac{-\ii}{2\pi(t-{\ii} s) }=\frac{1}{2\pi}\frac{s-{\ii}   t}{t^2+s^2}.\,\,\,\,\,\label{tfor21b}
\ee
Where the symbol $*$ denotes convolution. Hence  $\bm{\mathcal V}({\bf r},t,s)$ is obtained by time-averaging  the 
physical function  $ \bm{\mathfrak  V}({\bf r},t)$ over the low pass Cauchy  filter $C_s(t)=-{\ii}/[2\pi (t-{\ii}s)]$, whose real and imaginary parts, (being a Kramers-Kr\H{o}nig pair), are a Lorentzian of width $\Delta t=2s$ and an odd function resulting from the product of this Lorentzian by $-t/s$. Hence  $2s$ constitutes  the {\it minimum time interval} with which the quantity $\bm{\mathcal V}({\bf r},t,s)$ can be {\it resolved}, and $\Delta t $ is a {\it time resolution scale} for the analytic signals $\bm{\mathcal V}({\bf r},\tao)$ associated to the real quantities $\bm{\mathfrak V}({\bf r},t)$ \cite{kaiser}. In antenna and circuit theory, $s$ is known as {\it reactive time}, (measured in second reactive, sr) \cite{kaiser}.

Introducing the complex derivatives:
\be
 \partial_\tao=\frac{1}{2}(\partial_t+{\ii}\partial_s)\, ,\,\,\,\, \,\,\,\ \partial_\tao^*=\frac{1}{2}(\partial_t-{\ii}\partial_s) ;\,\,\,\, \,\,\,\,\label{tfor21aabb}\\
\partial_\tao\bm{\mathcal V}^*=\partial_t\bm{\mathcal V}
^*={\ii}\partial_s\bm{\mathcal V}^*,\,\,\,\partial_\tao^*\bm{\mathcal V}=\partial_t\bm{\mathcal V}=-{\ii}\partial_s\bm{\mathcal V},\,\,
\label{tfor21ab}
\ee
 the 
Maxwell equations: $\nabla \cdot{\mathfrak  E}=4\pi{\mathfrak \rho}$, $\nabla\cdot{\mathfrak  H}=0$, $\nabla \times \bm{\mathfrak  E}=-(1/c)\partial_t \bm{\mathfrak  H}$ and  $\nabla \times 
\bm{\mathfrak  H}=(1/c)\partial_t \bm{\mathfrak  H}+(4\pi/c) \bm{\mathfrak  J}$, yield for the analytic signals associated to the fields:
\be
\nabla \cdot \bm{\mathcal E}=4\pi\rho\,,\,\,\, \nabla \cdot \bm{\mathcal  H}=0\,,\,\,\,
\nabla \times \bm{\mathcal  E}=-\frac{1}{c}\,\partial_\tao^* \bm{\mathcal  H}\,,\,\,\, \nonumber \\
\nabla \times 
\bm{\mathcal  H}=\frac{1}{c}\,\partial_\tao^* \textcolor{red}{\bm{\mathcal  E}}+\frac{4\pi}{c} \bm{\mathcal  J}\,. \,\,\label{maxeqs}
\ee

In the Hilbert space of analytic signals we now  introduce the {\it complex  Lorentz force} $\bm{\mathcal F}$ on a system, surrounded by vacuum, with densities of  charge $\rho$ and current $ \bm{\mathcal J}$ occupying a volume $V_0$ contained in $V$. This force $\bm{\mathcal F}$ should be  identified with the complex source in the conservation of {\it complex linear mechanical momentum} ${\bf P}_{mech}$; viz.
\be
\bm{\mathcal F}({\bf r},t,s)\equiv\partial_t {\bf P}_{mech}({\bf r},t,s)=\frac{1}{2}\int_{V}d^3 r \,(\rho^*\,\bm{\mathcal E}+\frac{1}{c}\bm{\mathcal J}^*\times\bm{\mathcal B}). \,\,\,\,\,\,\,\,\,\label{tfor31}
\ee
Which substituting $\rho^*$ and $\bm{\mathcal J}$ through the Maxwell equations (\ref{maxeqs}), leads to
\be
\bm{\mathcal F}({\bf r},s,t)\equiv\partial_t {\bf P}_{mech}=
\frac{1}{8\pi}\int_{V}d^3 r \,[\bm{\mathcal E}(\nabla\cdot\bm{\mathcal  E}^*)\nonumber \\
+
\bm{\mathcal B}^*(\nabla\cdot\bm{\mathcal B})-\bm{\mathcal B}\times(\nabla\times \bm{\mathcal B}^*)-\frac{1}{c}(\partial_\tao \bm{\mathcal E}^*)\times\bm{\mathcal B}].\,\,\, \label{tfor31a}
\ee
We now recall (\ref{tfor21aabb}) and (\ref{tfor21ab})  using  the identities: $\partial_\tao\bm{\mathcal E}^*=\partial_t \bm{\mathcal E}^*$ and   $\partial_\tao^*\bm{\mathcal B}=\partial_t\bm{\mathcal B}$. Then, 
$(\partial_\tao\bm{\mathcal E}^*)\times\bm{\mathcal B}=\partial_t(\bm{\mathcal E}^*\times\bm{\mathcal B})-\bm{\mathcal E}^*\times(\partial_t\bm{\mathcal B})=\partial_t(\bm{\mathcal E}^*\times\bm{\mathcal B})-\bm{\mathcal E}^*\times(\partial_\tao^*\bm{\mathcal B})$. Therefore using the third equation (\ref{maxeqs}) we obtain:
\be
\bm{\mathcal F}({\bf r},s,t)\equiv\partial_t {\bf P}_{mech}=\frac{1}{8\pi}\int_{V}d^3 r \,[\bm{\mathcal E}(\nabla\cdot\bm{\mathcal E}^*)+
\bm{\mathcal B}^*(\nabla\cdot\bm{\mathcal B})\nonumber \\
-\bm{\mathcal B}\times(\nabla\times\bm{\mathcal B}^*)
-\bm{\mathcal E^*}\times(\nabla\times\bm{\mathcal E}) \nonumber \\
-\frac{1}{c}\partial_t(\bm{\mathcal E}^*\times\bm{\mathcal B})].\,\,\,\,\,\,\,\,\, \label{tfor32}
\ee
Now, the scaled complex Poynting momentum is:  $\bm{\mathcal G}({\bf r},t,s)=(1/c^2)\bm{\mathcal  S}({\bf r},t,s)=(1/8\pi c)[\bm{\mathcal  E}({\bf r},t,s)   \times \bm{\mathcal  B}^*({\bf r},t,s) ]$. Therefore, operating on the analytic signals in the four first terms of the integrand of (\ref{tfor32}), and from  the identity: $ {\bf a}^*\times(\nabla\times {\bf a})=a_j^* \partial_i a_j - a_j^* \partial_j a_i$, $ (i,j=1,2,3)$, we finally obtain for the complex Lorentz force the following {\it conservation equation} of the {\it scaled complex linear momentum} 
\be
{\mathcal F}_i({\bf r},s,t)\equiv\partial_t {\bf P}_{mech\, i}=-\int_{V}d^3 r \,\partial_t {\mathcal G}_i^*+\int_{\partial V}d^2 r \, \bm{\mathcal T}_{ij}n_j\nonumber \\
+\frac{\ii}{8\pi}\int_{V}d^3 r\,\mbox{Im}[{\mathcal B}_j^* \partial_i {\mathcal B}_j -
{\mathcal E}_j^* \partial_i  {\mathcal E}_j]. \,\,\,\,\,\,\,\,  \label{tfor33}
\ee
Where   $\mbox{Im}$ denotes imaginary part,  ${ n}_j$ is the $jth$ Cartesian component of the unit outward normal to the surface $\partial V$ of $V$, and the scaled   CMST is
\be
\bm{\mathcal T}_{ij}({\bf r},s,t)=\frac{1}{8\pi}[{\mathcal E}_i  {\mathcal E}_j^* + {\mathcal B}_i ^*{\mathcal B}_j -\frac{1}{2}\delta_{ij}(|\bm{\mathcal E}|^2 + |\bm{\mathcal B}|^2 )]. \,\,\,\,\,\,\,\,\,\,\,\, \label{tfor34}
\ee
There is a remarkable appearence in  (\ref{tfor33}) of the orbital (or canonical) momentum densities due to the electric and  magnetic fields, {\rr (in this connection, we remark that after Belinfante, the terms orbital  \cite{belinfante1} and canonical  \cite{belinfante2} are indistinctly employed for $\bm{\mathcal P}_{e}^O$, $\bm{\mathcal P}_{m}^O$, and $\bm{\mathcal P}^O=(1/2)(\bm{\mathcal P}_{e}^O+\bm{\mathcal P}_{m}^O)$ \cite{berry,nietoPRR,bliokh1,bliokh2,bekshaevNJP,nori1,xu,nori2,gao,zheng})}:
\be
(\bm{\mathcal P}_{e}^O)_i({\bf r},s,t)=\frac{1}{8\pi \omega}\mbox{Im}  [{\mathcal E}_j^* \partial_i {\mathcal E}_j],\,\,\, \,\,\,\,\nonumber \\
 (\bm{\mathcal P}_{m}^O)_i({\bf r},s,t)= \frac{1}{8\pi \omega}\mbox{Im}[ {\mathcal B}_j^* \partial_i {\mathcal B}_j ], \,\,\,\,\,\, (\omega=kc). \label{tfor35}
\ee
With which the conservation law (\ref{tfor33}) reads
\be 
\partial_t[ {\bf P}_{mech\, i}+\int_{V}d^3 r  \bm{\mathcal G}_i^*]= \int_{\partial V}d^2 r  \bm{\mathcal T}_{ij} n_j \nonumber \\
+{\ii}\omega\int_{V}d^3 r [\bm{\mathcal P}_{m}^O - \bm{\mathcal P}_{e}^O]_i.\, \,\,\,\,\,\,\,\,\, \label{tfor36}
\ee
Equation (\ref{tfor36}) is one of  the main results of this work. Its real and imaginary parts are
\be 
\bm{\mathcal F}_i^R({\bf r},t,s)\equiv\partial_t {\bf P}_{mech \, i}^R=-\int_{V}d^3 r  \partial_t \bm{\mathcal G}_i^{R}+ \int_{\partial V}d^2 r   \bm{\mathcal T}_{ij}^R n_j, \,\,\,\,\,\,\,\,\, \label{FR}\\
 \bm{\mathcal T}_{ij}^R=\frac{1}{8\pi}\mbox{Re}[ {\mathcal E}_i   {\mathcal E}_j^* +  {\mathcal B}_i ^* {\mathcal B}_j] -\frac{1}{2}\delta_{ij}(| \bm{\mathcal E}|^2 + | \bm{\mathcal B}|^2 ); \,\,\,\,\,\,\,\,\,\,\,\,\,\,\,\,\,\,\label{TR}
\ee
and
\be 
 \bm{\mathcal F}_i^I({\bf r},t,s)\equiv\partial_t {\bf P}_{mech\, i}^I=\int_{V}d^3 r \,\partial_t \bm{\mathcal G}_i^{I}
+ \int_{\partial V}d^2 r \, {\mathcal T}_{ij}^I n_j\nonumber \\
+\omega\int_{V}d^3 r\, [ \bm{\mathcal P}_{m}^O-  
\bm{\mathcal P}_{e}^O]_i\,\,; \,\,\, \,\,\,\,\,\, \label{tfor37b}
\ee
with the scaled {\it  imaginary},  or {\it reactive, Maxwell stress tensor} (IMST) ${\mathcal T}_{ij}^I$:
\be
 \bm{\mathcal T}_{ij}^I({\bf r},t,s)=\frac{1}{8\pi}\mbox{Im}[ {\mathcal E}_i {\mathcal E}_j^* +  {\mathcal B}_i ^* 
{\mathcal B}_j].\, \,\,\, \,\,\, \,\, \, \,\,\, \,\,\, \,\,\, \,\,\, \,\,\, \,\,\, \,\,\, \,\,\, \,\,\, \,\,\, \,\,\, \,\, \label{tfor37bb}
\ee
The superscripts $R$ and $I$ denote real and imaginary parts. It should be reminded that, although only explicitely written in the extreme left, all quantities in the above equations are functions of ${\bf r}$, $t$ and $s$. However, since there are no $s$-derivatives in  (\ref{tfor36})-(\ref{tfor37b}), they  also hold  in the limit $s\rightarrow 0$ and have physical meaning even if they are not scaled, then becoming instantaneous ones. 
This is in contrast with the energy  in the complex Poynting vector theorem \cite{kaiser}. While the reactive power is determined in  volt-ampere reactive (var), we note that the ILF is measured in newtons. Thus when $s\rightarrow 0$ Eq. (\ref{TR}) becomes the familiar time-dependent conservation equation, in terms of  analytic signals, for the time  variation of  instantaneous linear momentum ${\bf P}_{mech}^R({\bf r},t)$.

Nonetheless, the novel Eq.  (\ref{tfor37b})  represents a quite different process: the interaction wave-object yields  a change of  an additional scaled  linear momentum  ${\bf P}_{mech}^I({\bf r},t,s)$, giving rise to an {\it imaginary Lorentz force} (ILF) $\bm{\mathcal F}_i^I({\bf r},t,s)$ on the body to which a time change of reactive   momentum  of the field is substracted,   irrespective of whether or not the incident wave possess it  (like e.g. an evanescent or a two-wave interference field); so that even if the incident wave has no IPM, like propagating plane waves and beams,   it arises in the scattering process. We shall insist on this below.

The ILF  also stems from the flux, given by $ \bm{\mathcal T}_{ij}^I$,  into the volume $V$ that surrounds the object,  of  $\omega\int_{V}d^3 r\, [\bm{\mathcal P}_{m}^O - \bm{\mathcal P}_{e}^O]$ given by the difference between the magnetic and electric {\rr orbital (or canonical) momenta} (\ref{tfor35}) of the field,  and that we define as the  {\it reactive strength of orbital (or canonical) momentum} (ROM) stored in $V$. 
 (We wish to remark the difference of the ROM concept with that of reactive orbital momentum that we introduced in \cite{nietoPRR},  Eq. (26),  as the difference between the imaginary parts of the magnetic and electric orbital momenta). 

Since  Eq. (\ref{FR}) also holds for $s\rightarrow 0$, it rules instantaneous quantities too. Then, as we shall show below, $\int_{\partial V}d^2 r \, {\mathcal T}_{ij}^I n_j$  for monochromatic fields is associated to an alternating flux of ROM, flowing from the body and returning to it. 

 For these reasons, we call $\bm{\mathcal F}^I$ the {\it reactive force} on the object of volume $V_0$. These quantities, and $\bm{\mathcal F}^I$ in particular,  have zero time-average, (like the reactive Poynting vector associated to  the alternating flow of reactive power),  in contrast with the {\it active force} $\bm{\mathcal F}^R$ which constitutes the time-averaged force sensed by the object, and observed in most experimental observations. Like in circuit theory \cite{kaiser,akagi, czarnecki}, as $s$ diminishes there is an increase of modes of the standing flow going back and forth from $V_0$, which corresponds to the ROM. 
The size $s$ of  the Cauchy filter $C_s(t)$ in (\ref{tfor21b}), goes from $ t= \infty$ to $t= minimum$ {\it time}  $scale$  \cite{kaiser} which in our case is $0$, at which no more modes  linked to the bouncing  flow associated to the IMST, appear. 

The arise of the {\rr canonical} momenta  in the IMST law (\ref{tfor37b}),  defining the ROM,  is quite illuminating since it highlights the prominent role of these canonical momenta, (rather than of the Poynting momentum) in the generation of optical forces \cite{nietoPRR,bliokh1,nieto1, nori1}. This will further  be discussed below.  

Therefore we have obtained in (\ref{tfor37b}) a fundamental  law for the instantaneous ($s\rightarrow 0$)\rr---\bb and also for the scaled $s\neq 0$\rr---\bb force that confers another physical property to the IPM, $\bm{\mathcal G}^{I}$, in the realm of momentum conservation  in light-matter interactions, which transcends that previously found in connection with self-forces on magnetoelectric particles \cite{nietoPRR,nieto1,nieto2}, assigned to  specific  illuminating fields, like e.g. evanescent, standing, and cylindrical vector beams \cite{bliokh1, bliokh2,xuPRL2019}. 

This exchange of imaginary momentum  produces the reactive  flow  IMST and  ILF, (therefore it should be  instantaneously observed with a time-varying incident field, like e.g. an ultrashort  pulse), which gives rise to an accretion of  {\it  reactive strengtht of canonical momentum}, ROM,  around the  body. As such,  this ROM  is not associated to the scattered field radiated into the far-zone, i.e. to the ``radiated'' RMST, but to  reactive power, and thus it remains inside and in the near-field of the object volume $V_0$. In this regard, we remark that this field imaginary momentum exchange is opposite to that of field (i.e. Poynting) time-averaged 
momentum, thus substracting, rather than adding, to the variation of imaginary linear momentum.  \rr This means that although the light wave loses field momentum $\bm{\mathcal G}^{R}$ on interaction, it may gain reactive field momentum $\bm{\mathcal G}^{I}$, even when it does not exist in the illuminating field. This is the case of e.g. an incident plane wave, discussed later on. \bb

As a consequence, we may expect the ROM, and thus the  flow  related to the  IMST and  ILF, to severely affect the flow given by the RMST into the far-zone, namely, the standard time-averaged electromagnetic optical force, RLF, sensed by the body. The ILF is, as seen in the next sections, characterized by the total ROM
stored in and out the particle, and its flow; in analogy with the imaginary work in terms of  the  reactive
energy in and around the particle and its flow: the reactive momentum IPM \cite{nietoPRR}.  

In this context, while the time-averaged force, i.e. the RLF, is the flow whose density is the RMST ``radiated'' into the far zone,  the reactive force, ILF, is the stored ROM inside and in the near-field  due to the flow  IMST.  Thus the complex flow, CMST, into the far and near zones constitutes a novel way  of understanding the significance of both the standard RLF as well as the ILF. Hence the stored ROM will constitute a hindrance to the time-averaged force, RLF, sensed by the particle. As we shall see, the external  stored ROM is due to the interference of the incident and scattered fields. 

\rr Figure 1 at the end of next section depicts this process for monochromatic fields for which it  is more easily illustrated. Let us see next the CMST for these fields.
\bb

\section{Time-harmonic fields. The  imaginary Maxwell stress tensor and the reactive force   in the space-frequency domain} 
 \subsection{Complex stress tensor and reactive strength of orbital momentum}
For the spatial parts of time-harmonic  fields whose analytic signals  are $\bm{\mathcal  E}({\bf r},\tao)={\bf E}({\bf r})\exp(-{\ii}\omega t)$,
 $\bm{\mathcal  B}({\bf r},\tao)={\bf B}({\bf r})\exp(-{\ii}\omega t)$ and   $\bm{\mathcal  J}({\bf r},\tao)={\bf J}({\bf r})\exp(-{\ii}\omega t)$, ($\tao=t$), ($\omega=kc$),  Eq. (\ref{tfor36})  becomes
\be 
\bm{\mathcal F}_i= \int_{\partial V}d^2 r \, T_{ij} n_j+{\ii}\omega\int_{V}d^3 r\, ({\bf P}_{m}^O -{\bf P}_{e}^O)_i.\,  \,\,\,\,\,\,\,\,\,\,\,\,  \label{bcfor8}
\ee
Which is the {\it complex Maxwell stress tensor  theorem} in the space-frequency domain; the CMST being
\be
T_{ij}=\frac{1}{8\pi}[E_i  E_j^* + B_i ^*B_j -\frac{1}{2}\delta_{ij}(|{\bf E}|^2 + |{\bf B}|^2 )].  \label{bcfor6}
\ee
With the electric and magnetic {\rr canonical} momentum densities:
\be
({\bf P}_{e}^O)_i=\frac{1}{8\pi \omega}\mbox{Im}  [E_j^* \partial_i E_j],\,\,\,  ({\bf P}_{m}^O)_i= \frac{1}{8\pi \omega}\mbox{Im}[ B_j^* \partial_i B_j ] \,\,. \,\,\,\,\,\,\,\,\,\,\,\,\,\,\,\,\label{bcfor7}
\ee
 The real part of (\ref{bcfor8}) is the well-known conservation law of linear momentum for monochromatic fields \cite{chaumet1}:
 \be
<\frac{d{\bf P}_{mech}}{dt}>\equiv<\bm{\mathcal F}_i> 
\equiv\mbox{Re} \{\bm{\mathcal F}\}_i \nonumber \\
\equiv\frac{1}{2}\int_{V}\mbox{Re} [ \rho^*{\bf E}({\bf r})+\frac{\bf J^*}{c}\times{\bf B}({\bf r})]_i\,d^3 r
\nonumber \\
=
 \int_{\partial V}d^2 r \, T_{ij}^R n_j=  \int_{\partial V}d^2 r \, <T_{ij}>n_j, \label{bcfor9}
\ee

Which expresses the total time-averaged electromagnetic  optical force,  $<{\bf F}>$, on the system of charges and currents of the object of volume $V_0$  contained in the integration  volume $V$.  

On the other hand, the imaginary part of  (\ref{bcfor8}) reads:
\be
 \bm{\mathcal F}_i^I\equiv\frac{1}{2}\int_{V}\mbox{Im} [ \rho^*{\bf E}({\bf r})+\frac{\bf J^*}{c}\times{\bf B}({\bf r})]_i\,d^3 r \nonumber \\
= \int_{\partial V}d^2 r \, T_{ij}^I ({\bf r})n_j+\omega\int_{V}d^3 r\, ({\bf P}_{m}^O - {\bf P}_{e}^O)_i . \, \,\,\, \,\,\, \,\, \label{bcfor10}
\ee
From (\ref{bcfor6}) the imaginary, or reactive,  Maxwell stress tensor (IMST) is:
\be
T_{ij}^I=\frac{1}{8\pi}\mbox{Im}\{E_i  E_j^* + B_i ^*B_j \}. \label{bcfor10b}
\ee
The first term of the right side of Eq. (\ref{bcfor10})  expresses $T_{ij}^I$ as the flow density IMST into $V$ across its surface $\partial V$. Because of this, $T_{ij}^I$ is a   reactive quantity. Furthermore, as 
stated above, the second term is the ROM {\it  stored} in $V$, namely in and around the body volume $V_0$. Then,  the {\it  reactive Lorentz force} (ILF), $\bm{\mathcal F}^I$, is  the  here uncovered optical
force, acting on the object, related to the alternating  instantaneous one,   as a result of the inward  flow   IMST  $T_{ij}^I$ and accretion of reactive strength of orbital momentum, $\omega\int_{V}d^3 r\, ({\bf P}_{m}^O - {\bf P}_{e}^O)$, in and around the object volume $V_0$. 

Analogously, while Eq. (\ref{bcfor9})  yields the  time-averaged optical  $<\bm \Gamma>$ on the object, of lever arm ${\bf r}$: \,${\bf r}\times \bm{\mathcal F}^R$ \cite{nieto_torque,qiu_torque}, Eq. (\ref{bcfor10}) leads to the  {\it reactive torque}: $\bm {\Xi}^I={\bf r}\times \bm{\mathcal F}^I$, [see Appendix A of the Supp. Mat.)].

\subsection{The instantaneous Maxwell stress tensor}
At this point it is convenient to introduce the  instantaneous  Maxwell stress tensor, that gives rise to the alternating instantaneous ILF,  built by the   fields $\bm{\mathfrak   E}({\bf r},t)=\mbox{Re}\{\bm {\mathcal E}({\bf r},t)\}$ and  $\bm{\mathfrak   B}({\bf r},t)=\mbox{Re}\{\bm {\mathcal B}({\bf r},t)\}$. Like for other instantaneous time-harmonic quantities \cite{harrington, balanis}, it is immediate to obtain:
\be
{\mathfrak   T}_{ij}({\bf r},t)=\frac{1}{4\pi}\{{\mathfrak  E}_i({\bf r},t) {\mathfrak  E}_j({\bf r},t)+
{\mathfrak   B}_i({\bf r},t) {\mathfrak    B}_j({\bf r},t) \nonumber \\
-\frac{1}{2}\delta_{ij}[{\mathfrak    E}_i^2({\bf r},t)+{\mathfrak    B}_i^2({\bf r},t)]\}
=\nonumber \,\,\,\, \,\,\\
<{ T}_{ij}({\bf r})>+ \frac{1}{8\pi}\mbox{Re}\{ [{  E}_i({\bf r}) {  E}_j({\bf r})+
{  B}_i({\bf r}) {  B}_j({\bf r})  \nonumber \\
-\frac{1}{2}\delta_{ij}(|{  E}_i({\bf r})|^2+|{  B}_i({\bf r})|^2)]
\exp(-2{\ii}\omega t)\}, \nonumber\,\,\,\, \,\,
\ee

 However by a simple calculation   we gain insight  expressing it as:
\be
{\mathfrak   T}_{ij}({\bf r},t)=
{ T}_{ij}^R({\bf r})(1+\cos 2\omega t)+ { T}_{ij}^I({\bf r})\sin 2\omega t \nonumber \\
+\frac{1}{4\pi}\{ {  E}_i^R({\bf r}) {  E}_j^I({\bf r})+{  B}_i^I({\bf r}) {  B}_j^R({\bf r})
  \nonumber \\
-\frac{1}{2}\delta_{ij}[{  E}_i^R({\bf r}) {  E}_j^I({\bf r})+{  B}_i^I({\bf r}) {  B}_j^R({\bf r})]\}\sin 2 \omega t 
 \nonumber  \\-\frac{1}{4\pi}\{{  E}_i^I({\bf r}) {  E}_j^I({\bf r})+{  B}_i^I({\bf r}) {  B}_j^I({\bf r}) \nonumber \\
-\frac{1}{2}\delta_{ij}[{  E}_i^I({\bf r}) {  E}_j^I({\bf r})+{  B}_i^I({\bf r}) {  B}_j^I({\bf r}) ]\}\cos 2\omega t . \,\,\,\,\,\ \,\,\,\label{geninst} 
\ee
While the term with   ${ T}_{ij}^R({\bf r})\equiv <{ T}_{ij}({\bf r})>$ does not change sign with time, as expected from the instantaneous MST part associated with the time-averaged  flow of momentum,   the term containing the   reactive Maxwell stress tensor, ${ T}_{ij}^I({\bf r})$,    alternates its sign  at frequency $2\omega$ following the variation of $\sin2\omega t$. This is in accordance with the  interpretation of the imaginary part of (\ref{bcfor8}) being related to the ROM  flux, i.e. the IMST,  going back and forth from the object volume $V_0$ \rr with zero time-average. (Note that the Fourier decomposition (\ref{tfor21a}) into time-harmonic components  shows this zero time-average  of the ROM flow also for general time-varying fields). 
\bb
 In addition, there is an alternating generally non-zero  contribution to this instantaneous flow  ${\mathfrak   T}_{ij}({\bf r},t)$ in the terms within the curly brackets  of (\ref{geninst}).  Obviously only the ${ T}_{ij}^R$ term remains on time-averaging in  (\ref{geninst}).

\subsection{The external reactive strength of orbital momentum}
One may obtain additional discernment on the role played by the terms of the conservation  law (\ref{bcfor10}). Let us take the volume $V$ to be $V_{\infty}$ of a large sphere of radius $r$  such that $kr\rightarrow\infty$. The  flow IMST   across its surface   $\partial V_{\infty}$ is zero since the CMST is real in the far zone \cite{nieto1}, (this may also be easily seen considering the far zone expression of the scattered fields. Then only the diagonal elements of the  CMST contribute to the Lorentz force and, as such, this reduces to a real quantity; namely, only $<\bm{\mathcal F}>$ is obtained in this region). Then in the volume $V_0$  of the scattering body within $V$ we have
\be
\bm{\mathcal F}^I\equiv\frac{1}{2}\int_{V_0}\mbox{Im} [ \rho^*{\bf E}({\bf r})+\frac{\bf J^*}{c}\times{\bf B}({\bf r})] \,d^3 r= \nonumber \\
 \omega\int_{V_{\infty}} d^3 r\, ({\bf P}_{m}^O - {\bf P}_{e}^O) . \, \,\,\, \,\,\, \,\, \label{bcfor10aa}
\ee

Equation (\ref{bcfor10aa}) shows  that the ILF, i.e. the source term in the left side of  (\ref{bcfor10}), is given by the overall ROM. In addition, introducing (\ref{bcfor10aa}) into (\ref{bcfor10}), we obtain
\be
 \int_{\partial V}d^2 r \, T_{ij}^I n_j=\omega\int_{V_{\infty}-V}d^3 r\, ({\bf P}_{m}^O - {\bf P}_{e}^O)_i
. \, \,\,\, \,\,\, \,\, \label{bcfor10bb}
\ee 
 Equation (\ref{bcfor10bb}) is important  because it illustrates that the part of the ILF due to the flow IMST    into the volume $V$ surrounding the body volume $V_0$,  is given by the total outside ROM stored  between $V_{\infty}$ and $V$. In particular, $V$ may approach the source  volume $V_0$. We also see from (\ref{bcfor10bb}) that in the far-field ($FF$) region: ${\bf P}_{m}^{O\,FF} = {\bf P}_{e}^{O\,FF} $ since  the right side of (\ref{bcfor10bb}) is zero as $V$ approaches $V_{\infty}$.

\rr Now we are in position to have the overall perspective of the CMST theorem in both the far-field and near-field regions, and to understand how the  flow IMST builds up the ROM inside and outside the particle, with the ILF being generated. Figure 1 illustrates the process. It should be recalled that the illuminated object behaves like an antenna, so that the interaction also conveys a reactive work on the charges,  building-up reactive power which hinders the efficiency of  radiated  scattered energy \cite{jackson,nietoPRR,ziolkowski}. As we shall see in section VI,  the ILF and stored ROM (which  are the dynamical analogues of the reactive work and reactive power, respectively) counteract against the RLF, (i.e. the dynamical analogue of the radiated energy). In consequence, a large ROM, and hence a strong ILF, conveys a decrease of ``radiative" RMST, i.e. of RLF. Conversely,  low ROM and ILF are linked to a larger RLF.\bb

\bb\begin{figure}[htbp]
\begin{centering}
\centerline{\includegraphics[width=0.5\columnwidth]{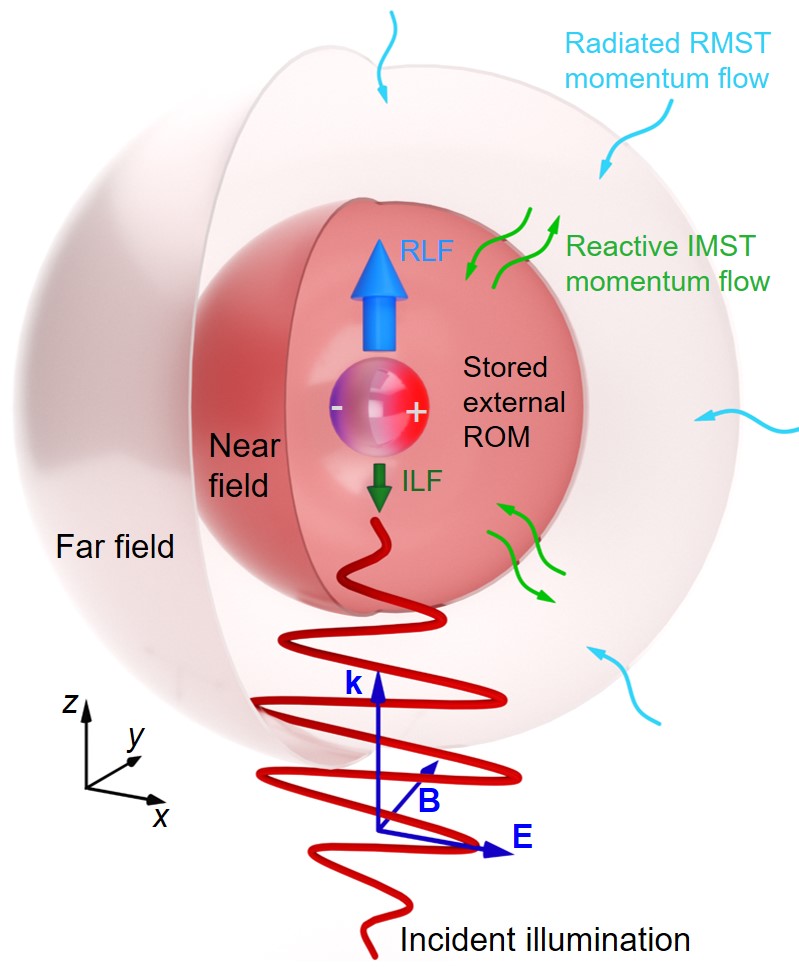}}
\end{centering}
\caption{Outline of the physical process, (the near-field region is enlarged  and the radiative area has been shrunk to ease reading). \rr When an electromagnetic wave, (for example a monochromatic plane wave, as illustrated here), impinges on a body,  charges are separated, thus inducing  multipoles.  The object  feels a time-averaged Lorentz force, RLF, associated to the field (i.e. Poynting) momentum flux, RMST, flowing into a  far-field  surface, (e.g. spherical as shown). However, there also appears a flow, IMST, of imaginary field  momentum, related to momentum flux going back and forth across a near-field surface, with zero time-average, which accumulates reactive strength of orbital momentum, ROM, stored both inside and in the near-field of the object, and meanwhile
yields a reactive force, ILF, on the body.
\bb}  
\end{figure}

\section{The canonical and spin momenta and the complex  Lorentz force}
In this section we establish a  connection  of the real and imaginary parts of the \rr canonical \bb and spin momenta and of the complex Lorentz force introduced above. 

\subsection{The imaginary Lorentz force and the spin momenta. The reactive strength of Poynting momentum. Implications for the reactive Maxwell stress tensor}
The ILF $\bm{\mathcal F}^I$ may be expressed  by writing the flow  IMST   in terms of the real electric and magnetic spin  (i.e. Belinfante)  momenta, which is of conceptual interest. This is obtained using 
the identity: $ {\bf a}^*\times(\nabla\times {\bf a})=a_j^* \partial_i a_j - a_j^* \partial_j a_i$, from which we have:  $\mbox{Im}\{- {\bf a}^*\times(\nabla\times {\bf a})+{\bf a}( \nabla\cdot {\bf a}^*)\}=\mbox{Im} \{\partial_j (a_i a_j^*) - a_j^* \partial_i a_j\}$. On the other hand, since the Levi-Civita tensor holds: $\epsilon_{ijk}\epsilon_{klm}=\delta_{il}  \delta_{jm}  -\delta_{im} \delta_{jl}$, we have $(1/2)\nabla\times\mbox{Im}\{{\bf a}^* \times {\bf a}\}=\mbox{Im}\{\partial_j (a_i^* a_j)\}$. 

Hence, using the above identity, we may write these spin momentum densities \cite{nietoPRR,bliokh1} as:
\be
{\bf P}_{e}^S=\frac{1}{16\pi \omega}\nabla\times\mbox{Im}\{{\bf E}^* \times {\bf E}\}=\frac{1}{8\pi \omega}\mbox{Im}\{\partial_j({ E}_i^* { E}_j)\}
\,\,,\, \,\,\,\,  \nonumber \\
  {\bf P}_{m}^S=\frac{1}{16\pi \omega}\nabla\times\mbox{Im}\{{\bf B}^* \times {\bf 
B}\}=\frac{1}{8\pi \omega}\mbox{Im}\{\partial_j({ B}_i^* { B}_j)\} \,.\,\, \,\,\,\,\, \,\,\,\,\label{bcfor4A2}
\ee
Therefore  the ILF, Eq. (\ref{bcfor10}), reads
\be
\bm{\mathcal F}_i^I=\frac{1}{8\pi}\int_{V}\mbox{Im} [{\bf E}( \nabla\cdot {\bf E}^*)+{\bf B}^*( \nabla\cdot {\bf B})-  {\bf E}^*\times(\nabla\times {\bf E})\nonumber \\
-{\bf B}\times (\nabla\times {\bf B}^*)]_i \,d^3 r= 
\frac{1}{8\pi }\int_{V}\mbox{Im}[-\partial_j (E_i^* E_j)\nonumber \\
 - E_j^* \partial_i E_j+\partial_j (B_i^* B_j^*) + B_j^* 
\partial_i B_j] d^3 r \,  ,  \, \,\,\,\,\, \,\,\,\,\,    \label{bcfor4A1}
\ee
which  finally becomes 
\be
\bm{\mathcal F}^I=\omega\int_{V}d^3r  [({\bf P}_m^S-{\bf P}_e^S)+ ({\bf P}_{m}^O - {\bf P}_{e}^O)]
. \, \,\,\, \,\,\, \,\, \label{bcfor4A221}
\ee
Equation (\ref{bcfor4A221}), which is other of the main results of this work, provides a physical meaning of the ILF   in terms of the spin and  angular momenta as a {\it reactive strength of Poynting momentum} whose density is : $\omega  [({\bf P}_{m}^O+{\bf P}_m^S)-({\bf P}_e^O  +{\bf P}_{e}^s)]$.  And
\be
 \int_{\partial V}d^2 r \,  T_{ij}^I n_j=\omega\int_{V}d^3r ({\bf P}_m^S-{\bf P}_e^S)_i; \, \,\,\,\,\,\nonumber \\
\mbox{or:} \,\,\,\,
\nabla\cdot T_{ij}^I=\omega({\bf P}_m^S-{\bf P}_e^S)_i  \,  .     \label{bcfor4A22}
\ee
I.e. the {\it reactive strength, of spin momentum} (RSM), $\omega\int_{V}d^3r ({\bf P}_m^S-{\bf P}_e^S)$, in $V$ is  equivalent to the incoming flow IMST  across $\partial V$.

In addition,  from Eqs. (\ref{bcfor10bb}) and  (\ref{bcfor4A22}) we have:
\be
\int_{V}d^3r ({\bf P}_m^S-{\bf P}_e^S)
 =\int_{V_{\infty}-V}d^3 r\, ({\bf P}_{m}^O - {\bf P}_{e}^O) , \, \,\,\, \,\,\, \,\, \label{bcfor4b1}
\ee
And thus we obtain that as $V$ approaches ${V_{\infty}}$, one has that 
\be
\int_{V_{\infty}}d^3r ({\bf P}_m^S-{\bf P}_e^S)=0 \,. \, \,\,\, \,\,\, \,\, \label{bcfor4b11}
\ee
We emphasize that when  $V_0$ is a scattering volume, the above equations hold for the total field given by the sum of the incident and  scattered fields.

{\it Equation (\ref{bcfor4b1}) is an important  novel balance law for the formation of RSM  in a finite volume $V$ that contains the body volume $V_0$, as it equals  the  accumulation of ROM in the whole space outside $V$, and so it builds the ILF}. In particular it holds when $V\rightarrow V_0$.  On the other hand, as regards (\ref{bcfor4b11}) we shall later show that  the overall spin momentum of the field emitted, or scattered, by a dipolar particle is  zero.

In particular, in free space and for surface waves, one knows that $ \int_{V_{\infty}}d^3r {\bf P}^S=(1/2)\int_{V_{\infty}}d^3r ({\bf P}_m^S+{\bf P}_e^S)=0$ \cite{bliokh1,bekshaevNJP},   therefore (\ref{bcfor4b11}) will imply  that in the whole space $V_{\infty}$ the overall electric and magnetic spin momenta of a free-field are zero, and thus according to (\ref{bcfor4b1}) this amounts to $V=0$. This is compatible with the evidence  that in this case no scattering object is present.

\subsection{The time-averaged Poynting momentum with sources and the imaginary Lorentz force}
At this point  we must remark the compatibility  of  the ILF, Eq. (\ref{bcfor4A221}), with    the formulation of the (real)  Poynting momentum with sources.  

Let us, first, employ the third Maxwell equation to eliminate ${\bf B}$ in the definition of the time-averaged Poynting momentum  $<{\bf g}>=<{\bf S}>/c^2=(1/8\pi c)\mbox{Re} \{{\bf E}\times{\bf B}^*\}$:
\be
<{\bf g}>={\bf P}_e^S+{\bf P}_e^O +\frac{1}{8\pi \omega}\mbox{Im}\{{\bf E}( \nabla\cdot {\bf E}^*)\}= \nonumber \\
{\bf P}_e^S+{\bf P}_e^O +\frac{1}{2 \omega}\mbox{Im}\{\rho^*{\bf E}\} . \label{bcfor4A3}
\ee
Correspondingly, by eliminating ${\bf E}$ in the definition of  $<{\bf g}>$, we obtain in terms  of the magnetic momenta: 
\be
<{\bf g}>=-\frac{1}{2c\omega}\mbox{Im}\{{\bf J}^* \times {\bf B}\}+
{\bf P}_m^S+{\bf P}_m^O \nonumber \\
+\frac{1}{8\pi \omega}\mbox{Im}\{{\bf B}( \nabla\cdot {\bf B}^*)\}= \nonumber \\
{\bf P}_m^S+{\bf P}_m^O -\frac{1}{2c\omega}\mbox{Im}\{{\bf J}^* \times {\bf B}\} , \label{bcfor4A4}
\ee

Adding  (\ref{bcfor4A3}) and (\ref{bcfor4A4}) we get an expression for the time-averaged field momentum valid in a space which is not source-free, but that  contains  charge and current distributions:

\be
<{\bf g}>={\bf P}^S+{\bf P}^O +\frac{1}{4 \omega}\mbox{Im}\{\rho^*{\bf E}-\frac{1}{c}{\bf J}^* \times {\bf B}\}. \label{bcfor4A10}
\ee
Where we have employed the (real, i.e. time-averaged)  {\rr canonical} and spin momenta \cite{nietoPRR,bliokh1,bekshaevNJP}:
\be
{\bf P}^{O}=\frac{1}{2}({\bf P}_{e}^{O} +{\bf P}_{m}^{O}),\,\,\, \,\,\,
{\bf P}^{S}=\frac{1}{2}({\bf P}_{e}^{S} +{\bf P}_{m}^{S})\, . \label{bcfor4A11}
\ee
In free-space   (\ref{bcfor4A10}) turns into the well-known decomposition of  $<{\bf g}>$ as the sum of the spin and {\rr canonical} momentum densities. In Appendix B of Supp. Mat.we show the derivation of  (\ref{bcfor4A3})  and  (\ref{bcfor4A3}) from first Lagrangian principles.

{\it Note the compatibility of  (\ref{bcfor4A221})  with (\ref{bcfor4A3}) and  (\ref{bcfor4A4}), since substracting  (\ref{bcfor4A3}) from (\ref{bcfor4A4}), and integrating in $V$, we immediately obtain  (\ref{bcfor4A221})}.

\subsection{Imaginary momenta and the time-averaged force}
 The above time-averaged spin and {\rr canonical} momenta are the real parts of  complex {\rr canonical} and spin momenta \cite{nietoPRR,bliokh1}: $\tilde{\bf P}^O ={\bf P}^{O}+{\ii}{\bf P}^{O\,I}$ and  $\tilde{\bf P}^S ={\bf P}^{S}+{\ii}{\bf P}^{S\,I}$.

The imaginary parts,  ${\bf P}^{O\,I}$ and ${\bf P}^{S\,I}$,  play an important role complementary to that of the above real parts, namely in the time-averaged force. To see it, we now proceed with the reactive Poynting momentum ${\bf g}^I=(1/8\pi c)\mbox{Im} \{{\bf E}\times{\bf B}^*\}$ in a way analogous to that leading to Eqs. (\ref{bcfor4A3}) and (\ref{bcfor4A4}) . It is straightforward to obtain:
\be
({\bf g}^I)_i=\frac{1}{8\pi \omega}\{-\mbox{Re}[ \partial_j({ E}_i ^{*} { E}_j)]+\frac{1}{2}\delta_{ij}\partial_j|{\bf E} |^2\}\nonumber \\
+\frac{1}{2 \omega}\mbox{Re}\{\rho^*{\bf E}\}_i, \,\,\, \label{bcfor5A1}
\ee
and
\be
({\bf g}^I)_i=\frac{1}{8\pi \omega}\{\mbox{Re} [\partial_j( B_i ^{*} { B}_j)]-\frac{1}{2}\delta_{ij}\partial_j|{\bf B} |^2\}\nonumber \\
-\frac{1}{2 \omega}\mbox{Re}\{\frac{1}{c}{\bf J}^* \times {\bf B}\}_i. \,\,\,\,\, (i,j=1,2,3), \label{bcfor5A2}
\ee
Substracting (\ref{bcfor5A1}) from (\ref{bcfor5A2}) we obtain the well-known time-averaged force density, [cf. Eq.(\ref{bcfor9})], 
\be
\frac{1}{2 }\mbox{Re}\{\rho^*({\bf E})_i + \frac{1}{c}({\bf J}^* \times {\bf B})_i\}=\nonumber
\rr\{\bb\frac{1}{8\pi }\partial_j\mbox{Re} ({ E}_i  { E}_j^{*}+B_i ^{*} { B}_j)\\-\frac{1}{2}\delta_{ij}\partial_j(|{\bf E} |^2 +
|{\bf B} |^2)\}.  \,\,\,\,\,\,\,\,\,\,\,\,\, \label{bcfor5A3}
\ee
On the other hand, adding (\ref{bcfor5A1}) and (\ref{bcfor5A2}) we derive a representation for the  reactive  Poynting momentum density:
\be
({\bf g}^I)_i=\frac{1}{8\pi \omega}\partial_j \{ \frac{1}{2}\mbox{Re} ( B_i ^{*} { B}_j-{ E}_i  { E}_j^{*})-\frac{1}{4}\delta_{ij}(|{\bf B} |^2 -|{\bf E} |^2)\}\nonumber \\
+\frac{1}{4 \omega}\mbox{Re}\{\rho^*{\bf E}-\frac{1}{c}{\bf J}^* \times {\bf B}\}_i
.  \,\,\,\,\,\,\,\,\,\,\,\,\, \label{bcfor5A4}
\ee
We now introduce the imaginary spin curl and orbital momenta, viz.:
\be
({\bf P}_e^{S\,I})_i=\frac{1}{8\pi \omega }\mbox{Re}\{\partial_j ({ E}_i^{*}  { E}_j)\}\,,\,\,\nonumber \\
({\bf P}_m^{S\,I})_i=\frac{1}{8\pi \omega}\mbox{Re}\{\partial_j ({ B}_i ^{*} { B}_j)\}\,, \nonumber\\
({\bf P}_e^{O\,I})_i=-\frac{1}{8\pi \omega }\partial_j \frac{1}{2}\delta_{ij}|{\bf E} |^2=-\frac{1}{8\pi \omega} \frac{1}{2}\partial_i|{\bf E} |^2 \,,\nonumber \\
({\bf P}_m^{O\,I})_i=-\frac{1}{8\pi \omega } \frac{1}{2}\partial_i|{\bf B} |^2 \,.\,\,\,\,\,\,\,\,\,\,\,\,\label{bcfor5A5}
\ee
We emphasize that the definition of the spin momenta as $\partial_j(\cdot)$ in  (\ref{bcfor5A5}) [cf.   also  (\ref{bcfor4A2})] is useful as it permits a direct introduction of the  CMST, as well as of the complex spin momenta, whose imaginary parts yield:
\be
{\bf P}^{S\,I}=\frac{1}{2}({\bf P}_e^{S\,I} +{\bf P}_m^{S\,I})\,;\,\,\,\,\,\,\,\,\,\,
{\bf P}^{O\,I}=\frac{1}{2}({\bf P}_e^{O\,I} +{\bf P}_m^{O\,I}),    \,\,\,\,\,\,\,\,\,\  \label{bcfor5A6}
\ee
and the reactive spin and orbital momenta \cite{nietoPRR}:
\be
\bm{\mathcal P}^{S}=\frac{1}{2}({\bf P}_m^{S\,I} -{\bf P}_e^{S\,I})\,;\,\,\,\,\,\,\,\,\,\,
\bm{\mathcal P}^{O}=\frac{1}{2}({\bf P}_m^{O\,I} -{\bf P}_e^{O\,I}).    \,\,\,\,\,\,\,\,\,\  \label{bcfor5A6b}
\ee
So that the imaginary spin and orbital momenta introduced in Eqs. (\ref{bcfor5A5}) and (\ref{bcfor5A6})  permit us to write the time-averaged force, (\ref{bcfor5A3}), as
\be
<\bm{\mathcal F}>=2\omega\int_Vd^3 r\ ({\bf P}^{S\,I}+{\bf P}^{O\,I}), \,\,\,\,\,\,\,\,\,\  \label{bcfor5A7}
\ee
which is the desired representation of the RLF in terms of the imaginary momenta. While the imaginary Poynting momentum, (\ref{bcfor5A4}), reads
\be
{\bf g}^I=\bm{\mathcal P}^{S}+\bm{\mathcal P}^{O}
+\frac{1}{4 kc}\mbox{Re}\{\rho^*{\bf E}-\frac{1}{c}{\bf J}^* \times {\bf B}\}
.  \,\,\, \label{bcfor5A8}
\ee
Equation (\ref{bcfor5A8}) is the representation of ${\bf g}^I$ through the reactive  spin and orbital momenta in  presence of sources; while (\ref{bcfor5A7}) and  (\ref{bcfor4A221}) formulate the real and imaginary Lorentz forces in terms of the respective imaginary and real parts of the spin and {\rr canonical} momenta. These momenta  are therewith  shown to be the ultimate dynamic quantities  that characterize   the complex optical force. We shall later make use of the fact that in a scattering configuration they correspond to the total fields; namely, incident plus scattered.

\subsection{Example 1: Evanescent wave}
Let us consider the monochromatic reactive wavefield consisting of the evanescent wave created  by total internal reflection  at a plane interface $z=0$ separating  air in $z\geq 0$ from a dielectric in the half-space $z< 0$. The plane of incidence being $OXZ$. The complex spatial parts of the electric and magnetic vectors in $z\geq 0$, are expressed in a Cartesian coordinate basis $\{\hat{\bf x},\hat{\bf y},\hat{\bf z}\}$ as  \cite{nietoPRR, nietoOL}:
\be
{\bf E}= \left(-\frac{{\ii}q}{k } T_{ \parallel}, T_{\perp},
\frac{K}{k } T_{\parallel}\right) \exp({\ii}Kx-qz), \,\,\,\nonumber \\
{\bf B}=  \left(-\frac{{\ii}q}{k } T_{ \perp}, -
T_{\parallel}, \frac{K}{k } T_{ \perp}\right)  \exp({\ii}Kx-qz) .\,\,\,\, \label{evan}
\ee 
For TE or $s$ (TM or $p$) - polarization , i.e. ${\bf E}$ (${\bf B}$) perpendicular to the plane of incidence $OXZ$, only those components with the transmission coefficient $T_{ \perp}$,
($T_{\parallel}$) are chosen. $ K$  denotes the component, parallel to the interface, of the wavevector ${\bf k}= ( K,0,{\ii}q)$, $q=\sqrt{ K^2 -k ^2}$.

The densities of energy, $w=w_e+w_m$, reactive power, $w_{react}=2 \omega (w_m-w_e)$, Poynting 
momentum, spin and {\rr canonical} momentum of this wave in $Z\geq 0$ are straightforwardly obtained from  (\ref{evan}) and  well-known \cite{nietoPRR,bliokh1}.

Since in this case \cite{nietoPRR} ${\bf P}^{S\,I}=-{\bf P}^{O\,I}$,   the global time-averaged force density  in $z\geq 0$ due to the evanescent wave is, following  (\ref{bcfor5A7}),  equal to zero. This is to be expected since no body exists in $z\geq 0$; and such a force  $<\bm{\mathcal F}>$, Eq. (\ref{bcfor5A7}),  with $V$ being  the  source-free space $z\geq 0$, is [cf. Eq. (\ref{bcfor5A3})] also equal to the flow  RMST on the surface $\partial V$. Since this flow is the same whatever the contour surrounding the sources is, we can take $\partial V$ in the far-zone,  ($kz\rightarrow\infty$), at which the evanescent wave is zero, and thus it does not contribute to this flux.

Furthermore, note that the evanescent wave CMST divergence is
\be
\nabla\cdot T_{ij}=\partial_j T_{ij}=\partial_zT_{i3}=[{\ii}\frac{Kq^2}{4\pi k^2}(|T_{\parallel}|^2-|T_{\perp}|^2),0,0] e^{-2qz}\nonumber \\
=-{\ii}\frac{K}{\omega}(w_{react},0,0). \,\,\,\,\,\,\,\,\,\,\,\,\,\,\label{evanreeners5}
\ee
The extreme right of (\ref{evanreeners5}) has been written in terms of  $w_{react}$  \cite{nietoPRR}. The real part of  (\ref{evanreeners5}) is zero, in agreement with the above  statement. Nevertheless, the imaginary part is
\be
\nabla\cdot T_{ij}^I=\partial_zT_{i3}^I=\omega({\bf P}_m^{S}-{\bf P}_e^{S})=-\frac{K}{\omega}(w_{react},0,0), \,\,\,\,\,\,\,\,\,\,\,\label{evanreeners6}
\ee
which is associated [cf. Eq. (\ref{geninst})] to a back and forth flow of momentum  in the  $x$-direction of propagation of the evanescent wave, without any net transfer of momentum to any body in $z\geq  0$. Besides we know \cite{nietoPRR} that
\be
-\frac{K}{\omega}(w_{react},0,0)=-\omega({\bf P}_m^{O}-{\bf P}_e^{O}). \label{evanreeners7}
\ee
Hence, replacing  the first term of the right side of Eq. (\ref{bcfor10}) by (\ref{evanreeners6}) and (\ref{evanreeners7}), we get for the reactive force density: $\frac{1}{2}\mbox{Im} [ \rho^*{\bf E}({\bf r})+\frac{\bf J^*}{c}\times{\bf B}({\bf r})]_i=0$, wich agrees with the absence of object in  $z\geq 0$.

In addition, like the {\rr canonical} momentum, (cf. Eq.(\ref{bcfor4A11}) and \cite{nietoPRR}):  ${\bf P}^{O}=\frac{K}{\omega}\,[w \, ,0\, ,0 ]$, the flow  $\nabla\cdot T_{ij}^I=\omega({\bf P}_m^{S}-{\bf P}_e^{S})$ is superluminal, ($K>k$). However, while ${\bf P}^{O}$ has an $x$-component characterized by the electromagnetic energy density $w$, the flow density $\nabla\cdot T_{ij}^I$ is, according to  (\ref{evanreeners6}), governed by the reactive power,  $w_{react}$, of the evanescent wave.

\section{Near-field nature of the imaginary stress tensor}
As discussed above, (see also Appendix D), the flow IMST is zero in the far-zone because in that region only the diagonal part of the CMST contributes to this flow, and hence it  is real. Next we prove that only evanescent components exist in the flow IMST, thus showing its near-field nature as a reactive quantity \cite{nietoPRR}.

Using Eq.  (\ref{bcfor4A22}), we use for simplicity a framework such that the sources are in
$z < 0$ and thus the integration is done on the $z = z_0\geq 0$ plane. The electric field  ${\bf E}({\bf r})$ propagating into the half-space $z \geq 0$ is represented by its angular spectrum of plane-wave components \cite{nietoPRR}. Using the subscripts $h$ and $e$ for homogeneous and evanescent components of complex amplitudes: electric, ${\bf e}({\bf K})$, and magnetic, ${\bf b}({\bf K})$,  the integrated IMST divergence  (\ref{bcfor4A22}) per unit length on $z$  is, (cf. Appendix C of Supp. Mat.):
\be
\omega\int_{-\infty}^{\infty} d^2 {\bf R} ({\bf P}_m^S-{\bf P}_e^S)\nonumber \\
={\pi}\int_{K> k}d^2 {\bf K} \, q_e \exp(-2q_e z_0) \mbox{Im}\{b_{e\,z}^*({\bf K}){\bf b}_{e\,\perp}({\bf K}) \nonumber \\
 -e_{e\,z}^*({\bf K}){\bf e}_{e\,\perp}
({\bf K}) \} .\,\,\,\,\,\,\,\,\label{bcfor6b2}
 \ee
Where 
\be
{\bf b}_{\perp}({\bf K})=(b_x({\bf K}), b_y({\bf K}),0 ),\,\,\,\, {\bf e}_{\perp}({\bf K})
=(e_x ({\bf K}),  e_y  ({\bf K}) , 0 ). \nonumber
\ee

Equation (\ref{bcfor6b2}) shows a contribution from only the evanescent part of the angular spectrum of the fields and, as such, this integral is a near-field quantity. Notice the characteristic exponential $z$-decay as $z=z_0$ increases, and that  the transversal components, ${\bf E}_{\perp}$ and ${\bf B}_{\perp}$, contribute to this IMST flux in such a way that the component of this flow  along $OZ$, $\int_{-\infty}^{\infty} d^2 {\bf R} ({\bf P}_m^S-{\bf P}_e^S)\cdot  \hat{\bf z}$,  is zero, as should be,  in the decay direction. Hence, like the imaginary Poynting momentum \cite{nietoPRR}, although in the near-field and intermediate regions the IMST  contributes locally to the ILF,  beyond it conveys no force.

\section{Reactive forces on an electric and  magnetic dipole}

\subsection{The imaginary  Maxwell stress tensor and the reactive force on a magnetoelectric dipolar particle}

The flow CMST  on a dipole is  obtained in Appendix D of Supp. Mat.. Calculations are done in the near-field zone of the emitter. We remark, however, that unless one makes the integration surface  $\partial V$,  enclosing the dipole, to strictly  shrink into its center point, the flux RMST that yields the RLF  is  better obtained in the far-zone as done in \cite{nieto1}, where it also acquires a physical significance as the ``radiated" MST flow, independent of the distance to the dipole, like the radiated energy \cite{nietoPRR}.

The reactive part, the IMST, stems   from the near-field zone of the dipolar fields on the  integration surface $\partial V$, Eq. (\ref{bcfor10}). Hence there is no contribution from the far-field. As stated above, the integration of the CMST in the far-zone is  real. We address dipolar particles in the wide sense \cite{nieto_torque}, namely whose  polarizabilities are defined  by the first electric and magnetic Mie coefficients.

The flow IMST  depends on the integration surface, like the reactive Poynting vector flux \cite{nietoPRR}. Introducing the result of Appendix D (see Supp. Mat.) into Eq. (\ref{bcfor10}), the reactive force on a small dipolar magnetoelectric particle is

\be
\bm{\mathcal F}_k^I = \mbox{Im}\{ \int_{\partial V_0}d^2 r \, T_{kj}^{(mix)} n_j
\}+\omega\int_{V\rightarrow V_0}d^3 r\,  ({\bf P}_{m}^{O} - {\bf P}_{e}^{O}) . \nonumber \\
\,\, \,\, (j,k=1,2,3). \,\,\,\,\,\,\,\,\,\,\,\,\,\,\,\,\,\,
\label{bcfor17}
\ee
Where as seen in Appendix D:
\be
\mbox{Im}\{ \int_{\partial V_0}d^2 r \, T_{kj}^{(mix)} n_j\}=
\mbox{Im}\{[\frac{1}{10}(1-{\ii}ka)\,\,\,\,\,\,\, \,\,\,\,\,\,\,\,\,\,\,\,\,\,\,\,\,\,\,\,\,\,\,\,\,\,\,\nonumber \\
-\frac{1}{30}k^2a^2]\exp({\ii}ka)\,{p_j}\,[\partial_k E_j^{*}+ \partial_j E_{k}^{*}]
 \,\,\,\,\,\,\,\ \,\,\,\,\,\,\, \,\,\,\,\,\,\,\,\,\,\,\,\,\,\nonumber \\  
+\frac{1}{3}(1 + {\ii}ka)\exp (-{\ii}ka)\,{p_j^*}\partial_j E_{k}\,\,\,\,\,\,\, \,\,\,\,\,\,\,\,\,\,\,\,\,\,\,\,\, \,\,\,\,\,\,\,\,\nonumber \\
+\frac{1}{6}(k^2a^2-{\ii}ka)\exp(-{\ii}ka)p_j^*(\partial_k E_{j}-\partial_j E_{k})\}
\,\ \,\,\,\,\,\,\, \,\,\,\,\,\,\,\,\,\,\,\,\, \,\,\,\,\,\,\,\nonumber \\
-\mbox{Im}\{[\frac{1}{10}(1-{\ii}ka)-\frac{1}{30}k^2a^2]\exp({\ii}ka){m_j^*}\,[\partial_k B_j
 \,\,\,\,\,\,\,\ \,\,\,\,\,\,\, \,\,\,\,\,\,\,\,\,\,\,\,\,\,\nonumber \\  
+ \partial_j B_{k}]+\frac{1}{3}(1 + {\ii}ka)\exp (-{\ii}ka)\,{m_j}\partial_j B_{k}^*\,\,\,\,\,\,\,\,\,\,\,\,\,\,\,\,\,\,\,\,\nonumber\\
+\frac{1}{6}(k^2a^2-{\ii}ka)\exp(-{\ii}ka)m_j(\partial_k B_{j}^*-\partial_j B_{k}^*)\}
.  \,\,\,\,\ \,\,\,\,\,\,\,\,\,\, \,\,\,\,\,\,\,\,\,\,\label{12}
\ee
 The superscript $(i)$ has been omitted in (\ref{12}),  understanding that ${\bf E}$, ${\bf B}$ represent the incident field. Equation (\ref{12}) is the flow IMST across the minimum sphere $\partial V_0$, of radius $a$, that contains the dipole. If this is a magnetoelectric spherical particle, $\partial V_0$ is the particle surface taken from outside.  In contrast with the  RMST and time-averaged force, RLF, (see Appendix D), the  flow IMST gives no interference between the induced electric and magnetic dipolar moments ${\bf p}$ and ${\bf m}$. The superscript $(mix)$ in Eq. (\ref{bcfor17}) expresses quantities due to the interference of the incident and scattered fields, (cf. Appendiz D). 

As the exterior volume  $V$ tends to $V_0$, the integral of (\ref{bcfor17})  is :
\be 
\int_{V_0}d^3 r\, [{\bf P}_{m}^{O}- {\bf P}_{e}^{O}]=\int_{\partial V_0}d^2 r\, [{\bf P}_{m}^{O\,(mix)}- {\bf P}_{e}^{O\,(mix)}] \,\,\,\,\nonumber \\
+\int_{ V_0}d^3 r\, [{\bf P}_{m}^{O\,(in)}- {\bf P}_{e}^{O\,(in)}]\, . \,\,\, \,\,\,\,\,\, \label{bcfor8e11}
\ee
The integral on $\partial V_0$ is due to the contribution of the external ROM  in $V-V_0$ that as $V\rightarrow V_0$  
shrinks into $\partial V_0$. The  superscript $(in)$ denotes that the canonical momenta in (\ref{bcfor8e11}) are those of the field inside the sphere. The interference ROM density  is:
$({\bf P}_{e}^{O\,(mix)})\,_i-({\bf P}_{m}^{O\,(mix)})_i=\frac{1}{8\pi \omega}\mbox{Im}  [E_j^{(i)\,*} 
\partial_i E_j^{(s)}+E_j^{(s)\,*} \partial_i E_j^{(i)}- B_j^{(i)\,*} \partial_i B_j^{(s)}-B_j^{(s)\,*} \partial_i 
B_j^{(i)} ]$. In the next subsection [cf. Eq. (\ref{bcfor8e})]  we prove that $\omega\int_{V_0}d^3 r\,[({\bf P}_{e}^{O\,(s)})\,_i-({\bf 
P}_{m}^{O\,(s)})_i]=\frac{1}{8\pi }\int_{V_0}d^3 r\,\mbox{Im}  [E_j^{(s)\,*} \partial_i E_j^{(s)}- 
B_j^{(s)\,*} \partial_i B_j^{(s)} ]$, that would also appear in the right side of  (\ref{bcfor8e11}), is zero.

We next consider the simplified approximation of Eq. (\ref{12}), (cf. Appendix D):
\be
\mbox{Im}\{ \int_{\partial V_0}d^2 r \, T_{kj}^{(mix)} n_j\}=
-\frac{1}{10}\mbox{Im}\{ [p_j\,\partial_k E_{j}^{*}+ p_j\,\partial_j E_{k}^{*}\nonumber \\
-m_j\,\partial_k B_{j}^{*}- m_j\,\partial_j B_{k}^{*}]\}, \,\,\,\,\,\,\,\,\, (j,k=1,2,3). \,\,\,\,\,\,\,\,\,\,\,\,\label{13}
\ee
[If no contribution of the radiative part of the scattered field ${\bf E}^{(s)}$ were considered in the interference terms, one would have an expression like (\ref{13}) but with a factor $1/15$, rather than $1/10$, (see Appendix D)].

Let the electric  and magnetic plarizabilities of the object  be $\alpha_e$ and $\alpha_m$, so that ${\bf p}=\alpha_e {\bf E}^{(i)}$,  ${\bf m}=\alpha_e {\bf B}^{(i)}$, (body chirality is outside our aim here),  and assume $(1-{\ii}ka)\exp({\ii}ka)\simeq 1$.  Equation (\ref{13}) may then be written as
\be
\mbox{Im}\{ \int_{\partial V_0}d^2 r \, T_{kj}^{(mix)} n_j\}=
-\frac{1}{20} \{{\alpha}_e^I\,\nabla|{\bf  E}|^2 - {\alpha}_m^I\, \nabla|{\bf B}|^2 \}\nonumber \\
+\frac{4\pi \omega}{5}\{{\alpha}_e^R\, {\bf P}_{e}^{O} - {\alpha}_m^R\, {\bf P}_{m}^{O}\}\,\,\,\,\,\,\, \,\,\,\,\,\,\,\,\, \nonumber \\
-\frac{1}{10} \mbox{Im}\{\alpha_e   ({\bf E}\cdot \nabla)  {\bf E}^{*}  -\alpha_m ({\bf B}\cdot \nabla)  {\bf B}^{*}\}.
\,\, \,\,\,\,\,\, \label{bcfor201a}
\ee
The  two terms of the third curly bracket of (\ref{bcfor201a}) come from the effect of the near field on the particle;  as such,  they have a resemblance with those of the  time-averaged force on a quasistatic electric or magnetic dipole: $(1/2)\mbox{Re} \{\alpha_e  ({\bf E}\cdot \nabla)  {\bf E}^{*}\}$ and  $(1/2)\mbox{Re} \{\alpha_m  ({\bf B}\cdot \nabla)  {\bf B}^{*}\}$ \cite{boyer} .

We note that the first and second curly brackets  of (\ref{bcfor201a}),   i.e. the gradient and {\rr canonical} momentum terms, appear with the real and imaginary parts of the polarizabilities interchanged with respect to  those  of the time-averaged electric and magnetic forces (cf. Appendix D, see also \cite{nieto1}). Except when particles are strongly resonant, the imaginary parts of the polarizabilities  are a factor $k^3 a^3$ smaller than the real parts \cite{nieto1}.  Thus the conservative  gradient part of the reactive force   would appear with a much smaller weight than its  non-conservative   component. By contrast, as we shall illustrate, in resonance  such gradient reactive forces would be the strongest if the particle is not highly absorbing. 

Introducing    the Belinfante spin momentum and using the vector identities for divergenceless fields: $\mbox{Re}[ ({\bf E}\cdot \nabla)  {\bf E}^{*} ]=
\frac{1}{2}\nabla|{\bf  E}|^2-k\,\mbox{Im} [{\bf E}\times{\bf B}^* ]$\,;\,\,\,
$\mbox{Im}[ ({\bf E}\cdot \nabla)  {\bf E}^{*}]=\frac{1}{2}\nabla \times \mbox{Im}({\bf E}^*\times{\bf E})$;  with analogous relations for the ${\bf B}$-vector. Then the simplified approximation to the IMST, Eq. (\ref{bcfor201a}),  may be expressed in terms of the conservative gradient components, the real and imaginary field (Poynting) momenta  and the {\rr canonical momenta}, as
\be
\mbox{Im}\{ \int_{\partial V_0}d^2 r \, T_{kj}^{(mix)} n_j\}=
 -\frac{1}{10} \{\alpha_e^I\nabla|{\bf  E}|^2 - \alpha_m^I \nabla|{\bf B}|^2 \} \nonumber \\
+\frac{4\pi \omega}{5}\{(\alpha_m^R - \alpha_e^R)<{\bf g}>
+(\alpha_e^I -\alpha_m^I){\bf g}^I \, \,\,  \,\, \,\,\,\,\,\,\nonumber \\
+2 (\alpha_e^R{\bf P}_e^O - \alpha_m^R {\bf P}_m^O)\}  . \, \,\,  \,\, \,\,\,\,\,\, \label{bcfor201}
\ee
The sign of the gradient component of the IMST depends on that of the first curly bracket. In the next examples we shall show that the accretion of ROM  in and around the particle, influences the ``radiated'' time-averaged RMST into the   far-zone, and hence the RLF. 

In Appendix E of Supp. Mat. we heuristically intend to derive the ILF on an electric and a magnetic dipole following the procedure employed in \cite{patric2000} for the RLF. It is intriguing that, for example  on a purely electric dipolar particle hit by a linearly polarized plane wave, that ILF  is five times  the expression given by Eq. (\ref{13}) or its equivalent (\ref{bcfor201}).  In the Example 2, below, we shall analyze this latter case, showing the correctness of  the ILF obtained in Appendix E.

However, the ILF on a purely magnetic dipolar particle obtained in Appendix E presents a sign opposite to that of  Eq. (\ref{13}).  In fact, the polarization current density  holds: $\nabla\cdot\bm{\mathcal J}=\nabla\cdot d\bm{\mathcal P}/dt$ (see e.g. \cite{Zwangill}, Section 14.2.2), and thus  one may add to the  relationship: $\bm{\mathcal J}= d\bm{\mathcal P}/dt$ any term $\nabla\times \bm{\mathcal M}$. Hence, given the lack of experimental evidence yet,  we have an indeterminacy of such a term. Further research may clarify whether this ambiguity  causes the failure of the equations of  Appendix E  to describe the reactive force on resonant dipolar particles, (see Examples 4 and 5 below), while this term $\nabla\times \bm{\mathcal M}$ appears to be zero for non-resonant dipoles, as seen in Example 2. Hence, taking into account that the complex Maxwell stress tensor theorem and its consequences is an unexplored territory, the formulation of Appendix E on dipolar particles deserves further experimental and theoretical research.

\subsection{Further physical consequences. The overall spin momentum. The interior and external reactive  strength of orbital momentum}
As stated in Appendix D of the Sup. Mat., the  far-zone flow IMST on a dipole, is zero term by term since only the diagonal elements of the CMST, which are real,  contribute to it, giving the time-averaged force, (see also \cite{nieto1}). So we infer that, in addition to (\ref{bcfor4b11}), we have
\be
\int_{V_{\infty}}d^3r \, {\bf P}_m^S=\int_{V_{\infty}}d^3 r \, {\bf P}_e^S=0. \label{bcfor4b2}
\ee
Equation  (\ref{bcfor4b2}) states  {\it  the vanishing of the overall  electric and magnetic  spin momenta of the total field, (i.e. incident plus scattered) produced by a dipole}. It includes the object volume $V_0$ of charges and currents. This is a generalization of a previous result \cite{bliokh1} for  the spin momentum of a free field or of an evanescent wave \cite {bliokh1}. As seen from  Appendix D, Eq.  (\ref{bcfor4b11}), but not Eq. (\ref{bcfor4b2}), also holds for the scattered field.

Since, as we have seen in connection with Eq. (\ref{bcfor10bb}),  as $V=V_{\infty}$ one has that  ${\bf P}_{m}^{O\,FF}= {\bf P}_{e}^{O\,FF}$, it is likely that (\ref{bcfor4b2}) holds for the total field from any arbitrary body. However, we have not been able to prove it here. Note, nonetheless, that although the spin momentum of the total field is zero in the whole space with sources, this momentum is not  a virtual quantity locally. Indeed the Belinfante spin momentum gives rise to forces near surfaces, like it happens with the transversal force, which depends on the reactive helicity of evanescent waves created on $\partial V_0$ \cite{nietoPRR,bliokh1}.

 The fields ${\bf E}^{(in)},\,{\bf B}^{(in)}$ inside the object volume $V_0$ cannot produce any  net force  on the body. Therefore  using Eq. (\ref{bcfor8}) for these fields one has: $ \int_{\partial V_0}d^2 r \, T_{ij}^{(in)} n_j=-{\ii}\omega\int_{V_0}d^3 r\, [{\bf P}_{m}^{O\,(in)} - {\bf P}_{e}^{O\,(in)}]$.

On the other hand, we may write an equation like  (\ref{bcfor10}) for the scattered field in an arbitrary volume $V$ enclosing the dipole
\be 
\mbox{Im}\{\bm{\mathcal F}_i^{(s)}\}= \nonumber \\
\int_{\partial V}d^2 r \, \mbox{Im}\{T_{ij}^{(s)}\} n_j+{\ii}\omega\int_{V}d^3 r\, [{\bf P}_{m}^{O\,(s)}- {\bf P}_{e}^{O\,(s)}]_i\,. \,\,\, \,\,\,\, \label{bcfor8a}
\ee
Where $\mbox{Im}\{\bm{\mathcal F}_i^{(s)}\}$ is a force on the particle due to the scattered field only, certainly different from the actual reactive force $\bm{\mathcal F}_i^I$  stemmed from the total field.

Taking into account that, as seen in Eq. (\ref{bcfor8a}), if $V=V_{\infty}$ one has $ \int_{\partial V_{\infty}}d^2 r \,\mbox{Im}\{T_{ij}^{(s)}\} n_j=0$ since  $ \int_{\partial V_{\infty}}d^2 r \,T_{ij}^{(s)} n_j$  is real, (see also \cite{nieto1}), the above equation (\ref{bcfor8a}) becomes
\be
\mbox{Im}\{\bm{\mathcal F}_i^{(s)}\}={\ii}\omega\int_{V_{\infty}}d^3 r\, [{\bf P}_{m}^{O\,(s)} - {\bf P}_{e}^{O\,(s)}],\, \,\,\,  
\label{bcfor8b}
\ee
which is similar  to (\ref{bcfor10aa}). Then introducing (\ref{bcfor8b}) into (\ref{bcfor8a}) we obtain in terms of the ROM of the scattered field
\be 
 \int_{\partial V}d^2 r \, \mbox{Im}\{T_{ij}^{(s)}\} n_j={\ii}\omega\int_{V_{\infty}-V}d^3 r\, [{\bf P}_{m}^{O\,(s)}- {\bf P}_{e}^{O\,(s)}]_i\,, \,\,\,\,\,\,\,\,\, \,\,\,\,\, \label{bcfor8d}
\ee
which is analogous to (\ref{bcfor10bb}).
Now, according to Appendix D,  on a dipolar particle: $\int_{\partial V_0}d^2 r \, \mbox{Im}\{T_{ij}^{(s)}\} n_j=0$, where $V_0$ is the smallest sphere enclosing the dipole, which for a dipolar particle we  take as its volume. Then  (\ref{bcfor8d}) reduces to
\be 
\int_{V_{\infty}-V_0}d^3 r\, [{\bf P}_{m}^{O\,(s)}- {\bf P}_{e}^{O\,(s)}]=0\,, \,\,\,  \label{bcfor8e}
\ee
Therefore we conclude that the ROM of the field scattered from a dipolar particle is zero, independently of whether it is a Rayleigh one or dipolar in the  wide sense \cite{nieto1,nieto2}. Hence, {\it the ILF on a dipolar particle, Eq. (\ref{bcfor17}), comes  exclusively from those terms of both the IMST and   ROM that contain interference  of the incident and scattered fields}. We think that  this likely  occurs with  an arbitrary body, although we have not proven it here.

\subsection{Example 2: Linearly polarized plane wave impinging on a low refractive index small dielectric particle}

Consider a low refractive index dielectric particle, like a polystyrene sphere, (refractive index $n_{PS} = 1.59$, radius $a = 50$ nm), illuminated by a linearly polarized propagating plane wave:
\be
{\bf E}= E_0(1,0,0)e^{{\ii}kz}, \,\,\, \,\,\,\,\, {\bf B}=E_0(0,1,0)e^{{\ii}kz} , \label{ILFex1}
\ee
$E_0$ being a constant. The incident  momenta are:
\be
{\bf P}_e^O={\bf P}_m^O=<{\bf g}>=\frac{1}{8\pi c}E_0^2 \,\hat{\bm{z}}. \label{ILFex2}
\ee
For the time-averaged and the reactive forces we test Eqs. (E3) and (E4) of Appendix E in the Supp. Mat., which yield
\be
<\bm{\mathcal F}_e>=\frac{1}{2}\alpha_e^I \,k\,E_0^2 \,\hat{\bm{z}},\,\,\, \,\,\,\,\,
\bm{\mathcal F}_e^I=\frac{1}{2} \alpha_e^R\, k\,E_0^2 \,\hat{\bm{z}} . \label{ILFex3}
\ee

\begin{figure*}[htbp]
\begin{centering}
\includegraphics[width=16cm]{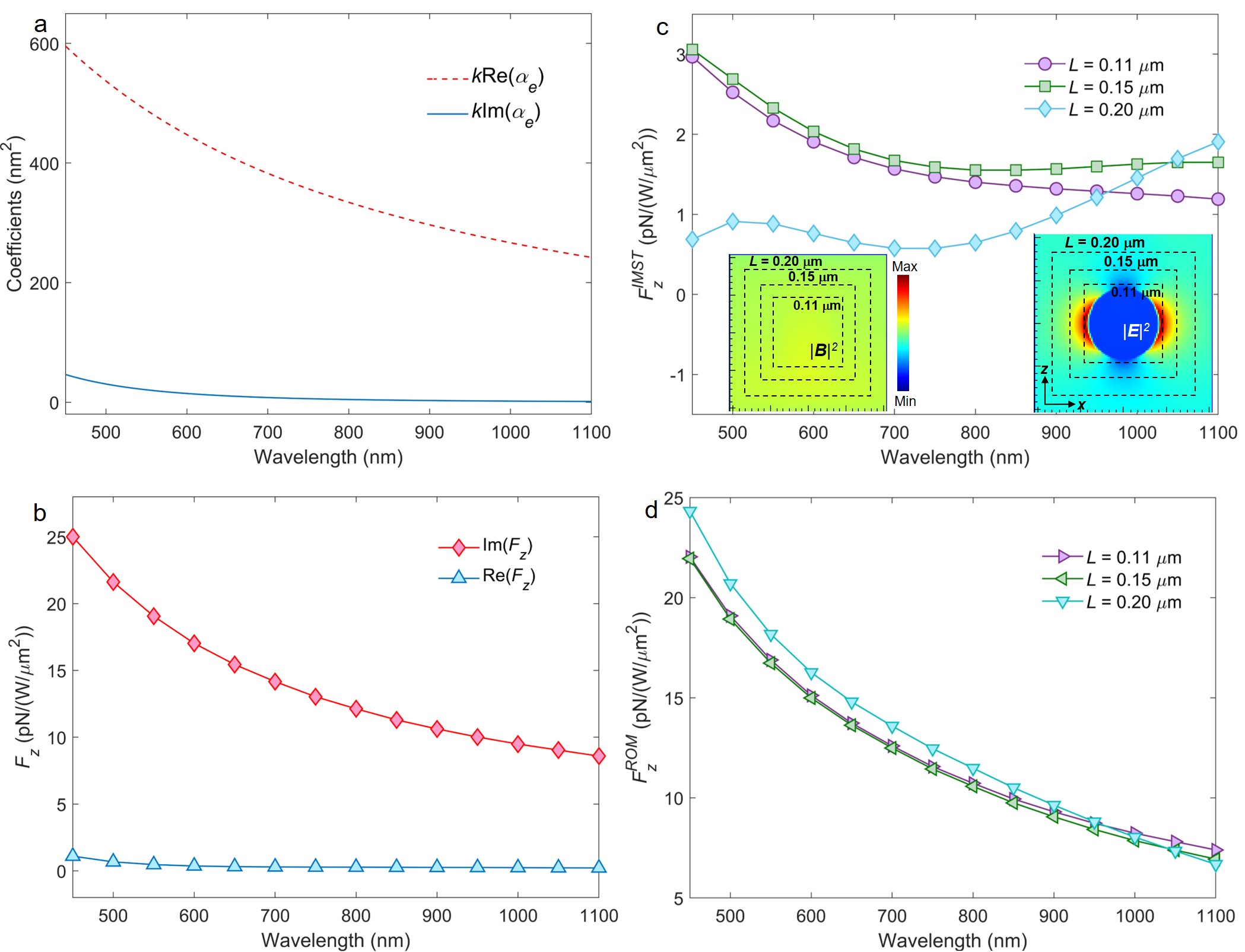}
\par\end{centering}
\caption{Linearly polarized propagating plane wave incident on a  polystyrene sphere of radius $a=50$ nm and refractive index $n_{PS} = 1.59$.  (a) Polarizability calculated via Mie theory.  (b) Numerical results, per unit of incident power density, $E_0=1$, for the ILF $F_z^I$, and $<F_z>$,  on the sphere. (c) IMST component, $F_z^{IMST}$, of  $F_z^I$, calculated with different cube integration contours. Insets show the spatial distribution on  the $x=0$ plane  of the electric and magnetic field intensities in and around the sphere excited at  $\lambda=800 \rm\,nm$. White broken line squares illustrate the integration contours. (d) ROM component, $F_z^{ROM}$,  of the ILF, calculated by subtracting  $F_z^{IMST}$ from  $F_z^I$.} 
\end{figure*}
Note that, as mentioned above, under plane wave illumination on a purely electric dipolar particle the ILF $\bm{\mathcal F}_e^I$, Eq. (\ref{ILFex3}),  is five times its IMST component given by the approximate equation   (\ref{bcfor201}). The numerical computation methods in this and following examples is the same, and indicated in Appendix G of the Supp. Mat..

The real and imaginary parts of the electric polarizability, $\alpha_e^R$ and $\alpha_e^I$,  as well as  $<\bm{\mathcal F}_e>$ and $\bm{\mathcal F}_e^I$, are plotted versus $\lambda$ in the visible and near-infrared in Figs. 2(a) and 2(b). The ILF, $\bm{\mathcal F}_e^I$,  of (\ref{ILFex3}) agrees well with its numerical calculation through the charge and current volume integral of the left side of Eq. (\ref{bcfor10}) and following the procedure described in Appendix G \cite{lumerics}. For example for $\lambda = 450$ nm Fig. 2(a) reads $k\alpha_e^R= 600 \times 10^{-18} \,{\rm m^2}$  per unit of incident power density.  This yields an ILF per unit power: $25\,{\rm pN/(W/\mu m^2)}$, which coincides with its value in Fig. \textcolor{red}{2}(b). In the same way, there is agreement between the RLF of  (\ref{ILFex3}) according to Fig. \textcolor{red}{2}(a) and its numerical computation of Fig. \textcolor{red}{2}(b), as it should.

As seen, the ILF is totally dominant upon the time-averaged force, specially at higher frequencies which lie in the visible region. This illustrates the physics described in this work; namely, underneath a weaker time-averaged force,  there is a relatively large build-up of ROM and thus of reactive force. This is shown in Fig. 2(c)  of  the IMST component, $F_z^{IMST}$, numerically obtained from the IMST, Eq. (\ref{bcfor10b}), evaluated from the total field through an FDTD computation via the Mie series, and Fig. 2(d) of the ROM component, $F_z^{ROM}$,  obtained by substracting  $F_z^{IMST}$ from  $F_z^I$. We observe that $F_z^{IMST}\ll F_z^{ROM}$, so that the ILF is largely due to its stronger ROM component.

The aforementioned counteraction of both the ILF and ROM on the RLF, evident in these figures, indicates the observability of the ILF and ROM, as well as the reactive nature of this near-field radiation pressure versus the ``far-field'' flow  RMST that yields the time-averaged force. Thus, underlying the weak RLF, $<\bm{\mathcal F}_e>$,  exerted by an incident wave on a low refractive index dielectric particle, there are a dominant  ROM and reactive force. This has an analogy  with the detrimental effect of the reactive work,  associated to the reactive power,  over the time-averaged work and  power radiated by an emitter into the far-zone \cite{geyi,nietoPRR,balanis}.

\rr Although the incident optical angular momentum gives rise to a reactive optical torque, as shown in Appendix B of Supp. Mat., and twisted structured light may have an effect on the ILF, the angular momentum of a beam similar to a circularly polarized (CP) plane wave does not alter the reactive force, $\bm{\mathcal F}^I$, of Eq. (\ref{ILFex3}), which therefore  also  applies for CP incidence.  The same as it happens with the time-averaged force. As a matter of fact, it is straightforward to see that the incident canonical momentum, Eq. (\ref{ILFex2}), is insensitive to the  spin in this case. 

This is illustrated in Fig. S1 of Appendix F of the Supp. Mat., which compares the electric field spatial distribution and the reactive force from a linearly polarized (LP) plane wave  with those pertaining to circular polarization (CP); both impinging on the above discussed PS particle. The angular momenta of more complex structured light fields may, however, have an effect on the ILF. This question is left open for future research.  \bb

\subsection{Example 3: Linearly polarized Gaussian beam incident on a small particle, electrically dipolar}

We address a beam with Gaussian transversal profile, incident on an electrically dipolar dielectric particle
\be
{\bf E}= E_0\, e^{-\frac{R^2}{\sigma^2}}(1,0,0)e^{{\ii}kz}, \,\, \,\,\,\,\,(R=\sqrt{x^2+y^2}),  \label{ILFex4}
\ee
Now the gradient component of the RLF and ILF appear. We have
\be
{\bf P}_e^O={\bf P}_m^O=\frac{1}{8\pi c}E_0^2\, e^{-2\frac{R^2}{\sigma^2}} \,\hat{\bm{z}}; \label{ILFex5}
\ee
and 
\be
\nabla|{\bf E}|^2=-\frac{4}{\sigma^2} E_0^2\, e^{-2\frac{R^2}{\sigma^2}}(x,y,0). \label{ILFex5}
\ee
So that
\be
<\bm{\mathcal F}_e>=E_0^2\, e^{-2\frac{R^2}{\sigma^2}}[-\frac{\alpha_e^R}{\sigma^2} (x,y,0)+
\frac{\alpha_e^I}{2} \,k\,(0,0,1)],\,\,\, \,\,\,\,\,\nonumber\\
\mbox{Im}\{\bm{\mathcal F}_e\}=E_0^2\, e^{-2\frac{R^2}{\sigma^2}}\,[\,\frac{4\alpha_e^I}{\sigma^2} (x,y,0)+
{\alpha_e^R} \,k\,(0,0,1)] , \,\,\, \,\,\,\,\,\label{ILFex7}
\ee
which shows the Hooke behavior of the conservative  gradient component as expected in optical tweezer set-ups, along with the pushing  nature of the ILF gradient component, as well as of the RLF at wavelengths where $ \alpha_e^R <0$, case in which the  scattering ILF becomes pulling. Therefore, this example illustrates the exchange of $\alpha_e^R$ and $\alpha_e^I $ in the RLF and ILF in an optical manipulation set-up through their gradient and scattering  components.

\begin{figure*}[htbp]
\begin{centering}
\includegraphics[width=16cm]{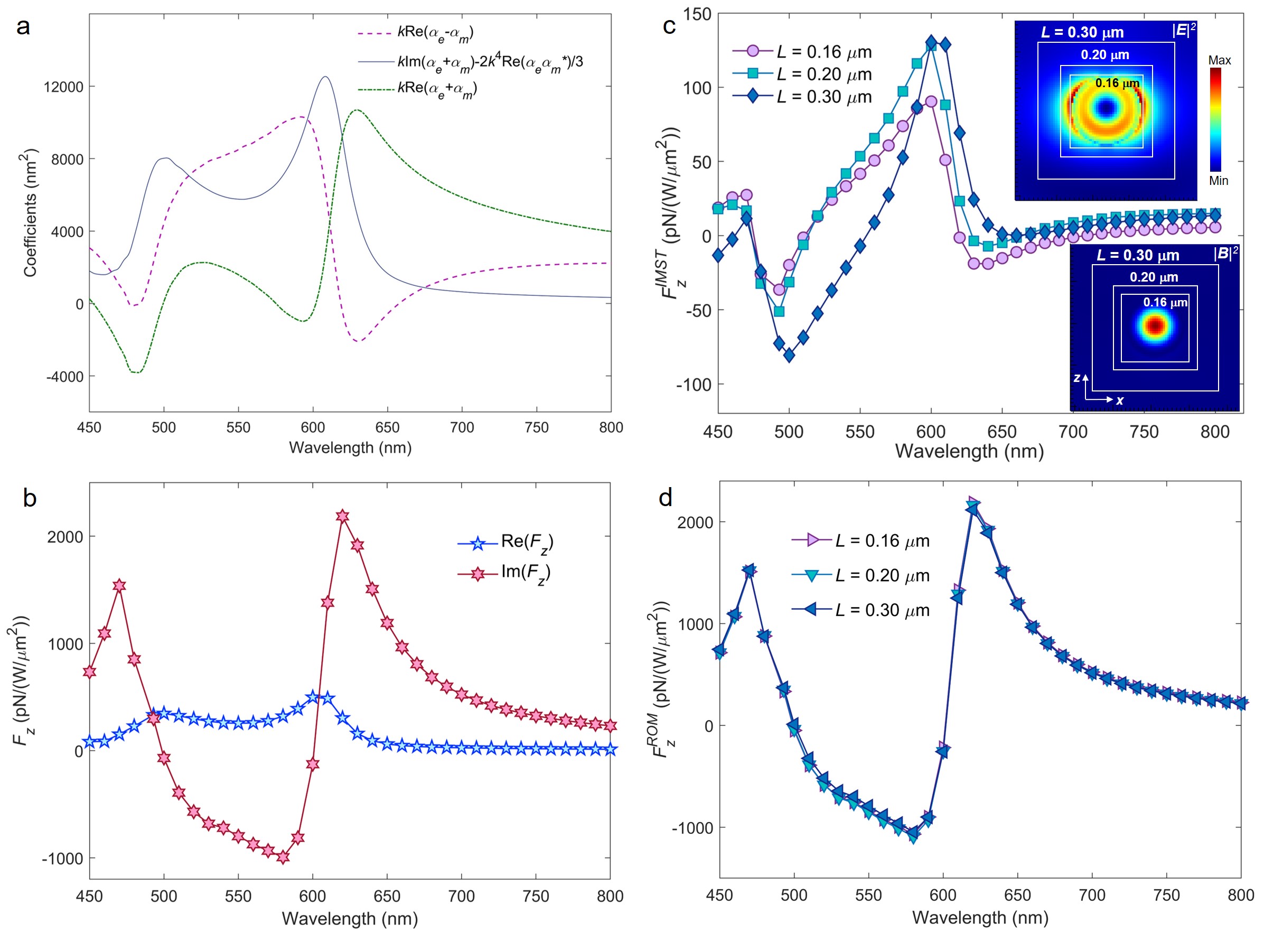}
\par\end{centering}
\caption{Linearly polarized propagating plane wave incident on a  Si sphere of radius $a=$75 nm. $E_0=1$. (a) Combined polarizabilities calculated via Mie theory. The expression of  $<F_z>$ given by these polarizabilities according to \cite{nieto1}  is shown in the full line, while the broken one depicts the theoretical  flow IMST, $F_z^{IMST}$, component  of the reactive force,  $F_z^I$, according to Eq. (\ref{bcfor201}).   (b) Numerical results for the time-averaged force $<F_z>$ and reactive force  $F_z^I$ on the sphere. $E_0=1$. (c) IMST component,  $F_z^{IMST}$, of  the ILF  numerically calculated with different cube integration contours. Insets show maps   on the  $x=0$ plane of $|{\bf E}|^2$ and $|{\bf B}|^2$ at  $\lambda=610$ nm which corresponds to the magnetic dipole resonance. White broken line squares illustrate the integration contours. (d) ROM component, $F_z^{ROM}$, of the ILF calculated by subtracting   $F_z^{IMST}$ from $F_z^I$.} 
\end{figure*}

\subsection{Example 4: Linearly polarized plane wave  incident on a  magnetoelectric high refractive index dipolar  particle}

Magnetoelectric particles are of great importance in nanophotonics \cite{won,nietoSi,kivshar,staude}. Given the well-known scaling property of high index spheres, results in a certain range of wavelengths are transposable to another spectrum in  particles with a different refractive index. It is, therefore, relevant to study this case as regards the build up of ROM and reactive forces.

We  consider the plane wave of Example 2 incident on a Si sphere of radius 75 nm. Figure 3(a)  depicts the RLF, the predicted IMST component, $\mbox{Im}\{ \int_{\partial V_0}d^2 r \, T_{kj}^{(mix)} n_j\} = (4\pi\omega/5)(\alpha_e^R - \alpha_m^R)$, according to the approximation:  (\ref{bcfor201}), and the factor $k(\alpha_e^R + \alpha_m^R)$ that yields the ILF according to Eq. (E21) of the Supp. Mat.

 The RLF, $F_z^R$, and ILF, $F_z^I$, numerically computed from the charge and current integral of the left side of Eqs. (\ref{bcfor9}) and (\ref{bcfor10}) as indicated in Appendix G, are shown in Figure 3(b). Now the effect of the electric and magnetic dipole resonances of the particle in the proximities of 500 and 610 nm, respectively, is observed.  Comparing Figs. 3(a) and 3(b) we note that Eq. (E21) does not yield the correct ILF;  neither Eq. (E20). In contrast with Example 2 of a non-resonant particle.

The shape of the computed spectrum of the flow IMST component $F_z^{IMST}$, shown in Fig. 3(c), is  similar to the theoretical ILF (\ref{ILFex3}) which follows the broken line of Fig. 3(a) according to Eq.  (\ref{bcfor201}).  The IMST force most similar to that of  (\ref{bcfor201}) is that obtained on the cube contour $\partial V$ with $L=0.16\;\rm\mu m$, almost tangent to the particle. On the other hand, a comparison of Figs. 3(b) and 3(d) shows an almost total contribution of the ROM to the ILF, manifesting the small weight of  $F_z^{IMST}$ in the reactive force on this resonant particle.

Figures. 3(b) and 3(c) indicate that $F_z^{IMST}$ is of the same order of magnitude as the time-averaged force in the whole range of wavelengths, while the ILF,  $F_z^{I}$, is much stronger than $<F_z>$, and  enhanced near the electric and magnetic resonances, although its sharp variation makes it to be near zero at the resonant wavelengths where the ``radiated'' force,  $<F_z>$, is maximum.  This latter feature keeps an analogy with the reactive power and helicity \cite{nietoPRR, ziolkowski} which are near-zero at resonant wavelengths at which there is maximum radiated power. \rr As a noted illustration, a photoinduced force microscopy experiment \cite{wika} detects the resonant  force  signal, on excitation of the magnetic dipole of a Si resonator, associated with the enhancement  of its quality factor and external stored reactive power. \bb

Actually, we observe  that $F_z^{I}$   is almost entirely due to the  large ROM contribution, $F_z^{ROM}$, and its enhancements, as depicted in Fig. 3(d). This once again illustrates the detrimental effect of this dominant ``reactance'' ILF on the ``dynamic radiative efficiency'' constituted by the time-averaged radiation force  RLF.  Now, according to (\ref{bcfor10aa}) the ILF equals the overall ROM, while we see from (\ref{bcfor10bb})    that $F_z^{IMST}$ is given by the external ROM, (which as remarked above is due to interference of the incident and scattered fields).

\begin{figure*} [htbp]
\begin{centering}
\includegraphics[width=16cm]{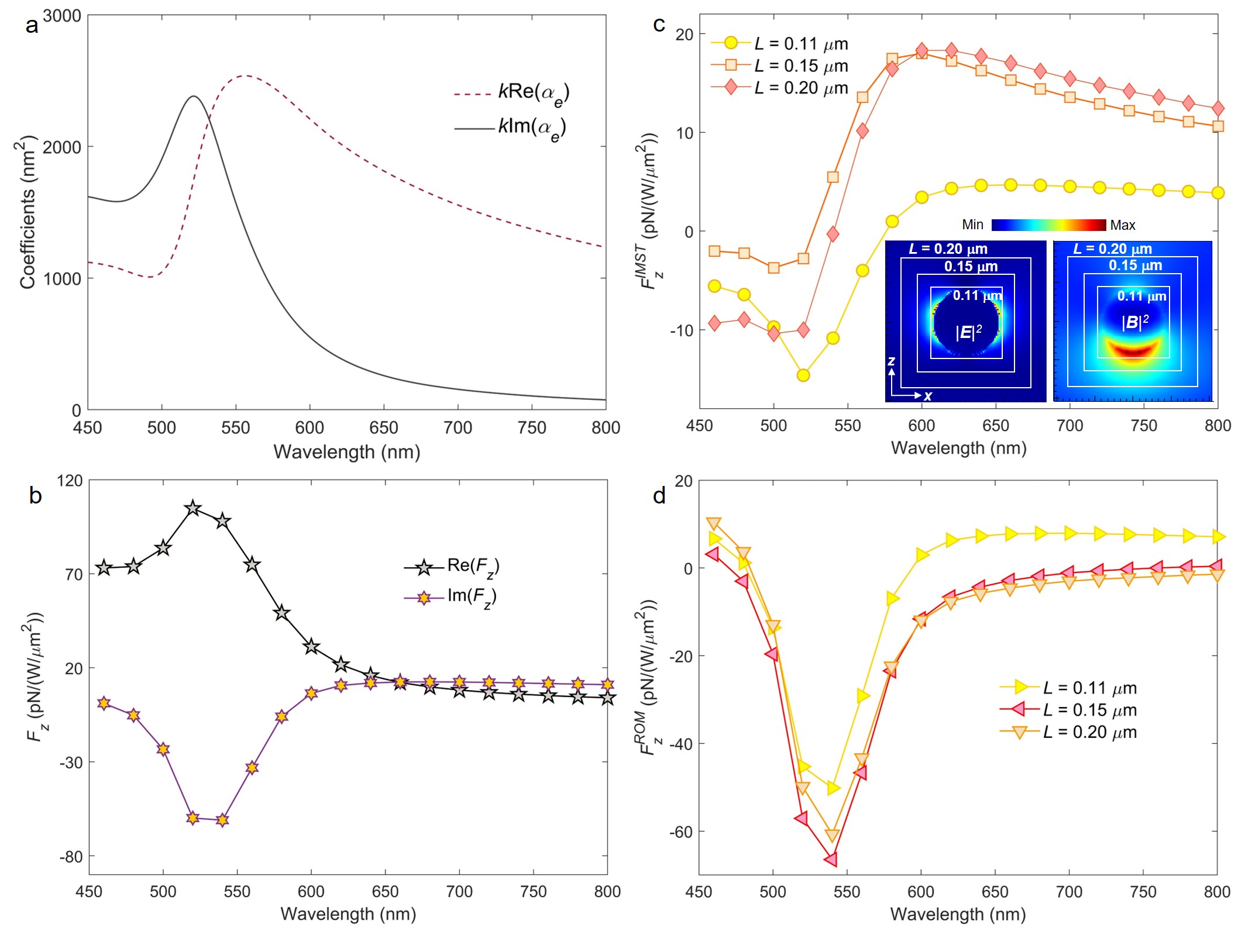}
\par\end{centering}
\caption{Linearly polarized propagating plane wave incident on an  Au sphere of radius $a=50\,{\rm nm}$. $E_0=1$. (a) Polarizability  calculated via Mie theory. (b) Numerical results for the time-averaged force $<F_z>$ and reactive force  $F_z^I$ on the sphere. $E_0=1$. (c) IMST component $F_z^{IMST}$ of the ILF  calculated with different cube integration contours. Insets show the spatial distribution on the  $x=0$ plane of the  electric and magnetic field intensities at the wavelength of 540 nm. White broken  line squares illustrate the integration contours. (d) ROM component, $F_z^{ROM}$, of the ILF calculated by subtracting $F_z^{IMST}$ from $F_z^{I}$.}
\end{figure*}

Therefore in a high index magnetoelectric particle, most of the ROM is built inside it; the reactive ILF  therewith mainly being due  to this internal ROM and its contribution, $F_z^{ROM}$, is completely dominant upon the time-averaged force.

\begin{figure*} [htbp]
\begin{centering}
\includegraphics[width=16cm]{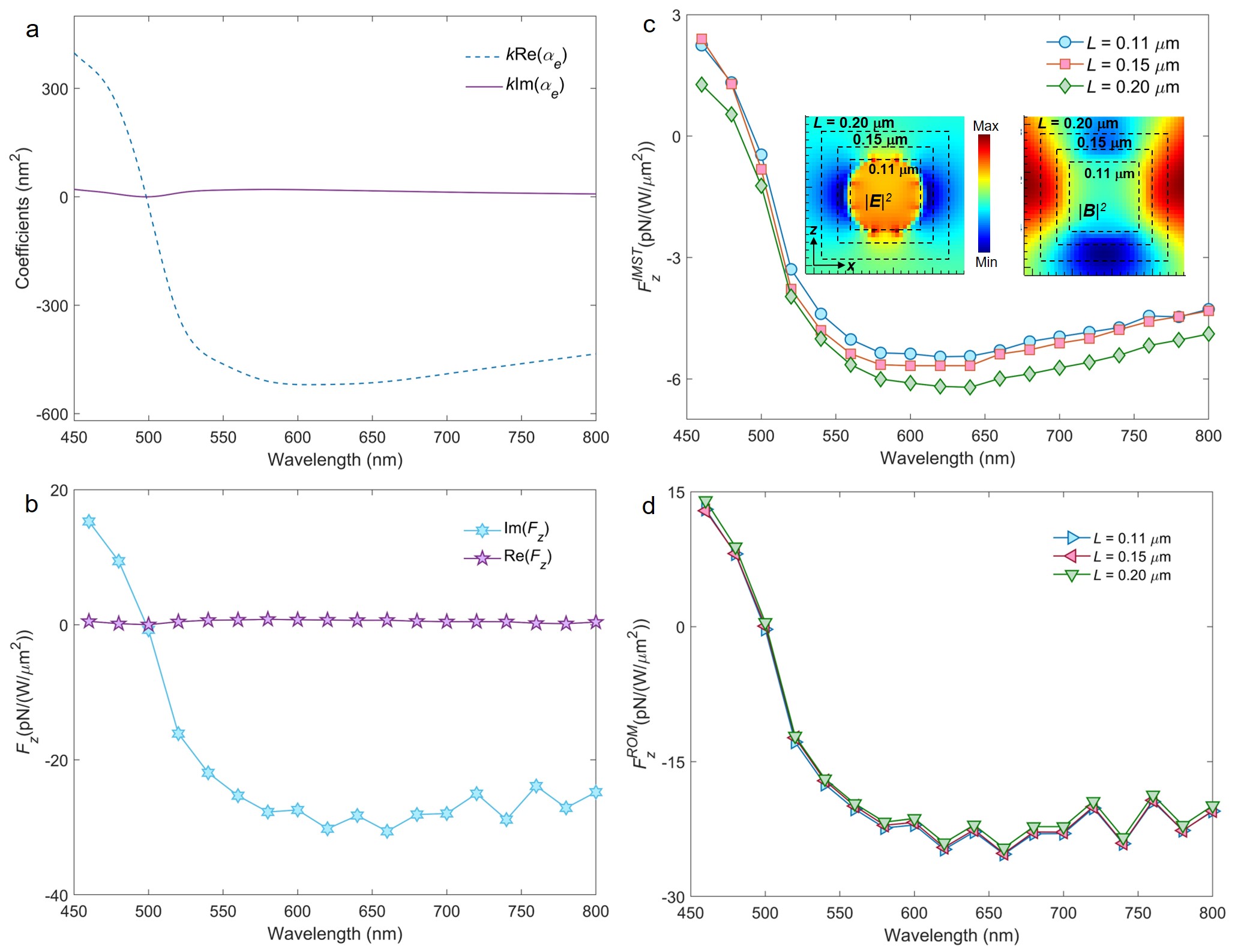}
\par\end{centering}
\caption{Same as Fig.4 for a  sphere of a hypothetical  material, that we call $\rr{\rm Au_3}\bb$, whose refractive index has a real part identical to that of  Au, but its imaginary part is artificially set to zero.}
\end{figure*}
\begin{figure*} [htbp]
\begin{centering}
\includegraphics[width=17cm]{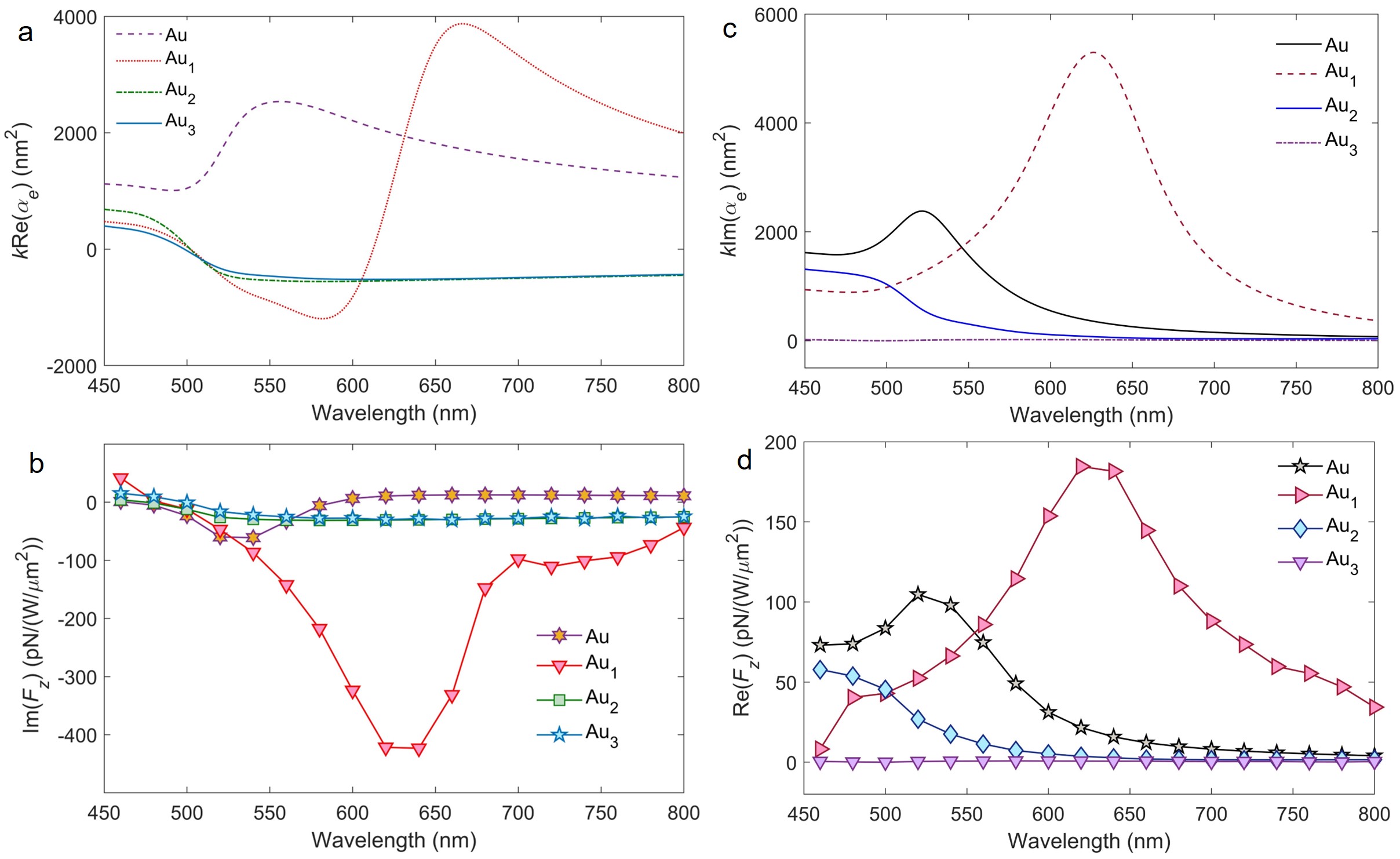}
\par\end{centering}
\caption{(a) Real part of polarizability of a sphere  of radius $a=$ 50 nm, from Mie theory of: Au and three hypothetical  materials denoted as $\rr{\rm Au_1}\bb$, $\rr{\rm Au_2}\bb$ and $\rr{\rm Au_3}\bb$, whose refractive index real parts are equal to those of Au, but their imaginary parts are artificially set to: A half  of Au imaginary part ($\rr{\rm Au_1}\bb$), the same as the real part ($\rr{\rm Au_2}\bb$), and zero ($\rr{\rm Au_3}\bb$). (b) Numerical results of ILF,  $F_z^I$, on each of  these spheres illuminated by a linearly polarized plane-wave with amplitude $E_0=1$.  (c) Imaginary part  of polarizability of the four spheres. (d) Numerical calculation of RLF, $<F_z>$,  on each of  these  spheres under the same incident wave as in (b).}
\end{figure*}

\subsection{Example 5: Linearly polarized plane propagating wave  incident on a plasmonic  particle}
  As seen in the previous examples, almost all contribution to the ILF is due to the internal ROM. Absorption in the volume   distribution of charges and currents changes this; it  may extract energy from the reactive power  \cite{harrington,banerjee,marengo}. A decrease  of reactive quantities would happen even in situations where both the RLF and ILF are resonant. To illustrate it, we now consider the above plane wave incident on  an plasmonic  Au spherical particle of radius $a=$ 50 nm, whose polarizability is shown in  Fig. 4(a).  A computation, as described in Appendix G of Supp. Mat.,  yields  the time-averaged force, $<F_z>$ and reactive force,  $F_z^I$, on this particle, [cf. Fig. 4(b)]. On the other hand,  Figs. 4(c) and 4(d) depict     $F_z^{IMST}$ and $F_z^{ROM}$, respectively.

We see that while $<F_z>$ follows the theoretical expression (\ref{ILFex3}) obtained from Appendix E, $F_z^I$ does not, specially in the region of the  plasmon resonance peak where in spite of the pulling ILF enhancement,  the RLF is stronger. Moreover, at $\lambda<$ 450 nm  the RLF is totally dominant versus a very small ILF, while the opposite occurs at $\lambda>$ 650 nm, in accordance with the expected \rr antagonistic \bb role of both kind of forces.
 Also, as the internal electric field  is practically zero, although the inner ${\bf B}$ does not vanish, the  weight of the   internal  ROM,   i.e.  of $F_z^{ROM}$ [cf. Fig. 4(d)] is now much smaller than in the non-lossy particles of the previous examples.  The contribution of $F_z^{IMST}$ [cf. Fig. 4(c)] to the ILF is  now   non-negligible  versus that of the internal ROM plus the ROM in the four corners between the integration cube and the particle surface. So  we see that field losses  in Au are responsible of diminishing the internal ROM contribution to the ILF versus that of the IMST.

To get a closer look to the effect of absorption, Figs. 5(a)-5(d) depict the same as before for a hypothetical particle ($\rr{\rm Au_3}\bb$) whose refractive index $\hat{n}_{Au3}$ has a real part $n_{Au3}$ identical to that of Au, $n_{Au}$, but whose imaginary part $\kappa_{Au3}$ has artificially been set to zero. We see that now, as no resonance is present, all quantities behave as in a dielectric sphere, [compare with Figs. 2(a)-2(d)]. The ILF follows the heuristic theoretical equation (\ref{ILFex3}) and the antagonic role of both forces is now retrieved. It is interesting, nevertheless, that  close to the wavelength where the ILF changes sign, both forces coincide being practically zero. 

In this connection, Figs. 6(a)-6(d) show a direct comparison of the polarizability, RLF and ILF for the Au sphere and three hypothetical particles whose refractive index real part is the same as that of Au,  $\hat{n}_{Au}$ but with the imaginary  part  artificially chosen as: a half that of Au ($\rr{\rm Au_1}\bb$), the same as the real part ($\rr{\rm Au_2}\bb$), and zero ($\rr{\rm Au_3}\bb$). In the $\rr{\rm Au_1}\bb$ case the plasmon resonance is much stronger than in Au, but although the RLF is accordingly stronger, the pulling ILF more than doubles the RLF. The \rr antagonistic \bb role between both forces is observed in the $\rr{\rm Au_2}\bb$ and $\rr{\rm Au_3}\bb$ cases.
\newline
\rr
\section{Recapitulation: Discussion on the complex stress tensor theorem in a dielectric medium}

If the medium surrounding the illuminated body is not vacuum, but has a non-unity refractive index $n$, there is an ongoing debate on the form of the field momentum. Thus one may ask how this will affect the above formulation of the CMST theorem and the ILF.

Among all proposals for the  Poynting momentum and Maxwell stress tensor, we make here a discussion of this question by focusing on the Abraham and Minkowski forms, which are those whose respective ranges of validity have been more countersigned by experiments \cite{qiu1,milonni}. Further study on other formulations is outside the aims of this paper.

The Minkowski, $\bm{\mathfrak  G}^M$,  and Abraham, $\bm{\mathfrak G}^A$,  field momentum densities of an electromagnetic wave in an isotropic, linear, homogeneous nonabsorbing and nondispersive  medium of time-independent refractive index $n=\sqrt{\epsilon\mu}\neq 1$ are, in terms of the real electric and magnetic vectors, $\bm{\mathfrak E}$ and $\bm{\mathfrak H}=(1/\mu)\bm{\mathfrak B}$, in Gaussian units employed in this work \cite{milonni}:
\be
\bm{\mathfrak G}^M=\frac{\epsilon\mu}{4\pi c}\bm{\mathfrak E}({\bf r},t)\times\bm{\mathfrak H}({\bf r},t)
 ={\epsilon\mu}\,\bm{\mathfrak G}^A.\,\, \label{AMR}
\ee
For the fields represented by the analytic signals we thus introduce the Minkowski and Abraham complex Poynting momentum densities:
\be
\bm{\mathcal G}^M=\frac{\epsilon\mu}{8\pi c}\bm{\mathcal E}({\bf r},t)\times\bm{\mathcal H^*}({\bf r},t)
 ={\epsilon\mu}\,\bm{\mathcal G}^A; \,\, \label{AMC}
\ee
whose real and imaginary parts define the time-averaged and imaginary Minkowski and Abraham  Poynting momentum densities, respectively.

Then the complex Maxwell stress tensor theorem reads according to Minkowski:

\be
\bm{\mathcal F}_i=-\int_{V}d^3 r \,\partial_t \bm{\mathcal G}_i^{M\,*}+\int_{\partial V}d^2 r \, \bm{\mathcal T}_{ij}^M n_j\nonumber \\
+{\ii}\omega \int_{V}d^3 r [\bm{\mathcal P}_{m}^{O\,M} - \bm{\mathcal P}_{e}^{O\,M}]_i\, . \,\,\,\,\,\,\,\,  \label{MCSTT}
\ee
With the  Minkowski complex stress tensor:
\be
\bm{\mathcal T}_{ij}^M=\frac{1}{8\pi}[\epsilon{\mathcal E}_i  {\mathcal E}_j^* +\frac{1}{\mu} {\mathcal B}_i ^*{\mathcal B}_j -\frac{1}{2}\delta_{ij}(\epsilon|\bm{\mathcal E}|^2 +\frac{1}{\mu} |\bm{\mathcal B}|^2 )], \,\,\,\,\,\,\,\,\,\,\,\, \label{tMforM34}
\ee
and Minkowski orbital momenta:
\be
(\bm{\mathcal P}_{e}^{O\,M})_i=\frac{\epsilon}{8\pi \omega}\mbox{Im}  [{\mathcal E}_j^* \partial_i {\mathcal E}_j],\,\,\, \,\,\,\,\nonumber \\
 (\bm{\mathcal P}_{m}^{O\,M})_i= \frac{1}{8\pi \mu\omega}\mbox{Im}[ {\mathcal B}_j^* \partial_i {\mathcal B}_j ] \,\,\, . \label{tMforM35}
\ee
Whereas according to Abraham it should be:
\be
\bm{\mathcal F}_i+\bm{\mathcal F}_i^A=-\int_{V}d^3 r \,\partial_t \bm{\mathcal G}_i^{A\,*}+\int_{\partial V}d^2 r \, \bm{\mathcal T}_{ij}^M n_j\nonumber \\
+{\ii}\omega \int_{V}d^3 r [\bm{\mathcal P}_{m}^{O\,M} - \bm{\mathcal P}_{e}^{O\,M}]_i\, ; \,\,\,\,\,\,\,\,  \label{ACSTT}
\ee
where $\bm{\mathcal F}_i^A=(\epsilon\mu-1)\int_{V}d^3 r \,\partial_t \bm{\mathcal G}_i^{A\,*}$, and we recall that in Eqs. (\ref{MCSTT}) and (\ref{ACSTT}) $\bm{\mathcal F}_i=\frac{1}{2}\int_{V}d^3 r \,(\rho^*\bm{\mathcal E}+\frac{1}{c}\bm{\mathcal J}^*\times\bm{\mathcal B}).$

Although on using (\ref{AMC}) in (\ref{ACSTT}) one inmediately retrieves  (\ref{MCSTT}), Eq. (\ref{ACSTT}) exhibits a first term on the right side that, in analogy with the first term in the right-hand side of (\ref{MCSTT}), suggests that $\bm{\mathcal G}_i^{A\,*}$ should be the complex  momentum density of the field, producing a complex Abraham force on the object given by the sum of the complex Lorentz force $\bm{\mathcal F}_i$ and $\bm{\mathcal F}_i^A$.

In consequence, the reactive force produced by a general time-dependent field on a body in a medium is given by the imaginary part of either Eq. (\ref{MCSTT}) or (\ref{ACSTT}), depending on the choice of Poynting momentum. The  reactive force that the Abraham momentum predicts does not coincide with the ILF, but contains the additional component $\mbox{Im}\{\bm{\mathcal F}_i^A\}$.

For time-harmonic fields, both $\bm{\mathcal G}_i^{M\,*}$ and $\bm{\mathcal G}_i^{A\,*}$ become time-independent,  and therefore (\ref{MCSTT}) and (\ref{ACSTT}) yield the same CMST equation, identical to (\ref{bcfor8}) with the Minkowski CMST pertaining to the embedding medium:
\be
T_{ij}^M=\frac{1}{8\pi}[\epsilon E_i  E_j^* + \frac{1}{\mu}B_i ^*B_j -\frac{1}{2}\delta_{ij}(\epsilon|{\bf E}|^2 + \frac{1}{\mu}|{\bf B}|^2 )], \,\,\,\,\, \,\,\,\,\label{AMbcfor6}
\ee
and the orbital momentum densities:
\be
(\bm{ P}_{e}^{O\, M})_i=\frac{\epsilon}{8\pi \omega}\mbox{Im}  [{ E}_j^* \partial_i { E}_j],\,\,\,\,\, \,\,\,\,\nonumber \\
 (\bm{ P}_{m}^{O\,M})_i= \frac{1}{8\pi \mu\omega}\mbox{Im}[ { B}_j^* \partial_i {\mathcal B}_j ] \,\,\, . \label{tMforM35}
\ee
Then the reactive force and the ILF coincide, being given by an equation identical to 
(\ref{bcfor10}) with the Minkowski IMST:  $
T_{ij}^{M\,I}=\frac{1}{8\pi}\mbox{Im}\{\epsilon E_i  E_j^* + \frac{1}{\mu} B_i ^*B_j \} $ and ROM expressed by the momenta (\ref{tMforM35}).

Hence, the Abraham-Minkowski debate influences the CMST theorem for arbitrary time-dependent fields, but  not for time-harmonic (or monochromatic) electromagnetic waves. All our conclusions on these latter wavefields remain valid and unaffected by this debate.

\bb

\section{Conclusions}

In summary,  we have formulated the existence of a complex force  in  light-matter interactions which splits into two, either scaled  or instantaneous. The real part is the standard time-averaged force, RLF,  due to transfer of  Poynting momentum. The imaginary part, ILF, established here, stems from the exchange of reactive (i.e. imaginary Poynting) momentum and is linked to the accretion of what we find as the  reactive strength of canonical (or orbital) momentum; and that, like the reactive energy, unavoidably appears as the incident wevefield hits the object.

Within the area of nanophotonics,  near-field effects and reactive quantities at the nanoscale are shown here to be of  importance  in connection with optical manipulation. Thus we highlight the main conclusions of this work:

(1) Since the Maxwell stress tensor is the basis of electromagnetic optical forces and binding,  with most current  detection on time averaging, the imaginary Maxwell stress tensor, its associated  reactive stress of orbital momentum,  and the reactive Lorentz force established here, constitute the other side of the  dynamical effects in light-matter interactions.

(2) The emergence of ROM and ILF is  associated to the appearance of reactive energy and reactive work,  recently remarked in nanoantennas. The ILF and ROM play a hindrance role versus the standard  RLF, so that  a large ROM storage conveys a loss of radiative force, i.e. of RLF, and vice-versa. This is illustrated with examples and makes the ROM and ILF  indirectly observable. 

\rr It is quite interesting that the electric and magnetic canonical momenta are the quantities characterizing the force ``reactance" constituted by the ROM, in analogy with the electric and magnetic energies defining the reactive energy. This fact emphasizes the capital role \cite{bliokh1} of the canonical momentum in the radiation pressure. \bb

(3) Our results show  that, in absence of body losses, the internal ROM contribution to the ILF is dominant versus that of the IMST (i.e. external ROM). However, absorption diminish the interior field energy and hence the internal ROM, so that although near  resonant  (e.g. plasmon) wavelengths, all reactive quantities are enhanced, the resulting resonant near-field ILF becomes not so much a hurdle to the resonant ``radiated'' force RLF.

(4) The picture  illustrated in this work    is paradigmatic to the complementary roles of ROM (and ILF) and the time-averaged force,  completing the whole picture of the optical force emergence.  As the RLF is characterized by the flow
RMST  across a far-zone surface sphere, and it is independent of its radius, (in analogy with the efficiency of power radiated into the far-field), the stored ROM, and its associated ILF,  are characterized by the flow IMST  across a near-field sphere that surrounds the body distribution of charges and currents. Therefore, the ROM acts as a ``reactance'' force  on the body.

(5) Associated with these dynamic concepts, and with the role of sources in the definition of the real and imaginary field (i.e. Poynting) momenta, is the characterization of the  RLF  by  the imaginary orbital and spin momenta, and of the ILF by  what we introduce as the reactive strength of Poynting momentum.

\rr
(6) While the CMST theorem, and hence the reactive force, of general arbitrary time-dependent light waves is affected by the choice of field momentum in the context of the Minkowski-Abraham debate, we have demonstrated that in the case of time-harmonic fields, this choice has no effect neither in the reactive force nor in the ROM, and like the RMST, the IMST is that of Minkowski.
\bb

Given the recent advances in optical manipulation at the nanoscale, increasing knowledge on the details of the generation and control of electromagnetic optical forces  is continuously required. Hence  these long time uncovered  reactive dynamic quantities should be relevant in practice. 

\rr Like in the design of RF antennas and nanoantennas  the radiation efficiency is increased by  diminishing the reactive power and reactive work, in the context of optical manipulation one may  act on the ROM and ILF in order to optimize the time-averaged optical force. So the  scenario established in this study contains a novel tool to handle the mechanical action of light on matter.

 Given the zero time-average of the ROM flow related to the instantaneous reactive force, their direct detection may enter the domain of femtosecond and attosecond optics on using subcycle pulse illumination \cite{keller,chang,tisch}.
 \bb

Although we have emphasized the nanophotonics domain, the fundamental physics of the complex Lorentz  force highlights its presence in the general electrodynamics realm, and suggests its existence in the interaction of sound, fluids \cite{bliokh_water}, and  other matter waves, thus opening a new landscape of possible dynamic phenomena of waves on matter.

\section*{Conflict of interest}
The authors declare no conflict of interest.
\newline
\newline

\section*{Acknowledgments}
MN-V work was  supported by Ministerio de Ciencia e Innovación  of Spain, grant PGC2018-095777-B-C21. X.X acknowledges the National Natural Science Foundation of China (11804119). A critical reading and  interesting comments from Dr. M. W. Puga, as well as important questions  from two anonymous reviewers,  are gratefully acknowledged.


\section{Supplementary Material for ``The complex Maxwell stress tensor theorem: The imaginary  stress tensor and the reactive strength of orbital momentum. A novel scenery underlying  optical forces''}    
\author{Manuel Nieto-Vesperinas}
\email{mnieto@icmm.csic.es}
\affiliation{Instituto de Ciencia de Materiales de Madrid, Consejo Superior de
Investigaciones Cient\'{i}ficas.\\
 Campus de Cantoblanco, Madrid 28049, Spain. www.icmm.csic.es/mnv}
\author{Xiaohao Xu}
\email{xuxhao_dakuren@163.com}
\affiliation{State Key Laboratory of Transient Optics and Photonics, Xi’an Institute of Optics and Precision Mechanics, Chinese Academy of Sciences, Xi’an 710119, China}
\affiliation{ Institute of Nanophotonics, Jinan University, Guangzhou 511443, China }

\maketitle
  \setcounter{equation}{0} 
\renewcommand{\figurename}{FIG. S}
  \setcounter{figure}{0}
\appendix 

\section{Reactive torque. A dipolar particle}
The  time-averaged optical torque $<\bm \Gamma>={\bf r}\times\mbox{Re} \{\bm{\mathcal F}\}$ on the object, of lever arm ${\bf r}$ is: 
\be
<\bm \Gamma>=\int_{S}d^2 r \, {\bf r}\times\mbox{Re} \{T_{ij}\} n_j. \label{bcfor11}
\ee
In turn, the  {\it reactive torque}: $\bm {\Xi}={\bf r}\times \mbox{Im} \{\bm{\mathcal F}\}$ reads:
\be
\bm {\Xi}=\int_{S}d^2 r \,{\bf r}\times\mbox{Im}\{ T_{ij}\}n_j+\omega\int_{V}d^3 r\, ({\bf L}_{m}^O - {\bf L}_{e}^O)
.   \label{bcfor12}
\ee
With the electric and magnetic time-averaged orbital angular momenta:
\be
{\bf L}_{e}^O= {\bf r}\times {\bf P}_{e}^O \,\,\, \,\,\,\,         {\bf L}_{m}^O={\bf r}\times
{\bf P}_{m}^O .  \label{bcfor13}
\ee 
We shall call {\it reactive strength of orbital angular momentum} density to the quantity $\omega({\bf L}_{m}^O - {\bf L}_{e}^O)$.

\section{The canonical and spin momenta with sources. Lagrangian derivation}
\subsection{The electric canonical and spin momenta}
It is well-known that the electric and magnetic classical fields governed by Maxwell’s equations hold dual symmetry in free-space \cite{jackson}. Their lack of duality in presence of sources, i.e. electric charges and currents,  has been studied by many authors  who, following P.A.M. Dirac  \cite{dirac},  postulate the existence of (so far unobserved) magnetic charges and currents that restore such symmetry, see e.g. \cite{cabibbo,schwinger,brandt,likang,bekshaevNJP}.

Without recurring to  magnetic sources, deriving a Lagrangian in dual space that leads to an energy-momentum tensor from which electromagnetic quantities fulfill  all conservation laws, and that yields a consistent decomposition of the energy flow (Poynting vector) into a {\rr canonical (or orbital)} momentum and a spin momentum, like for electromagnetic fields in free-space \cite{bekshaevNJP}, is problematic.

Here we introduce, notwithstanding,  potentials and an energy-momentum tensor, that lead to such a possible decomposition and, hence, a characterization of the canonical and spin momenta in presence of sources.

We write the Lagrangian for the electromagnetic field $F_{\alpha \beta}$=({\bf E},{\bf H}) with sources \cite{landau2}:
\be
{\cal L}=-\frac{1}{16\pi}F_{\alpha \beta}F^{\alpha \beta}-\frac{1}{c} j_{\gamma}A^\gamma. \label{lag}
\ee
Greek indices run as: $0 , 1 , 2 , 3$. 
Covariant and contravariant tensor indices are  related by \cite{jackson}: $T_{...}^{...\alpha}= g^{\alpha \beta}T_{...\beta}^{...}$ ; where $g^{\alpha \beta}=g_{\alpha \beta}$ is the Euclidean space metric tensor: $g^{\alpha \beta}=0$ when $\alpha\neq\beta$, $g^{0 0}=1$, $g^{11}=g^{22}=g^{33}=-1$ ;  $g^{\alpha \beta}g_{\beta \gamma}=\delta^{\alpha}_{\gamma}$. The coordinate vector is denoted as either $x^\alpha=(ct,{\bf r})$, or
$x_\alpha=g_{\alpha\gamma}x^\gamma=(ct,-{\bf r})$,  so that the scalar product of two 4-vectors is 
$A_{\gamma} B^{\gamma}=A^0B^0-{\bf A}\cdot{\bf B}$;  and $\partial_\alpha=\frac{\partial}{\partial 
x^\alpha}=(\frac{\partial}{c\,\partial t},\nabla)$,  $\partial^\alpha=\frac{\partial}{\partial x_\alpha}=(\frac{\partial}{c\,\partial t},-\nabla)$. 

 The current and potential  4-vectors are \cite{landau2}: $j_\gamma=(c\rho, -{\bf J})$ and $A^\gamma=(\phi, {\bf A})$. Also $F^{\alpha \beta}=(-{\bf E},{\bf H})$, $F_{\alpha \beta}=({\bf E},{\bf H})$.   $F_{\alpha \beta}=g_{\alpha\sigma}g_{\beta\tau}F^{\sigma\tau}$. $F^{\alpha \beta}=\partial^\alpha A^\beta-\partial^\beta A^\alpha$ and $F_{\alpha \beta}=\partial_\alpha A_\beta-\partial_\beta A_\alpha$ convey $ {\bf E}=-\frac{1}{c} \frac{\partial{\bf A}}{\partial t}-\nabla \phi$,   $ {\bf H}=\nabla  \times {\bf A}$.
 It is well-known that the Lagrange equations associated to  (\ref{lag}) lead to the second pair of Maxwell equations:  $\partial_\gamma F^{\alpha\gamma}=-\frac{4\pi}{c}j^\gamma$, namely: $\nabla\cdot {\bf E}=4\pi\rho$ and $\nabla\times {\bf H}=\frac{\partial{\bf E}}{\partial t}+\frac{4\pi}{c}{\bf J}$. 

As is known,  the Lagrangian (\ref{lag}) gives rise to the canonical energy-momentum tensor:
\be
\tilde{T}^{\alpha \beta}=\partial^\alpha A^\gamma \frac{\partial \cal L}{\partial(\partial_\beta A^\gamma)}-
g^{\alpha \beta} {\cal L}=
-\frac{1}{4 \pi} (\partial^{\alpha} A^{\gamma}) F^{ \beta}_{ \gamma}+g^{\alpha\beta} (\frac{1}{16 \pi}F_{\gamma \sigma}F^{\gamma \sigma} +\frac{1}{c} j_{\gamma}A^\gamma) . \label{tem}
\ee
The electric canonical (or orbital) momentum $P_e^{O \alpha}$ of the electromagnetic field is given by the component $\tilde{T}^{\alpha 0}/c$ \cite{bekshaevNJP}, with $\alpha \neq 0$ 
\be
P_e^{O \alpha}=\frac{1}{c}\tilde{T}^{\alpha 0}=
-\frac{1}{4 \pi c} (\partial^{\alpha} A^{\gamma}) F^{ 0}_{ \gamma}+g^{\alpha 0} (\frac{1}{16 \pi c}F_{\gamma \sigma}F^{\gamma 0} +\frac{1}{c^2} j_{\gamma}A^\gamma)
=-\frac{1}{4 \pi c} (\partial^{\alpha} A^{\gamma}) F^{ 0}_{ \gamma}
 ,  \,\, (\alpha\neq 0).\,\,\,\,\,\,\label{tem0}
\ee
I.e, since $F^{ 0}_{ 0}=0$ and $\alpha\neq 0$,  the $ith$ component of the cananonical momentum is:
\be
P_e^{O i}=\frac{1}{c}\tilde{T}^{i 0}=
-\frac{1}{4 \pi c} (\partial^{i} A^{j}) F^{ 0}_{ j}=-\frac{1}{4\pi  c} (\partial^{i} A^{j})g_{jk} F^{0 k}
\nonumber \\
=\frac{1}{4 \pi c} (\partial^{i} A^{j}) F^{0j}=\frac{1}{4 \pi c}E^{j } \partial_{i} A^{j}   ,  \,\, (i,j,=1,2,3).\,\,\,\label{tem0a}
\ee
Henceforth being understood  that   latin indices run as $1,2,3$.

Concerning the electric spin momentum $P_e^{S\, i}$ associated to the tensor  $\Delta{T}^{\alpha \beta}$ that added to the canonical energy-momentum tensor with sources (\ref{tem}) symmetrizes it  \cite{bekshaevNJP}, we choose
\be
\Delta{T}^{\alpha \beta}=\partial_\gamma\psi^{\alpha\beta\gamma}+\frac{1}{c} A^{\alpha}j^\beta -
 \frac{1}{16 \pi c}g^{\alpha\beta} j_{\gamma}A^\gamma
=\frac{1}{4 \pi} \partial_{\gamma} (A^{\alpha} F^{ \beta \gamma})+\frac{1}{c} A^{\alpha}j^\beta-
 \frac{1}{ c}g^{\alpha\beta} j_{\gamma}A^\gamma . \nonumber \\
\psi^{\alpha\beta\gamma}=\frac{1}{4 \pi} A^{\alpha} F^{ \beta \gamma} \, ,
\,\,\,\,\,\ \,\,\psi^{\alpha\beta\gamma}=
-\psi^{\alpha\gamma\beta},\,\,\,\, \partial_{\beta}\partial_\gamma\psi^{\alpha\beta\gamma}=0. \,\,\,\,\,\,\,\,\,\,\,\,\,\,\,\, 
\label{tem0b}
\ee
So that, since  $\alpha\neq 0$ and $F^{00}=0$, $P_e^{S\, i}$ would be
\be
P_e^{S i}=\frac{1}{c}\Delta{T}^{\alpha 0}=\frac{1}{c}\Delta{T}^{i 0} =\frac{1}{4 \pi c} \partial_{j} (A^{i} F^{ 0 j})+\frac{1}{c} A^{i}j^0= -\frac{1}{4 \pi c} \partial_{j}(E ^j A^{i})+\frac{1}{c} A^ij^0,  \,\, (i,j=1,2,3).\,\,\,\,\,\,\label{tem0c}
\ee
The sum of the canonical and spin momenta, $P_e^{O i}$  and $P_e^{S i}$, Eqs. (\ref{tem0a}) and (\ref{tem0c}), is that part ${\bf g}_e$ of the field momentum due to the electric field, i.e. 
\be
g_e^{ i}=\frac{1}{4 \pi c}[E^{j } \partial_{i} A^{j}- \partial_{j}(E ^j A^{i})]+\frac{1}{c} A^ij^0 ,  \,\, (i,j=1,2,3)\,\,\,\,\label{tem0d}
\ee
In order to make the link of  (\ref{tem0d})  with the expression \rr(32)\bb, we consider time-harmonic fields. Then the spatial parts hold
\be
{\bf A}=-\frac{{\ii}}{k}({\bf E}+\nabla \phi) , \,\, \,\,\,\,  \nabla \cdot {\bf E}=4\pi \rho,\,\, \,\,\,\, \label{tem0e}
\ee
After introducing the time average on the ${\cal O}$-operation as: $<A {\cal O} B>=(1/2)Re(A^* {\cal O} B)$, straightforward operations lead to the time-average of the electric field (i.e. Poynting) momentum
\be
<{\bf g}>= {\bf P}_e^S+{\bf P}_e^O +\frac{1}{2 \omega}\mbox{Im}\{\rho^*{\bf E}\} . \label{tem0f}
\ee
Equation (\ref{tem0f}) is identical to \rr Eq. (32)\bb. 

Note that we did not need to introduce any  choice of gauge in the 4-potential $A^\alpha$. This is due to the fact that our selected $\Delta{T}^{\alpha \beta}$, Eq. (\ref{tem0b}),  automatically  symmetrizes the energy-momentum tensor with sources, (\ref{tem}), as the term $\frac{1}{c} A^{\alpha}j^\beta$ in (\ref{tem0b}) cancels an identical term obtained from $ \frac{1}{4 \pi} \partial_{\gamma} (A^{\alpha} F^{ \beta \gamma})$ and the second Maxwell equation:  $\partial _\gamma F^{\beta\gamma}=-\frac{4\pi}{c} j^\beta$. Consequently, the  symmetrized energy-momentum tensor results
\be
{T}^{\alpha \beta}= \tilde{ T}^{\alpha\beta}+\Delta{ T}^{\alpha\beta}=-\frac{1}{4 \pi} F^{\alpha\gamma }F_{\gamma}^{\beta }+\frac{1}{16 \pi}g^{\alpha \beta}F^{\delta\sigma}F_{\delta\sigma }\,\,\,  , \,\, \,\,\, \label{tem0g}
\ee
which  has the same functional form as the symmetric free-space energy-momentum tensor \cite{jackson,landau2} and, as such, it  fulfils the conservation equation with sources
\be
\partial_\alpha  T^{\alpha\beta}=-\frac{1}{c}F^{\beta\gamma}j_\gamma.\,\, \,\,\, \label{tem0h}
\ee

\subsection{The magnetic canonical and spin momenta with sources}
Within the aforementoned limitations from the lack of duality between the electric and magnetic field in presence of
(non-magnetic) sources, we now pass on to addressing dual quantities. First,  we quote the dual field pseudotensor:
$G^{\alpha\beta}=\frac{1}{2}\epsilon^{\alpha\beta\gamma\delta}F_{\gamma\delta} $, where $
\epsilon^{\alpha\beta\gamma\delta}$ denotes the fourth-order Levi-Civita completely antisymmetric tensor. This (free-field) dual pseudotensor  holds: $G_{\alpha \beta}=\partial_\alpha C_\beta-\partial_\beta C_\alpha $, \,\, $G^{\alpha \beta}=\partial^\alpha C^\beta-\partial^\beta C^\alpha$.

We introduce the dual fields with sources
\be
G^{(s)\,\alpha \beta}=-G^{(s)\, \beta\alpha}=(-{\bf H}, {\bf E})=(\partial^\alpha C^\beta-\partial^\beta C^\alpha)+4\pi R^{\alpha\beta}.\,\,\,\,\,\,  \label{tem1a}
\ee
So that 
\be
G_{\alpha \beta}^{(s)}=({\bf H},{\bf E})=(\partial_\alpha C_\beta-\partial_\beta C_\alpha)+4\pi 
R_{\alpha\beta}=g_{\beta\tau}G^{(s)\,\sigma \tau}. \label{tem11a}
\ee
The superscript $(s)$ denotes fields in presence of sources. The dual potential   4-vector is  $C^\gamma=(\theta, {\bf C})$,  and the tensor 
\be
R^{\alpha \beta}=-R^{ \beta\alpha}=R_{\alpha \beta}=-R_{\beta\alpha }=(\bm{0},\bm{\Upsilon}), \,\,\,\,\,
\,\,\,\,\,\,\,  \bm{0}=(0,0,0),  \,\,\,\,\,\,\,\,\,\,\,\, \frac{\partial \bm{\Upsilon}}{\partial t}=-{\bf J}.  
\label{tem1b}
\ee
The ordering of ${\bf 0}$ and $\bm{\Upsilon}$ in $R^{\alpha \beta}$ (or in $R_{\alpha \beta}$), Eq. 
(\ref{tem1b}), is the same as that of ${\bf H}$ and ${\bf E}$ in   $G^{(s)\,\alpha\beta}$, respectively; (or in 
$G_{\alpha\beta}^{(s)})$.  The 4-vector potential is $C^\gamma=(\theta, {\bf C})$. Equations  (\ref{tem1a}) and 
(\ref{tem11a}) mean
\be
{\bf E}=-\nabla\times{\bf C}+4\pi \bm{\Upsilon},\,\,\,\,\,\, \,\,\,\,\,\,{\bf H}=-\frac{1}{c}\frac{\partial{\bf C}}{\partial t}-\nabla\theta \,\,\,\, . \,\,\, \label{tem1c}
\ee
Note that introducing the Maxwell equation: $\nabla\cdot {\bf E}=4\pi \rho$ into the first of Eqs.  (\ref{tem1c}) 
 one obtains the well-known continuity equation: $\nabla\cdot{\bf J}+\frac{\partial\rho}{\partial t}=0$. 

It should be remarked that  we introduced the tensor $R_{\alpha \beta}$  to make the fields $G_{\alpha \beta}$ (or $G^{\alpha \beta}$), written in vectorial form: $({\bf H},{\bf E})$ [or $(-{\bf H},{\bf E})$]  as Eqs. (\ref{tem1c}),  to fulfil the second pair of Maxwell equations with sources:     $\nabla\cdot {\bf E}=4\pi\rho$ and $\nabla\times{\bf H}=\frac{1}{c}\partial_t {\bf E}+\frac{4\pi}{c}{\bf J}$. Likewise, it is known that the vectorial form of the first pair of Maxwell equations: $\nabla\times{\bf E}=\frac{1}{c}\partial_t {\bf H}$ and $\nabla\cdot {\bf H}=0$ is obtained from   $F_{\alpha\beta}=\partial_\alpha A_\beta-\partial_\beta A_\alpha=({\bf E},{\bf H})$  (or from $F^{\alpha\beta}$) , expressed  as ${\bf E}=-\frac{1}{c}\frac{\partial{\bf A}}{\partial t}-\nabla\phi$ , \,\,\,\,${\bf H}=\nabla\times{\bf A}$.

However, we note that the Lagrange equations of the Lagrangian built from $F_{\alpha\beta}$, Eq. (\ref{lag}), yield  the second pair of Maxwell's equations written in tensor notation: $\partial_\gamma F^{\alpha\gamma}=-\frac{4\pi}{c}j^\gamma$, but not the first pair:  $\partial_\gamma G^{\alpha\gamma}=0$.  Such  first pair of Maxwell  equations is obtained from  the dual Lagrangian without sources \cite{jackson,landau2}
\be
\hat{\cal L}=-\frac{1}{16\pi}G_{\alpha \beta}G^{\alpha \beta}. \label{lagd}
\ee 
This asymmetry reflects the breackdown of electromagnetic duality in absence of magnetic sources. On these grounds,   even  in  presence of sources, we propose the  source-free dual Lagrangian  (\ref{lagd}) rather than the one built from the pseudotensor $G_{\alpha \beta}^{(s)}$ and $G^{(s)\,\alpha \beta}$ introduced above. This Lagrangian also yields the correct energy.

In this connection it is worth observing that had one employed in (\ref{lagd}) $G_{\alpha \beta}^{(s)}$ and $G^{(s)\,\alpha \beta}$ given by ({\ref{tem1a}) and (\ref{tem11a}), rather than their free-space expressions, the first pair of Maxwell equations would be obtained from such a Lagrangian if  the following extra condition holds:
\be
\partial_\beta R^{\alpha\beta}=0, \label{rcond}
\ee
 which according to (\ref{tem1b}) means that $\nabla\times \bm{\Upsilon}=0$, and hence $\nabla\times \bf{J}=0$. Thus the current density ${\bf J}$ would be longitudinal. We believe that this is an unnecessary and little realistic restriction.

Note that  ({\ref{rcond}) was obtained from the Lagrange equations using  in (\ref{lagd})
 Eqs. ({\ref{tem1a}) and ({\ref{tem11a}) instead of  $G_{\alpha \beta}$ and $G^{\alpha \beta}$,  as well as the equalities: $\frac{\partial [(\partial^\alpha C^\beta-\partial^\beta C^\alpha) R_{\alpha\beta}]  }{\partial(\partial^\alpha C^\beta)}=2R_{\alpha\beta}=2R^{\alpha\beta}=\frac{\partial [(\partial_\alpha C_\beta-\partial_\beta C_\alpha) R^{\alpha\beta}]  }{\partial(\partial_\alpha C_\beta)}$ since $R^{\alpha\beta}=R_{\alpha\beta}$ and $R_{\alpha\beta}=-R_{\beta\alpha}$.

The canonical energy-momentum tensor then  is
\be
\tilde{\hat{T}}^{\alpha \beta}=\partial^\alpha C^\gamma \frac{\partial \cal L}{\partial(\partial_\beta C^\gamma)}-g^{\alpha \beta}\hat {\cal L}=
-\frac{1}{4 \pi} (\partial^{\alpha} C^{\gamma}) G^{ \beta}_{ \gamma}+\frac{1}{16 \pi}g^{\alpha\beta}G_{\gamma \sigma}G^{\gamma \sigma} . \label{temmd}
\ee
The magnetic canonical momentum $\hat{P}_m^{O \alpha}$  is given by  $\tilde{T}^{\alpha 0}/c$, with $\alpha \neq 0$ 
\be
\hat{P}_m^{O \alpha}=\frac{1}{c}\tilde{\hat{T}}^{\alpha 0}=
-\frac{1}{4 \pi c} (\partial^{\alpha} C^{\gamma}) G^{ 0}_{ \gamma}+ \frac{1}{16 \pi c}g^{\alpha 0}G_{\gamma \sigma}G^{\gamma 0}
=-\frac{1}{4 \pi c} (\partial^{\alpha} C^{\gamma}) G^{ 0}_{ \gamma}
 ,  \,\, (\alpha\neq 0).\,\,\,\,\,\,\label{temm0}
\ee
And since $G^{ 0}_{ 0}=0$ and $\alpha\neq 0$,  the $ith$ component of the cananonical momentum finally is:
\be
\hat{P}_m^{O i}=\frac{1}{c}\tilde{\hat{T}}^{i 0}=
-\frac{1}{4 \pi c} (\partial^{i} C^{j}) G^{ 0}_{ j}=-\frac{1}{4\pi  c} (\partial^{i} G^{j})g_{jk} G^{0 k}
\nonumber \\
=\frac{1}{4 \pi c} (\partial^{i} C^{j}) G^{0j}=\frac{1}{4 \pi c}H^{j } \partial_{i} C^{j}   ,  \,\, (i,j,=1,2,3).\,\,\,\label{temm0a}
\ee

As for the magnetic spin momentum $\hat{P}_m^{S\, i}$ associated to the tensor  $\Delta\hat{T}^{\alpha \beta}$, we choose
\be
\Delta\hat{T}^{\alpha \beta}=\partial_\gamma\chi^{\alpha\beta\gamma} +\frac{1}{c^2}\epsilon_{\alpha\beta\gamma\delta}C^\gamma j^\delta
=\frac{1}{4 \pi} \partial_{\gamma} (C^{\alpha} F^{ \beta \gamma}) +\frac{1}{c^2}\epsilon_{\alpha\beta\gamma\delta}C^\gamma j^\delta . \nonumber \\
\chi^{\alpha\beta\gamma}=\frac{1}{4 \pi} C^{\alpha} F^{ \beta \gamma}\, ,
\,\,\,\,\,\ \,\,\chi^{\alpha\beta\gamma}=
-\chi^{\alpha\gamma\beta},\,\,\,\, \partial_{\beta}\partial_\gamma\chi^{\alpha\beta\gamma}=0. \,\,\,\,\,\,\,\,\,\,\,\,\,\,\,\, 
\label{temm0b}
\ee
As  $\alpha\neq 0$ and $G^{00}=0$,  we get
\be
\hat{P}_m^{S\, i}=\frac{1}{c}\Delta\hat{T}^{\alpha 0}=\frac{1}{c}\Delta\hat{T}^{i 0} =\frac{1}{4 \pi c} \partial_{j} (C^{i} G^{ 0 j})+\frac{1}{c^2} \epsilon_{i0kl}C^k j^l= -\frac{1}{4 \pi c} \partial_{j}(H ^j C^{i})+\frac{1}{c^2}\epsilon_{ikl}C^k j^l ,  \,\,).\,\,\,\,\,\,\label{temm0c}
\ee
The sum of the canonical and spin magnetic momenta, $\hat{P}_m^{O i}$  and $\hat{P}_m^{S i}$, Eqs. (\ref{temm0a}) and (\ref{temm0c}), is that part ${\bf g}_m$ of the Poynting momentum due to the magnetic field, namely 
\be
g_m^{ i}=\frac{1}{4 \pi c}[H^{j } \partial_{i} C^{j}- \partial_{j}(H ^j C^{i})]+\frac{1}{c^2}\epsilon_{ikl}C^k j^l  ,  \,\,\,\,(i,j,k,l=1,2,3). \,\,\,\,\label{temm0d}
\ee
Equation (\ref{temm0d})  coincides with \rr(33) \bb for time-harmonic fields. Note that then the spatial parts hold
\be
{\bf C}=-\frac{{\ii}}{k}({\bf H}+\nabla \theta) , \,\, \,\,\,\,  \nabla \cdot {\bf H}=0,\,\, \,\,\,\, \label{temm0e}
\ee
As before, after introducing the time average: $<A {\cal O} B>=(1/2)Re(A^* {\cal O} B)$, it is straightforward  to obtain
\be
<{\bf g}>= {\bf P}_m^S+{\bf P}_m^O -\frac{1}{2kc^2}\mbox{Im}\{{\bf J}^* \times {\bf H}\}  . \label{temm0f}
\ee
It should be remarked, however, that the electromagnetic duality breakdown requires that the fields $G^{ \alpha \beta}$ in the tensor $\Delta\hat{T}^{\alpha \beta}$, and  its corresponding magnetic spin momentum $\hat{P}_m^{S\, i}$,  meet some conditions. One  is that  $C^\alpha=(0, {\bf C})$. I.e. $\theta=0$.  Then: 
\be
{\bf H}=-\frac{1}{c}\frac{\partial{\bf C}}{\partial t}.  \,\,\, \label{temm1c}
\ee
So that $\nabla\cdot \partial_t{\bf C}=0$; and if we chose the Lorenz condition $\partial_\gamma C^\gamma=0$ , 
one would  also have $\nabla\cdot{\bf C}=0$. This is equivalent to a Coulomb gauge in the space of dual quantities. However in 
this space $\nabla\theta=0$ does not mean $\rho=0$ since $\nabla\cdot{\bf E}=4\pi\nabla\cdot\bm{\Upsilon}
=4\pi\rho$, and hence $\nabla\cdot{\bf J}=-\partial_t\rho$. Moreover, using (\ref{tem1c}) and the first Maxwell equation: $\nabla\times{\bf E}=-\frac{1}{c}\partial_t{\bf H}$ we get
\be
\nabla^2 {\bf C} -\frac{1}{c^2}\frac{\partial^2 {\bf C}}{\partial t ^2}=-4\pi \nabla\times\bm{\Upsilon},\label{temm0g}
\ee
which indicates that the source function of the wave equation for ${\bf C}$ is transversal. In the $F_{\alpha\beta}$-space, Coulomb's gauge involves  the source  in  Eq. (\ref{temm0g}) for ${\bf A}$ being a transversal current, and the charge is associated to a longitudinal current which acts as a source of  static (i.e. near) field. This does not happen in the  dual space, however, where nothing indicates that  the current density associated to the charge be longitudinal.

 On the other hand, although the source vector of the wave equation of  ${\bf C}$ is transversal, it does not coincide with a transversal current density. These features  of the potential $C^\gamma$ in the  space of dual fields $G_{\alpha\gamma}$ are in clear contrast with that of  the potential $A^\gamma$ in the space of $F_{\alpha\gamma}$.

Another aspect  of  the tensor $\Delta\hat{T}^{\alpha \beta}$ is that, added to the canonical dual tensor $\tilde{\hat{T}}^{\alpha \beta}$, it  does not yield a symmetrized  energy momentum tensor, in contrast with $\Delta{T}^{\alpha \beta}$, nor its divergence is zero. (Notice in this connection that neither the tensor $\Delta{T}^{\alpha \beta}$ of Eq. (\ref{tem0b}) posesses a zero divergence as it should unless $\partial_\gamma (j_\beta A^\beta)=0$). 
 These again are other symptoms of the electromagnetic duality breackdown in the presence of sources, and indicate that concerning symmetry of   $\tilde{\hat{T}}^{\alpha \beta}$  and null divergence of  $\Delta{T}^{\alpha \beta}$ and $\Delta\tilde{T}^{\alpha \beta}$,  there might be different choices  for these tensors, and even for the dual Lagrangian. We have selected here those that we were more strightforwardly able to find among those leading to the canonical and spin momenta of the electromagnetic field within a covariant formulation.

\section{The imaginary Maxwell stress tensor and the angular spectrum of the fields}
 First, we expand the fields into their angular spectra of plane waves \cite{mandel,nietolib}:
\be
{\bf E}({\bf r})=\int_{-\infty}^{\infty}d^2{\bf K}\,\, {\bf e}({\bf K}) \exp [{\ii}( {\bf K}\cdot {\bf R}+ k_z z)].  \label{bcfor19}
\ee
Where  ${\bf r}=({\bf R}, z)$, ${\bf R}=(x,y)$, ${\bf k}=({\bf K}, k_z)$, ${\bf K}=(K_x,K_y), |{\bf k}|^2=k^2$.  
\be
k_z=\sqrt{k^2- K^2}=q_h \, , \,K\leq k,\,\, (propagating\,\, waves).  \nonumber\\
k_z={\ii}\sqrt{K^2- k^2}={\ii}q_e \, , \,K> k, \,\,(evanescent \,\,waves).  \label{bcfor19bis}
\ee 
And an analogous expansion for  ${\bf B}({\bf r})$. Therefore, by splitting the field angular spectrum integral into homogeneous and evanescent  components, (with susbscript $h$ and $e$, respectively),
\be
\int_{-\infty}^{\infty} d^2 {\bf R} ({\bf P}_m^S-{\bf P}_e^S)=\frac{1}{16\pi \omega}\int_{-\infty}^{\infty}  d^2 {\bf R}\int_{K\leq k}d^2 {\bf K}\int_{K'\leq k} d^2 {\bf K}' \, \mbox{Im}\{\exp[-{\ii}({\bf K}-{\bf K}')\cdot {\bf R}+(q_h - q'_h)z)] \nonumber \\
\times [-{\ii}({\bf K}-{\bf K}' , q_h - q'_h) ]\times[{\bf b}_h^*({\bf K})\times{\bf b}_h({\bf K'})-{\bf e}_h^*({\bf K})\times{\bf e}_h({\bf K'})]\} \nonumber \\
+\frac{1}{16\pi \omega}\int_{-\infty}^{\infty}  d^2 {\bf R}\int_{K> k}d^2 {\bf K}\int_{K'> k} d^2 {\bf K}' \, \mbox{Im}\{\exp[-{\ii}({\bf K}-{\bf K}')\cdot {\bf R}+(q_e + q'_e)z)] \nonumber \\
\times [-{\ii}({\bf K}-{\bf K}') , -(q_e + q'_e) ]\times[{\bf b}_e^*({\bf K})\times{\bf b}_e({\bf K'})-{\bf e}_e^*({\bf K})\times{\bf e}_e({\bf K'})]\}. \,\,\,\,\,\,\,\,\,\,\,\,\label{bcfor6b1A}
\ee
Making $z=z_0\geq 0$,  the ${\bf R}$-integration yields $ (2\pi)^2\delta({\bf K}'-{\bf K})$. 
Then, after performing the  ${\bf K}'$-integral and expressing ${\bf b}_{\perp}({\bf K})=(b_x({\bf K}), b_y({\bf K}),0 ),\, {\bf e}_{\perp}({\bf K}) =(e_x ({\bf K}),  e_y  ({\bf K}) , 0 )$, we inmediately see that  the  the homogeneous (propagating) part of (\ref{bcfor6b1A}) is zero. Therefore the above becomes
\be
\int_{-\infty}^{\infty} d^2 {\bf R} ({\bf P}_m^S-{\bf P}_e^S)=\frac{\pi}{2\omega}
\int_{K> k}d^2 {\bf K} \,\exp (-2q_e z_0)
 \mbox{Im}\{ 
 [0,0 , -q_e ]\times[{\bf b}_e^*({\bf K})\times{\bf b}_e({\bf K})-{\bf e}_e^* ({\bf K})\times{\bf e}_e({\bf K})]\}
\nonumber \\
=\frac{\pi}{\omega}\int_{K> k}d^2 {\bf K} \, q_e \exp(-2q_e z_0) \mbox{Im}\{b_{e\,z}^*({\bf K}){\bf b}_{e\,\perp}({\bf K})  -e_{e\,z}^*({\bf K}){\bf e}_{e\,\perp}
({\bf K}) \} .\,\,\,\,\,\,\label{bcfor6b2A}
 \ee
In particular, we may take $z_0=0$.

\section{The  complex Maxwell stress tensor on a dipole}
Addressing the complex Maxwell stress tensor theorem, \rr Eq. (17)\bb, in the scattering from a  magnetoelectric dipole or particle, we split the fields  into incident ${\bf E}^{(i)}$, ${\bf B}^{(i)}$ and scattered ${\bf E}^{(s)}$, ${\bf B}^{(s)}$: ${\bf E}={\bf E}^{(i)}+{\bf E}^{(s)}$,  ${\bf B}={\bf B}^{(i)}+{\bf B}^{(s)}$.  The dipole scattered fields are
\be
{\bf E}^{(s)}({\bf r})=\frac{k^2}{\epsilon}[{\bf n}\times({\bf p}\times {\bf n})]\frac{e^{{\ii}kr}}{r}+\frac{1}{\epsilon}[3{\bf n}({\bf n}\cdot{\bf p})-{\bf p}](\frac{1}{r^3}-\frac{{\ii}k}{r^2})e^{{\ii}kr}
 \nonumber \\
-\sqrt{\frac{\mu}{\epsilon}}({\bf n}\times{\bf m})(\frac{k^2}{r} +\frac{{\ii}k}{r^2})e^{{\ii}kr}.\,\,\,\,\,\,\,\,\,\,\,\,\,\,\,\, ({\bf r}=r{\bf n}).  \label{app1} \\
{\bf B}^{(s)}({\bf r})={\mu k^2}[{\bf n}\times({\bf m}\times {\bf n})]\frac{e^{{\ii}kr}}{r}+{\mu}[3{\bf n}({\bf n}\cdot{\bf m})-{\bf m}](\frac{1}{r^3}-\frac{{\ii}k}{r^2})e^{{\ii}kr}\nonumber \\
+\sqrt{\frac{\mu}{\epsilon}}({\bf n}\times{\bf p})(\frac{k^2}{r} +\frac{{\ii}k}{r^2})e^{{\ii}kr}.\,\,\,\,\,\,\,\,\,\,\,\,(\epsilon=\mu=1).\,\,\,\, \label{app2}
\ee
We assume the surrounding medium to be vacuum or air, so that $\epsilon=\mu=1$. Introducing the above splitting into the first term of \rr(17)\bb, one obtains for the non-zero terms
\be
\int_{\partial V}d^2 r \, T_{ij}n_j=\frac{1}{8\pi}\int_{\partial V} d^2 r \{({\bf E}^{(i)\,*} \cdot {\bf n}){\bf E}^{(i)}
+({\bf B}^{(i)} \cdot {\bf n}){\bf  B}^{(i)\,*} \,\,\,\,\,\,\,\,\,\,\,\,  \,\,\,\,\,\,\,\,\,\,\,\,     \nonumber       \\ 
+({\bf E} ^{(i)\,*}\cdot {\bf n}){\bf E}^{(s)}+({\bf E}^{(s)\,*}\cdot {\bf n}){\bf E}^{(i)}+({\bf B}^{(i)} \cdot {\bf n}){\bf B}^{(s)\,*}+({\bf B} ^{(s)}\cdot {\bf n}){\bf B}^{(i)\,*} \,\,\,\,\,\,\,\,\,\,\,\,\,\,\,\,\,\,\,\nonumber \\
+[({\bf E}^{(s)\,*} \cdot {\bf n}){\bf E}^{(s)}
+({\bf B}^{(s)} \cdot {\bf n}){\bf  B}^{(s)\,*}]
-\,\mbox{Re}\{{\bf E}^{(i)\,*}\cdot {\bf E}^{(s)}+{\bf B}^{(i)}\cdot {\bf B}^{(s)\,*}\}
 +\frac{1}{2}(|{\bf E}^{(s)}|^2 +|{\bf B}^{(s)}|^2)]{\bf n} \} \,. \,\,\,\,\,\,\,\,\,\label{app4}
\ee
When one takes the real part of  (\ref{app4}) which yields the time-averaged force $<\bm{\mathcal  F}>$, the integral   (\ref{app4}) of  the first two terms with the incident field only, is zero; while  the last two terms of this real part of (\ref{app4}) are well-known \cite{nieto1,chaumet} to yield the electric-magnetic dipole interference force $<{\bf  F}_{em}>=-\frac{k^4}{3} \mbox{Re}\{{\bf p}\times {\bf m}^*\}$. Furthermore, since this real part is independent of the integration contour, by choosing ${\partial V}$ to be a sphere of large radius $r$ such that $kr\rightarrow \infty$,  there is no contribution of the fourth, sixth, seventh and eight  terms since the scattered field in this region of $S$ is known to be transversal to ${\bf n}$.  In addition, on this large surface ${\partial V}$ one inmediately sees  by applying Jones' lemma based on the principle of the stationary phase \cite{born,nieto1}, that the contribution of the  mixed incident-scattered field third and fifth terms  is also zero since ${\bf n}$ becomes equal to the incident wavevector,  with respect  to which the incident field is transversal.

 Therefore,  only the diagonal last four terms, which belong to the real part of the CMST, contribute to $<\bm{\mathcal  F}>$. This is the reason by which we emphasize through the text that the far-field flux IMST is zero. In fact,  taking the imaginary part of  (\ref{app4}) in the far-zone one obtains that all terms, one by one, are zero. This has important consequences for the total spin momentum, as shown in Section VI.B of the main text.

Taking the imaginary part  of (\ref{app4}), which depends on the choice of the  integration surface ${\partial V}$, the first integral  involving only ${\bf E}^{(i)}$ and  ${\bf B}^{(i)}$  is cancelled out by the incident reactive orbital momentum, even if the body is illuminated by an evanescent wave [cf. \rr Eq. (48)\bb].  
 We take as the dipole volume  and its boundary: $V_0$ and ${\partial V}_0$, respectively. They correspond to the smallest sphere of radius $a$ that encloses the dipole; or if this is a particle, ${\partial V}_0$ is the limiting outside sphere  circunscribed to its physical surface.  The contribution of the non-diagonal terms involving the scattered field yields
\be
\frac{1}{8\pi}\mbox{Im} \{\int_{\partial V_0}  d^2 r [({\bf E}^{(s)\,*} \cdot {\bf n}){\bf E}^{(s)}
+({\bf B}^{(s)} \cdot {\bf n}){\bf  B}^{(s)\,*}]\}=0. \label{scatt}
\ee
We note that the left side of  (\ref{scatt}) is:  $ -\frac{2k}{3a^3}\mbox{Re}\{{\bf p}^* \times {\bf m}\}$ when the  $r^{-1}$ dependent far-zone terms are excluded from ${\bf E}^{(s)}$ and ${\bf B}^{(s)}$ in (\ref{app1}) and (\ref{app2}).

Since the dipole is small versus the wavelength, we work with   ${\bf E}^{(i)}({\bf r})$ and  ${\bf B}^{(i)}({\bf r})$  expanded into a Taylor series around the origin of coordinates  which coincides with the dipole center:
\be
{\bf E}^{(i)}({\bf r})={\bf E}^{(i)}(0)+r[({\bf n}\cdot\nabla){\bf E}^{(i)}({\bf r})]_{r=0}\,,\,\,\,\,
{\bf B}^{(i)}({\bf r})={\bf B}^{(i)}(0)+r[({\bf n}\cdot\nabla){\bf B}^{(i)}({\bf r})]_{r=0}\,. \label{app3}
\ee
On surface integration of (\ref{app4}) there appear terms with factors $k^2a^2\exp(\mp {\ii}ka)$, $(1\pm {\ii}ka)\exp(\mp {\ii}ka)$ and   $(k^2a^2\mp {\ii}ka)\exp(\mp {\ii}ka)$. They stem  from those $(k^2/r)\exp(\mp {\ii}kr)$,   $(1/r^3\pm {\ii}k/r^2)\exp(\mp {\ii}kr)$ and  $(k^2/r \mp {\ii}k/r^2)\exp\mp ({\ii}kr)$, respectively, in the scattered fields (\ref{app1}) and (\ref{app2}). In compatibility with (\ref{app3}) and in order to obtain for the real part of (\ref{app4}) the correct well-known expression \cite{nieto1} which is  a quantity independent of $a$, (since we know that this real part should not depend on the integration contour), we should take these three factors as $ 0$, $1$ and $1$, respectively.  Further studies should confirm, however, whether this  always holds for the imaginary part of (\ref{app4}), or whether  one should include the full above $a$-factors in evaluating this imaginary part.

 In order to illustrate how the terms of (\ref{app4}) involving incident and scattered fields are evaluated, we show as an example the calculation  of the term
\be
-\frac{1}{8\pi}\mbox{Re}\int_{\partial V_0}  d^2 r \mbox{Re} ({\bf E}^{(i)\,*}\cdot {\bf E}^{(s)}){\bf n}
=-\frac{1}{8\pi}\mbox{Re}\int_{\partial V_0}  d^2 r ({\bf E}^{(i)\,*}({ 0})+r[({\bf n}\cdot\nabla){\bf E}^{(i)\,*}({\bf r})]_{{\bf r}=0}\cdot {\bf E}^{(s)}){\bf n}. \,\,\,\,\,\,\,\,\,\label{app5}
\ee
 Using spherical coordinates,  $d^2 r=r^2 d\phi \sin\theta d\theta$, 
 ${\bf n}=(\sin\theta\cos\phi, \sin\theta\sin\phi,\cos\theta)$. The part of ${\bf E}^{(s)}$ that contributes to  the force on the electric dipole and yields a non-zero  integral in  (\ref{app5}) reduces on using a framework such that  ${\bf p}=(0,0,p)$  in  (\ref{app1}), to
\be
-\frac{1}{8\pi}\mbox{Re}\int_{\partial V_0}  d^2 r \mbox{Re} ({\bf E}^{(i) \,*}\cdot {\bf E}^{(s)}){\bf n}
=-\frac{1}{8\pi}\mbox{Re}\int_0^{2\pi} a^2 (\frac{1}{a^3} - \frac{{\ii}k}{a^2})\exp ({\ii}ka) d\phi \sin\theta \,\nonumber 
\\
\times \int_0^{\pi}d\theta(\sin\theta\cos\phi, \sin\theta\sin\phi,\cos\theta)\,\{[3p\cos\theta(\sin\theta\cos\phi, \sin\theta\sin\phi,\cos\theta)\nonumber\\
-(0,0,p)]\cdot[1+a(\sin\theta\cos\phi\, \partial_x+ \sin\theta\sin\phi \, \partial_y
+\cos\theta\, \partial_z)](E_{i\,x}^* ({\bf r}),E_{i\,y}^* ({\bf r}),E_{i\,z}^* ({\bf r})\}_{r=a}
\nonumber \\
=-\frac{1}{8\pi}\mbox{Re}\{ (1 - {\ii}ka)\exp ({\ii}ka)\frac{4\pi p}{5}\,[\partial_z E_{i\,x}^*(0)+\partial_x E_{i\,z}^*(0) -\frac{5}{3}\partial_xE_{i\,z}^*(0)\,,\nonumber \\
\partial_z E_{i\,y}^*(0)\, \,
+\partial_y E_{i\,z}^*(0)-\frac{5}{3}\partial_yE_{i\,z}^*(0),\,
\partial_x E_{i\,x}^*(0)+\partial_y E_{i\,y}^*(0)+(3-\frac{5}{3})\partial_zE_{i\,z}^*(0)]\}\,\nonumber \\
=\frac{1}{5}\mbox{Re}\{\,p[\frac{1}{3}\partial_x E_{i\,z}^*(0)-\frac{1}{2}\partial_z E_{i\,x}^*(0)\, ,
\frac{1}{3}\partial_y E_{i\,z}^*(0)-\frac{1}{2}\partial_z E_{i\,y}^ *(0)\, ,-\frac{1}{6}\partial_z E_{i\,z}^*(0)]\}.\,\,\,\,\,\,\,\,\,\,\, \label{app6}
\ee
In all above expressions, and in subsequent calculations, we employ the shortened notation: $\partial_kE_{i l}(0)$ for  $[\partial_k E_l^{(i)}({\bf r})]_{{\bf r}=0}$, \,\, $(k,l=x,y,z)$. Notice that we considered $ka=0$ in obtaining the last line of   (\ref{app6})  by  shrinking $V_0$ to its center point. This is  justified  for the RMST only, since its  integration  is independent of  the sphere size.

By analogous calculations with the other terms of (\ref{app4}) contributing to the field scattered by the electric dipole,  one gets
\be
\frac{1}{8\pi}\int_{\partial V_0}  d^2 r ({\bf E}^{(i) \,*} \cdot {\bf n}){\bf E}^{(s)}=\nonumber \\\frac{p}{10}(1 - {\ii}ka)\exp ({\ii}ka)[\partial_x E_{i\,z}^*(0)\,+\partial_z E_{i\,x}^*(0)\, , \partial_y E_{i\,z}^*(0)\,+\partial_z E_{i\,y}^*(0)\,, 2 \partial_z E_{i\,z}^*(0)\,]\nonumber \\
-\frac{p}{30}k^2a^2\exp ({\ii}ka)[\partial_z E_{ix}^*+\partial_x E_{iz}^*\,,\partial_z E_{iy}^*+\partial_y E_{iz}^*\,, 2\partial_z E_{iz}^*\,]\,,\,\,\,\,\,\,\,\,\, \label{app7}
\ee
The second  term of (\ref{app7}), with the factor $k^2a^2$, is the contribution of the radiative part of  ${\bf E}^{(s)}$.
Also,
\be
\frac{1}{8\pi}\int_{\partial V_0}  d^2 r ({\bf E}^{(s)\,*} \cdot {\bf n}){\bf E}^{(i)}=\frac{p\,^*}{3}(1 + {\ii}ka)\exp (-{\ii}ka)\partial_z [E_{i\,x}(0)\, , E_{i\,y}(0)\,, E_{i\,z}(0)\,]. \label{app8}
\ee
In addition, in the magnetic part of the complex MST there is the term of (\ref{app2}) contributing to the magnetic field scattered by the induced electric dipole  ${\bf p}$, which is that of the intermediate-field region: $\frac{{\ii}ke^{{\ii}kr}}{r^2}{\bf n}\times {\bf p}$. By using the Maxwell equation $\nabla\times {\bf E}={\ii}k{\bf B}$ it yields
\be
-\frac{1}{8\pi}\int_{\partial V_0}  d^2 r \mbox{Re}\{{\bf B}^{(i)}\cdot {\bf B}^{(s)\,*}\}{\bf n}=\mbox{Re}\{\frac{p}{6}(\partial_x E_{i\,z}^*(0)-\partial_z E_{i\,x}^*(0)\, , \partial_y E_{i\,z}^*(0)-\partial_z E_{i\,y}^*(0), 0)\},\,\,\,\,\,\,\,\,\, \label{app9}
\ee
and
\be
\frac{1}{8\pi}\int_{\partial V_0}  d^2 r ({\bf B}^{(i)}\cdot {\bf n}){\bf B}^{(s)\,*}=\,\,\,\,\,\,\,\,\,\,\,\,\,\,\,\,\,\, \nonumber \\\frac{p^*}{6}(k^2a^2-{\ii}ka)\exp(-{\ii}ka)(\partial_x E_{i\,z}(0)-\partial_z E_{i\,x}(0)\, ,
\partial_y E_{i\,z}(0)-\partial_z E_{i\,y}(0), 0).\,\,\,\,\,\,\,\,\,
\label{app10}
\ee
Adding  (\ref{app6}) - (\ref{app10}), after taking their real part,   expressing ${\bf p}$ with arbitrary Cartesian components $p_j$,  $(j=1,2,3)$, one gets making $ka=0$ in (\ref{app7})  and (\ref{app8}), as well as  the $ka$-factor equal to one in (\ref{app10}), and  dropping the subscript  $i$ of the incident field
\be
<\bm{\mathcal F}_k>=\frac{1}{2}\mbox{Re}\{ p_{j}\,\partial_k E_{j}^*\},\,\,\,\,\,\,\,\,\, (j,k=1,2,3), \label{app11}
\ee
  Equation (\ref{app11}) is  the well-known time-averaged force on  an electric dipole \cite{patric2000,nieto1}. The  corresponding force on the magnetic dipole is obtained in an analogous way. Then the above near-field calculation yields the expression for the time-averaged force on a magnetoelectric dipole, which is well-known \cite{nieto1,chaumet},
\be
<\bm{\mathcal F}_k>=\frac{1}{2}\mbox{Re}\{ p_{j}\,\partial_k E_{j}^*+m_{j}\,\partial_k B_{j}^*
\}-\frac{k^4}{3}\mbox{Re}[{\bf p}\times{\bf m}^*]_k\,\,.\,\,\,\,\,\,\,\,\, (j,k=1,2,3), \label{app11a}
\ee
On the other hand,   taking the imaginary part in (\ref{app4}), and using  (\ref{app7}) - (\ref{app10}),  one obtains on the electric dipole
\be
\mbox{Im}\{ \int_{\partial V_0}d^2 r \, T_{kj}^{(mix)}({\bf p}) n_j\}=
\mbox{Im}\{[\frac{1}{10}(1-{\ii}ka)-\frac{1}{30}k^2a^2]\exp({\ii}ka)\,{p_j}\,[\partial_k E_j^{*}+ \partial_j E_{k}^{*}]
 \,\,\,\,\,\,\,\ \,\,\,\,\,\,\, \,\,\,\,\,\,\,\nonumber \\  
+\frac{1}{3}(1 + {\ii}ka)\exp (-{\ii}ka)\,{p_j^*}\partial_j E_{k}
+\frac{1}{6}(k^2a^2-{\ii}ka)\exp(-{\ii}ka)p_j^*(\partial_k E_{j}-\partial_j E_{k})\}\, 
. \,\, (j,k=1,2,3). \,\,\,\,\,\,\,\,\,\,\,\,\,\,\label{app12}
\ee
Where we have denoted as $T_{kj}^{(mix)}({\bf p})$  that part of the   CMST, Eq. (\ref{app4}),  that uniquely  involves the electric dipole moment ${\bf p}$.  The superscript $(mix)$ indicates that only interferences  incident-scattered field contribute to the IMST

Equation (\ref{app12}) is the flow IMST on an electric dipole. One approximation that simplifies (\ref{app12}) is  to consider all  parenthesis factors equal to $1$ and the term,  which comes from the far-field, with the $-(1/30)k^2 a^2$ factor as $-1/30$;  so that one gets an expression independent of $a$.
\be
\mbox{Im}\{ \int_{\partial V_0}d^2 r \, T_{kj}^{(mix)}({\bf p}) n_j\}=
-\frac{1}{10}\mbox{Im}\{ [p_j\,\partial_k E_{j}^{*}+ p_j\,\partial_j E_{k}^{*}]\}. \,\,\,\,\,\,\,\,\, (j,k=1,2,3). \,\,\,\,\,\,\,\,\,\,\,\,\label{app13}
\ee
Alternatively, one would obtain an expression akin to (\ref{app13}), but with a factor $-1/15$ instead of $-1/10$, by neglecting in (\ref{app12}) the term with the $-(1/30)k^2 a^2$. This is a plausible choice after making all parenthesis of (\ref{app12}) equal to one, which amounts to take $ka\simeq 0$; and hence there would be no clear justification of (\ref{app13}). However we should say that a similar question appears in the derivation of (\ref{app11}) and (\ref{app11a}) for the RLF that we know are correct.

The imaginary part of the surface integral $\mbox{Im}\{ \int_{\partial V_0}d^2 r \, T_{kj}^{(mix)}({\bf m}) n_j\}$  is derived in an analogous way, yielding an expression like (\ref{app12}) and (\ref{app13}) with  ${\bf m}^*$  and ${\bf B}$ replacing ${\bf p} $ and ${\bf E}^*$, respectively.  Then the simplified version according to (\ref{app13})  of the sum $\mbox{Im}\{ \int_{\partial V_0}d^2 r \, T_{kj}^{(mix)}({\bf p}) n_j\}+\mbox{Im}\{ \int_{\partial V_0}d^2 r \, T_{kj}^{(mix)}({\bf m}) n_j\}$ is
\be
\mbox{Im}\{ \int_{S}d^2 r \, T_{kj}^{(mix)} n_j\}=\frac{1}{8\pi}\int_{V}d^3 r \, \nabla \cdot \mbox{Im}\{{ E}_{k}{ E}_{j}^{*}+{ B}_{k}^{*}{ B}_{j}\}= \nonumber \\
-\frac{1}{10}\mbox{Im}[p_{j}\,\partial_k E_{j}^*+p_j\,\partial_j E_{k}^* - m_{j}\,\partial_k B_{j}^*-m_j\,\partial_j B_{k}^*]\},\,\,\,\,\,\,\,\,\, (j,k=1,2,3). \,\,\,\,\,\,\,\,\,\,\,\,\label{app14}
\ee
 The sum of  (\ref{app12}) and the analogous for $\{T_{kj}^{(mix)}({\bf m}) n_j\}$, or its simplified version (\ref{app14}) (with a factor $1/15$ rather than $1/10$ if the field radiative part of $r^{-1}$ dependence is  excluded), constitutes the proof of the IMST  of \rr Eq. (51)\bb.

\section{A heuristic obtention of the complex force from a time-harmonic field on an electric and a magnetic  dipole}
\subsection{Electric dipole}
We address the complex force from a time-harmonic electromagnetic field whose analytic signals are $\bm{\mathcal  E}({\bf r},t)=\mathbf{ E}({\bf r)}
\exp(-{\ii}\omega t)$, $\bm{\mathcal  B}({\bf r},t)=\mathbf{ B}({\bf r)}\exp(-{\ii}\omega t)$  on an electric dipole  at ${\bf 
r=0}$ of moment  $\bm{\mathcal  P}({\bf 0},t)=\mathbf{ p}\exp(-{\ii}\omega t)$, ${\bf p}=\alpha_e{\bf E}({\bf 0)}$, 
[$\bm{\mathcal  J}({\bf r},t)= {\bf J}({\bf r)}\exp(-{\ii}\omega t)={d}\bm{\mathcal  P}/{d t}$,   ${\bf J}({\bf r)}=-
{\ii}\omega{\bf p}\delta({\bf r})$]. In absence of magnetic charges, following \cite{patric2000}   we tentatively write:
\be
\bm{ F}_i=\frac{1}{2}[({\bf p}^*\cdot\nabla){\bf E}+{\ii}k{\bf p}^*\times{\bf B}]_i=\frac{1}{2}[\alpha_e {E_j}^*\partial_j{ E}_i+{\ii}k\alpha_e^*\epsilon_{ijk}{E_j}^*{ B}_k],\,\,\,\, (i,j,k=1,2,3).\label{dipa}
\ee
Since $B_k=-\frac{{\ii}}{k}\epsilon_{klm}\partial_lE_m$ and $\epsilon_{ijk}\epsilon_{klm}=\delta_{il}\delta_{jm}-\delta_{im}\delta_{jl}$, Eq. (\ref{dipa}) of the complex force becomes
\be
\bm{ F}_i=\frac{1}{2}\alpha_e^* {E_j}^*\partial_i{ E}_j,\,\,\,\,\,\,\,\,\, (i,j=1,2,3). \label{dipb}
\ee
Whose real and imaginary parts are: the well-known expression of the time-averaged force on an electric dipolar particle \cite{patric2000}:
\be
<\bm{ F}_i>=\frac{1}{2}\mbox{Re}\{\alpha_e {E_j}\partial_i{ E}_j^*\}, \label{dipc}
\ee
and 
\be
\mbox{Im}\{\bm{ F}_i\}=-\frac{1}{2}\mbox{Im}\{\alpha_e {E_j}\partial_i{ E}_j^*\}, \label{dipd}
\ee
that we suggest might rule the imaginary force.

\subsection{Magnetic dipole}
There exist  two possible expressions to be  obtained in the study of the RLF on a magnetic dipole of moment ${\bf m}({\bf r},t)$, depending on whether one models it as  a  close loop of electric current ($cld$)  or as a Gilbert dipole ($mcd$)  due to positive and negative magnetic charges \cite{boyer,vaidman,mcdonald}.

In the first case the time-dependent complex force exerted by a wavefield of  analytic signals  $\bm{\mathcal E}({\bf r},t)$, $\bm{\mathcal B}({\bf r},t)$  on a magnetic dipole  at ${\bf r=0}$ of moment  $\bm{\mathcal M}({\bf r},t)=\mathbf{ m}({\bf r)}\exp(-{\ii}\omega t)=\alpha_m \bm{\mathcal B}({\bf r)}$,  that we suggest following \cite{vaidman} for the RLF, is:

\be
 \bm{\mathcal F}_{cld}({\bf r},t)=\frac{1}{2}[\nabla(\bm{\mathcal M}^*\cdot \bm{\mathcal  B})-\frac{1}{c}\frac{\partial}{\partial t} (\bm{\mathcal M}^*\times \bm{\mathcal E})], \label{fmd1}
\ee
where $\frac{1}{c} (\bm{\mathcal M}^*\times \bm{\mathcal E})$ is the analogous,  in terms of the analytic signals associated to the fields, of.Shockley's {\it hidden momentum} \cite{shockley} 
 
Expanding the first term and using Maxwell's equation: $\nabla\times \bm{\mathcal B}={\partial \bm{\mathcal E}}/{\partial t} +\frac{c}{4\pi}\bm{\mathcal J}$, Eq. (\ref{fmd1}) acquires the form
\be
\bm{\mathcal F}_{cld}({\bf r},t)=\bm{\mathcal F}_{mcd}+\frac{2\pi}{c}\bm{\mathcal M}^*\times \bm{\mathcal J}, \label{fmd2}
\ee
where
\be
\bm{\mathcal F}_{mcd}({\bf r},t)=\frac{1}{2}[(\bm{\mathcal M}^*\cdot\nabla)\bm{\mathcal B}-\frac{1}{c}\frac{\partial \bm{\mathcal M}^*}{\partial t} \times \bm{\mathcal E} ] \label{fmd2a}
\ee
is the complex force on a Gilbert dipole in terms of the analytic signals. Therefore Eq.(\ref{fmd2}) is the relationship between the forces in the two models, $cld$ and $mcd$, of the magnetic dipole.

Now, the complex force $\bm{\mathcal F}_{cld}$ from a {\it time-harmonic} field 
$\bm{\mathcal E}({\bf r},t)=\mathbf{ E}({\bf r)}
\exp(-{\ii}\omega t)$, $\bm{\mathcal B}({\bf r},t)=\mathbf{ E}({\bf r)}\exp(-{\ii}\omega t)$  on a $cld$ magnetic dipole of moment  $\bm{\mathcal M}({\bf 0},t)$   is, since then the second term of (\ref{fmd1}) is zero,
\be
\bm{\mathcal F}_{cld}=\frac{1}{2}[\nabla({\bf m}^*\cdot{\bf B})+{\bf m}^*\times(\nabla\times {\bf B})]. \label{fmd3}
\ee
Where the ${\bf r}$-dependence of all quantities in  (\ref{fmd3}) is implicit. 
Expanding the double vector product of the second term and proceeding as in Appendix E.1, we obtain
\be
(\bm{\mathcal F}_{cld})_i=\frac{1}{2}{ m}_j^*\partial_i{ B_j} , \,\,\,\,\,(i,j=1,2,3). \label{fmd4}
\ee
Whose real part is the well-known  time-averaged force on a purely magnetic dipole \cite{nieto2,chaumet}:
\be
<\bm{\mathcal F}_{cld}>_i=\frac{1}{2}\mbox{Re}\{{ m}_j\partial_i{ B_j}^*\} .\label{fmd5}
\ee
And whose imaginary part is our tentatively proposed reactive force on a  magnetic dipole, 
\be
\mbox{Im}\{\bm{\mathcal F}_{cld}\}_i=-\frac{1}{2}\mbox{Im}\{{ m}_j\partial_i{ B_j}^*
\} ,  \,\,\,\,\,  { m_j}({\bf 0})=\alpha_m { B}_j({\bf 0}), \,\,\,\, (i,j=1,2,3). \label{fmd6}
\ee
On the other hand, the complex force of this time-harmonic wavefield on a Gilbert dipole of magnetic charge current density $\bm{\mathcal J}({\bf r},t)= {\bf J}_{mc}({\bf r)}\exp(-{\ii}\omega t)={d\bm{\mathcal M}}/{d t}$,   ${\bf J}_{mc}({\bf r)}=-{\ii}\omega{\bf m}\delta({\bf r})$, 
 is from (\ref{fmd2a}):
\be
\bm{\mathcal F}_{mcd}=\frac{1}{2}[({\bf m}^*\cdot\nabla){\bf B}-{\ii}{k}{\bf m}^* \times{\bf E}]=
\frac{1}{2}[({\bf m}^*\cdot\nabla){\bf B}+{\bf m}^* \times(\nabla\times {\bf B})-\frac {4\pi}{c}{\bf m}^* \times{\bf J}_{mc}]. \,\,\,\,\,  \label{fmd7}
\ee
Where the  Maxwell equation for $\nabla\times{\bf B}$ has been used to eliminate ${\bf E}$. Then, expressing the electric current density as: ${\bf J}_{mc} =-{\ii}\omega {\bf p}\delta({\bf r})$ and proceeding as before, we obtain:
\be
(\bm{\mathcal F}_{mcd})_i=\frac{1}{2}{ m}_j^*\partial_i{ B}_j+2{\ii}\pi k \delta({\bf r})({\bf m}^* \times{\bf p})_i.  \label{fmd8}
\ee
Notice that since $\int_{-\infty}^{\infty}d^3 {\bf r} \delta({\bf r})=1$, the second term of (\ref{fmd8}) has, like the first term, spatial dimension $L^{-2}$. Also we note that (\ref{fmd4}) and (\ref{fmd8}) hold (\ref{fmd2}) with the $1/2$ factor since $\frac{1}{2}\frac{4\pi}{c}{\bf m}^*\times{\bf J}=-2{\ii}\pi k \delta({\bf r})\,{\bf m}^* \times{\bf p}$.

Therefore, the real and imaginary parts of the force on the Gilbert dipole are respectively:
\be
<\bm{\mathcal F}_{mcd}>_i=\frac{1}{2}\mbox{Re}\{{ m}_j\partial_i{ B}_j^*\}+ 2\pi k \delta({\bf r})\mbox{Im}\{ {\bf p}\times{\bf m}^*\}_i ,  \label{fmd9}
\ee
and 
\be
\mbox{Im}\{\bm{\mathcal F}_{mcd}\}_i=-\frac{1}{2}\mbox{Im}\{{ m}_j\partial_i{ B}_j^*\}- 2\pi k \delta({\bf r})\mbox{Re}\{ {\bf p}\times{\bf m}^*\}_i , \,\,\,\, (i,j=1,2,3)  \label{fmd9}
\ee
At this stage we should remark that for a Gilbert dipole model the Maxwell equation is no longer $\nabla\times{\bf E}={\ii}k{\bf B}$, but  $\nabla\times{\bf E}={\ii}k{\bf B}-\frac{c}{4\pi}{\bf J}_{mc}$, where ${\bf J}_{mc}=-{\ii}\omega{\bf m} \delta({\bf r})$, as seen above. So instead of Eq. (\ref{dipb}), valid for an electric dipole without magnetic charges, the equation for the complex force on the electric dipole if one assumes the existence of magnetic charges is
\be
(\bm{\mathcal F})_{i}=\frac{1}{2}{ p}_j^*\partial_i{ E}_j +
2{\ii}\pi k \delta({\bf r})({\bf m} \times{\bf p}^*)_i.  \label{fmd10}
\ee
Whose real and imaginary parts are 
\be
<\bm{\mathcal F}>_{i}=\frac{1}{2}\mbox{Re}\{{ p}_j\partial_i{ E}_j^*\} -
2\pi k \delta({\bf r})\mbox{Im}\{{\bf p} \times{\bf m}^*\}_i.  \label{fmd11}
\ee
and 
\be
\mbox{Im}\{\bm{\mathcal F}\}_{i}=-\frac{1}{2}\mbox{Im}\{{ p}_j\partial_i{ E}_j^*\} -
2\pi k \delta({\bf r})\mbox{Re}\{{\bf p} \times{\bf m}^*\}_i , \,\,\,\, (i,j=1,2,3) . \label{fmd12}
\ee
\subsection{ELECTRIC AND MAGNETIC DIPOLE}
Therefore, we conclude that if   magnetic charges are assumed, the real and imaginary parts of the resulting electric and magnetic dipole forces are:
\be
<\bm{\mathcal F}>_{i}+<\bm{\mathcal F}_{mcd}>_i=<\bm{ F}>_{i}+<\bm{\mathcal F}_{cld}>_i\,=\frac{1}{2}\mbox{Re}\{{ p}_j\partial_i{ E}_j^*\} +\frac{1}{2}\mbox{Re}\{{ m}_j\partial_i{ B}_j^*\}, \label{fmd13}
\ee
and  our proposed expression:
\be
\mbox{Im}\{\bm{\mathcal F}\}_{i}+\mbox{Im}\{\bm{\mathcal F}_{mcd}\}_i=-\frac{1}{2}\mbox{Im}\{{ p}_j\partial_i{ E}_j^*\} -\frac{1}{2}\mbox{Im}\{{ m}_j\partial_i{ B}_j^*\}-4\pi k \delta({\bf r})\mbox{Re}\{{\bf p} \times{\bf m}^*\}_i.  \label{fmd14}
\ee
While in the model in which no magnetic charges are present, we obtain for the resulting reactive force
\be
\mbox{Im}\{\bm{F}\}_{i}+\mbox{Im}\{\bm{\mathcal F}_{cld}\}_i=-\frac{1}{2}\mbox{Im}\{{ p}_j\partial_i{ E}_j^*\} -\frac{1}{2}\mbox{Im}\{{ m}_j\partial_i{ B}_j^*\}.  \label{fmd15}
\ee
Which in contrast with the time-averaged force which is the same for the $cld$ amd $mcd$ models, (\ref{fmd15})  differs from (\ref{fmd14}) by  $4\pi k \delta({\bf r})\mbox{Re}\{{\bf p} \times{\bf m}^*\}_i$.

This procedure does not yield for the RLF (\ref{fmd13}) the interference term [cf. Eq. (\ref{app11a})]: $-\frac{k^4}{3} \mbox{Re}\{{\bf p}\times {\bf m}^*\}$ \cite{nieto1,chaumet} of the electric and magnetic dipoles. Therefore it is likely that the ILF so obtained, although valid for pure electric or magnetic dipoles, be not valid for magnetoelectric dipolar particles.

From Eq. (\ref{bcfor12}) we guess for the reactive spin torque:
\be
\bm {\Xi}^{spin}=-\frac{1}{5} \mbox{Im}[{\bf p}\times{\bf E}^{*}+{\bf m}\times{\bf B}^{*}]
+\frac{k}{c}\int_{V}d^3 r\, ({\bf L}_{e}^{O} - {\bf L}_{m}^{O})
.   \label{bcfor26}
\ee
Notice that, again, the  diagonal terms of the complex MST do not  contribute to the recoil, or scattering, component \cite{nieto1} of  $\bm {\Xi}^{spin}$.

Or in terms of the above quoted the electric and magnetic angular momenta: ${\mathscr{ F}}_e $ and ${\mathscr{ F}}_m $, we write
\be
\bm {\Xi}^{spin}=-\frac{k}{c}\{\frac{1}{5}( \alpha_e^I {\mathscr{ F}}_e + \alpha_m^I{\mathscr{ F}}_m)-\int_{V}d^3 r\, ({\bf L}_{e}^{O} - {\bf L}_{m}^{O}) \}.   \label{bcfor27}
\ee
Of course the angular orbital momenta, like we saw above for the orbital momenta, store power of the propagating plane wave components of the wavefields.

\rr

\section{The reactive force from an incident plane wave with spin angular momentum}

\begin{figure}[htbp]
\begin{centering}
\includegraphics[width=16cm]{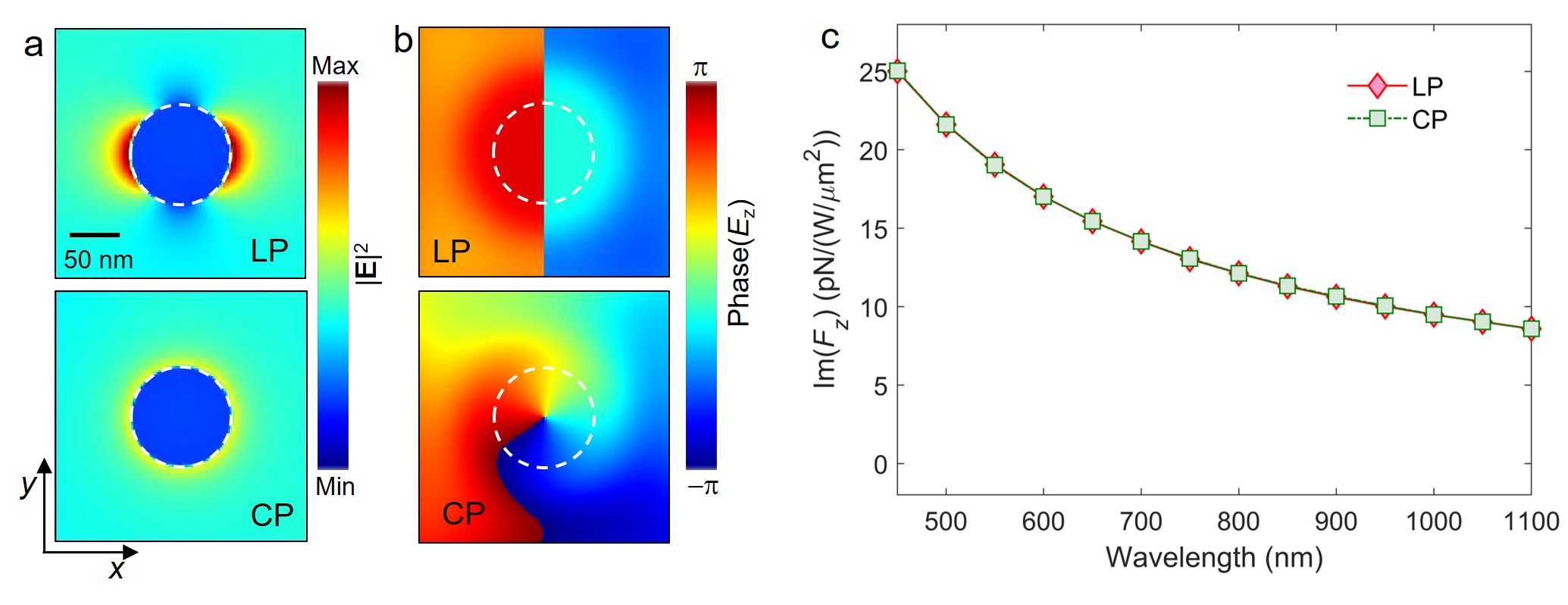}
\par\end{centering}
\caption{\rr Comparison of 
field distribution and reactive force for the PS particle of Fig.2 when the incident plane wave is linearly polarized (LP), [amplitude $E_0(1,0,0),\, E_0=1$], and circularly polarized (CP) [amplitude $E_0(1,{\rm i},0),\, E_0=1/\sqrt{2}]$, at the wavelength of 800\,nm. (a) Field intensity distribution. (b) Phase profile of the $z$-component of the electric field  in and around the particle after scattering,  for LP and CP illumination. Dashed lines outline the particle contour. The sharp vertical line in the upper right figure indicates that for LP illumination, after the light-particle interaction, the resulting electric field $z$-component has opposite signs on the left and right side of the $x = 0$ plane. (Notice that the space shown in the figures is the $XY$-plane where the particle center is located). (c) ILF spectrum for LP and CP light.\bb} 
\end{figure}

\rr Figure S1  compares results from a  linearly polarized (LP) plane wave  with those of circular polarization (CP), both incident on the PS particle of Example 2, and propagating along $OZ$. As an example, Fig. S1(a) depicts the field intensity in and around the PS particle for CP and LP illumination. Under LP light the field has a butterfly-shape intensity distribution, expected from a dipolar particle, while the intensity excited by CP light is uniform along the azimuthal direction, because of the rotational symmetry of the system. On the other hand, the interaction of particle with plane waves yields a longitudinal component of  the field,  whose phase  remarkably discriminates LP and CP illumination. As shown in Fig. S1(b), the LP light leads to a phase jump at the $x = 0$ plane; while CP illumination yields a phase distribution with a singular point, or vortex, which is known as the result of the spin-to-orbit coupling. However, despite these significant differences in the characteristics of the fields, the ILF produced by the CP illumination is  identical to that from LP light [cf. Fig. 2(b)], as illustrated in Fig. S1(c).  \bb

\section{ On the calculation of the $\bm{ROM}$}
Both   $<\bm{\mathcal F}>$ and $\mbox{Im}\{\bm{\mathcal F}\}$ are calculated from the complex Lorentz integral  in $V_0$, Eq.(6). The induced densities $\rho$ and  ${\bf J}$ are obtained from the constitutive relations on the total field
$ {\bf E}={\bf E}^{(i)}+{\bf E}^{(s)}$, $ {\bf B}={\bf B}^{(i)}+{\bf B}^{(s)}$ via the first and fourth Maxwell equations, respectively. 

In our numerical method, the total field is obtained using the commercial software package “FDTD Solutions” (Lumerical, Inc.). The simulation region is $0.22 \times 0.22 \times 0.22$ $ \rm\mu m^3$, and a uniform mesh with grid size of 5 nm was used. We compute the complex Lorentz force by the expression: $ \bm{\mathcal F}=\frac{1}{2}\int_{V} [ \rho^*{\bf E}({\bf r})+\frac{\bf J^*}{c}\times{\bf B}({\bf r})]\,d^3 r$, determining $\rho$ and ${\bf J}$  via the procedure described in \cite{lumerics}. Then we evaluate the IMST across the surface of a cube that encloses the spherical dipolar particle of volume $V_0$. Let the volume of this cube be denoted  $V_q$. The volume of the four corners between $V_q$ and $V_0$ is $V_{4c}$, its surface being $\partial V_{4c}$
         
Obviously since $ \rho=0$ and ${\bf J}=0$ outside $V_0$, we have from the total field  $ {\bf E}={\bf E}^{(i)}+{\bf E}^{(s)}$, $ {\bf B}={\bf B}^{(i)}+{\bf B}^{(s)}$ the following:
\be
0=\mbox{Im}\{\bm{\mathcal F}\}_{4c}= \int_{\partial V_{4c}}d^2 r \, \mbox{Im} \{T_{ij}\}n_j+{\ii}\omega\int_{V_{4c}}d^3 
r\, [{\bf P}_{m}^{O}- {\bf P}_{e}^{O}]_i\,, \,\,\,  \label{bcfor8g}
\ee
I.e.
\be
 \int_{\partial V_{4c}}d^2 r \, \mbox{Im}\{T_{ij}\} n_j=-{\ii}\omega\int_{V_{4c}}d^3 r\, 
[{\bf P}_{m}^{O}- {\bf P}_{e}^{O}]_i\, . \,\,\,  \label{bcfor8h}
\ee
But
\be
 \int_{\partial V_{4c}}d^2 r \, \mbox{Im}\{T_{ij}\} n_j= \int_{\partial V_q- \partial V_0}d^2 r \, T_{ij}^{} n_j  \,. \,\,\,  \label{bcfor8i}
\ee
Then from (\ref{bcfor8h}):
\be
{\ii}\omega\int_{V_{q}}d^3 r\, [{\bf P}_{m}^{O}- {\bf P}_{e}^{O}]_i={\ii}\omega\int_{V_{0}}d^3 r\, [{\bf P}_{m}^{O}- {\bf P}_{e}^{O}]_i - \int_{\partial V_{4c}}d^2 r \, \mbox{Im}\{T_{ij}\} n_j\,. \label{bcfor8j}
\ee
And from (\ref{bcfor8i})  and  (\ref{bcfor8j})   one derives:
\be
{\ii}\omega\int_{V_{q}}d^3 r\, [{\bf P}_{m}^{O}- {\bf P}_{e}^{O}]_i={\ii}\omega\int_{V_{0}}d^3 r\, [{\bf P}_{m}^{O}- {\bf P}_{e}^{O}]_i + \int_{\partial V_{0}-\partial V_{q}}d^2 r \, \mbox{Im}\{T_{ij}\} n_j\,. \label{bcfor8k}
\ee
Which taking into account \rr(57) \bb for $V= V_q$, leads to 
\be
{\ii}\omega\int_{V_{q}}d^3 r\, [{\bf P}_{m}^{O}- {\bf P}_{e}^{O}]_i=\mbox{Im}\{\bm{\mathcal F}_i\}
- \int_{\partial V_{q}}d^2 r \,\mbox{Im}\{T_{ij} \} n_j\,. \label{bcfor8l}
\ee
Equation (\ref{bcfor8l}) is the procedure we use to calculate the  ROM integrated in the volume $V_q$ from the two terms of its right side previously computed as described above.

\end{document}